\documentclass[12pt, draftclsnofoot, hidelinks, onecolumn]{IEEEtran}
\usepackage{epsfig,graphicx,subfigure,psfrag,amsmath,cases}
\usepackage{latexsym,amssymb,amsmath,epsfig,subfigure,algorithm,mathtools}
\usepackage{algorithmic}
\usepackage{color}
\usepackage{url}
\usepackage{scrtime}
\usepackage{cite}
\usepackage{subfigure}
\usepackage{bbding}
\usepackage{multicol}
\usepackage{bm}
\usepackage{xcolor}
\usepackage[framemethod=TikZ]{mdframed}
\usetikzlibrary{shadows}
\usepackage{environ}
\usepackage{varwidth}

\newlength{\MyMdframedWidthTweak}%
\NewEnviron{MyMdframed}[1][]{%
    \setlength{\MyMdframedWidthTweak}{\dimexpr%
        +\mdflength{innerleftmargin}
        +\mdflength{innerrightmargin}
        +\mdflength{leftmargin}
        +\mdflength{rightmargin}
        }%
    \savebox0{%
        \begin{varwidth}{\dimexpr\linewidth-\MyMdframedWidthTweak\relax}%
            \BODY
        \end{varwidth}%
    }%
    \begin{mdframed}[
        backgroundcolor=lightgray,
        shadow=true,
        shadowsize=4pt,
        roundcorner=5pt,
        userdefinedwidth=\dimexpr\wd0+\MyMdframedWidthTweak\relax,
        #1]
        \usebox0
    \end{mdframed}
}

\author{Zhiqiang Wei,~\IEEEmembership{Member,~IEEE,} Weijie Yuan,~\IEEEmembership{Member,~IEEE,} Shuangyang Li,~\IEEEmembership{Student Member,~IEEE,} Jinhong Yuan,~\IEEEmembership{Fellow,~IEEE,} and Derrick Wing Kwan Ng,~\IEEEmembership{Fellow,~IEEE}\vspace{-18mm}
	\thanks{Zhiqiang Wei, Weijie Yuan, Shuangyang Li, Jinhong Yuan, and Derrick Wing Kwan Ng are with the School of Electrical Engineering and Telecommunications, the University of New South Wales, Australia (email: zhiqiang.wei; weijie.yuan; shuangyang.li; j.yuan; w.k.ng@unsw.edu.au).}}

\title{Off-grid Channel Estimation with Sparse Bayesian Learning for OTFS Systems}

\newtheorem{proof}{proof}

\newtheorem{T-Prob}{Transformed Problem}
\newtheorem{proposition}{Proposition}

\DeclareMathOperator{\maxo}{maximize}
\DeclareMathOperator{\mino}{minimize}
\DeclareMathOperator{\diag}{\mathrm{diag}}

\newtheorem{Remark}{Remark}

\newcommand{\abs}[1]{\lvert#1\rvert}

\begin{document}
\maketitle
\begin{abstract}
This paper proposes an off-grid channel estimation scheme for orthogonal time-frequency space (OTFS) systems adopting the sparse Bayesian learning (SBL) framework.
To avoid channel spreading caused by the fractional delay and Doppler shifts and to fully exploit the channel sparsity in the delay-Doppler (DD) domain, we estimate the original DD domain channel response rather than the effective DD domain channel response as commonly adopted in the literature. 
OTFS channel estimation is firstly formulated as a one-dimensional (1D) off-grid sparse signal recovery (SSR) problem based on a virtual sampling grid defined in the DD space, where the on-grid and off-grid components of the delay and Doppler shifts are separated for estimation.
{In particular, the on-grid components of the delay and Doppler shifts are jointly determined by the entry indices with significant values in the recovered sparse vector.}
Then, the corresponding off-grid components are modeled as hyper-parameters in the proposed SBL framework, which can be estimated via the expectation-maximization method.
To strike a balance between channel estimation performance and computational complexity, we further propose a two-dimensional (2D) off-grid SSR problem via decoupling the delay and Doppler shift estimations.
In our developed 1D and 2D off-grid SBL-based channel estimation algorithms, the hyper-parameters are updated alternatively for computing the conditional posterior distribution of channels, which can be exploited to reconstruct the effective DD domain channel.
Compared with the 1D method, the proposed 2D method enjoys a much lower computational complexity while only suffers a slight performance degradation.
Simulation results verify the superior performance of the proposed channel estimation schemes over state-of-the-art schemes.
\end{abstract} 

\vspace{-4mm}
\section{Introduction}
Future wireless systems are envisioned to enable a wide range of emerging mobile applications, such as low-earth-orbit satellites (LEOS), autonomous cars, in-vehicle infotainment, and unmanned aerial vehicles (UAV) \cite{wei2020orthogonal}.
As the airborne, spaceborne, and vehicles are usually fast moving, the associated severe Doppler spread experienced by channels in high mobility propagation environments imposes new challenges on its air interface design.
Recently, a new modulation waveform, namely orthogonal time-frequency space (OTFS), has been proposed as a promising candidate for realizing high-mobility communications owing to its intrinsic advantages of handling the problems caused by Doppler effect\cite{Hadani2017orthogonal,li2020performance}.

In contrast to the existing one-dimensional (1D) time-domain or frequency-domain modulation techniques, such as orthogonal frequency-division multiplexing (OFDM), OTFS is a two-dimensional (2D) modulation scheme, where the information symbols are carried over 2D localized pulses defined in the delay-Doppler (DD) domain.
In fact, each information symbol multiplexed in the DD domain is spread across the whole time-frequency (TF) domain, allowing the potentials of exploiting the full diversity to achieve reliable communications against various channel impairments.
In addition, the channel in the DD domain exhibits various beneficial properties, such as separability, stability, and possibly sparsity \cite{wei2020orthogonal}, which can be exploited for efficient channel estimation and data detection.
In particular, processing the received signals in the Doppler domain of wireless channels allows us to separate the propagation paths experiencing an identical delay.
Besides, OTFS modulation can transform a time-variant channel in the TF domain into a 2D quasi-time-invariant channel in the DD domain and thus relieves the impacts caused by channel aging \cite{Hadani2017orthogonal,li2020performance}.
Additionally, the DD domain effective channel is generally sparse in an open-space rural propagation environment with a limited number of moving scatters.
As a result, the coupling relationship between data symbols and the channel in the DD domain is much simpler than the counterpart in the TF domain, which is crucial for accurate channel acquisition and efficient detection algorithm design.

Since OTFS multiplexes information symbols in the DD domain, it enables the possibility of channel probing via designing a new training signaling in the DD domain\cite{RavitejaOTFSCE,KollengodeMIMOOTFSDetectionCE}, which is a challenging but vital requirement for reliable detection.
In particular, channel spreading \cite{WeiWindowOTFS} caused by fractional Doppler and delay sacrifices the channel sparsity which degrades the channel estimation performance.
Usually, a guard space is inserted between the pilot and data symbols to mitigate the interference of unknown data symbols for channel estimation, which incurs a significant signaling overhead\cite{RavitejaOTFSCE}.
In fact, the channel estimation problem of OTFS was firstly studied in \cite{RavitejaOTFSCE}, where a single impulse was embedded in the DD domain and a threshold-based estimation method was proposed.
Then, its extension to multiple-input multiple-output OTFS (MIMO-OTFS) systems was investigated in \cite{KollengodeMIMOOTFSDetectionCE} via positioning the pilot impulses for different  transmit antennas with sufficient guard spacing in the DD domain.
However, their channel estimation performances are sensitive to the availability of guard space \cite{RavitejaOTFSCE,KollengodeMIMOOTFSDetectionCE}.
Indeed, to exploit the channel sparsity in the DD domain, more advanced channel estimation approaches based on compressed sensing techniques have been proposed in the literature.
For instance, the authors in \cite{ShenCEMassiveMIMO} proposed a three-dimensional (3D) structured orthogonal matching pursuit (SOMP) algorithm to estimate the delay-Doppler-angle domain channel via exploiting the 3D structured sparsity.
%
%But their proposed channel estimation model \cite{ShenCEMassiveMIMO} is only applicable to the integer Doppler case and thus the channel spreading caused by fractional Doppler has not been taken into account.
%
A further extension in \cite{LiPDMAOTFS} proposed a 3D Newtonized OMP (NOMP) algorithm, which can extract the fractional component in the Doppler and angle domains via Newton's method.
Yet, its channel estimation performance is mainly limited by the greedy OMP approach, which is generally far away from that achieved by the optimal estimation.
Besides, a most recent work in \cite{ZhaoSBLOTFS} applied the sparse Bayesian learning approach for the channel estimation in OTFS systems, which outperforms the OMP methods \cite{ShenCEMassiveMIMO,LiPDMAOTFS}.
Nevertheless, the proposed scheme in \cite{ZhaoSBLOTFS} only applies to the integer Doppler case and is essentially an on-grid method, where all the estimated Doppler shifts are constrained on the coarse DD domain grid, which is only valid in limited practical scenarios.

In practice, the DD domain channel sparsity may not always hold \cite{WeiWindowOTFS}, where conventional compressed sensing-based channel estimation cannot achieve a satisfactory performance.
In particular, due to the limited bandwidth and the limited frame duration, the exact Doppler frequency shift or delay shift may straddle a pair of finite-resolution bins, rather than falling exactly into a single bin, which is known as off-grid.
In this case, the effective DD domain channel is spread across all the Doppler and delay indices, when off-grid happens in the Doppler and delay domains, respectively.
As a result, although the \textit{original DD domain channel response} is sparse, the \textit{effective DD domain channel} may not be sparse due to channel spreading.
On the other hand, although increasing the sampling resolution in the DD domain is a viable approach to suppress the impacts caused by channel spreading, it consumes more bandwidth and introduces longer latency.
As a result, various approaches have been proposed as a remedy.
For instance, our previous work \cite{WeiWindowOTFS} proposed to apply a Dolph-Chebyshev (DC) window to suppress channel spreading and thus to enhance the effective DD domain channel sparsity, which substantially improves the channel estimation performance compared with \cite{RavitejaOTFSCE}.
Yet, \cite{WeiWindowOTFS} still estimates the effective DD domain channel, rather than the original DD domain channel response, and thus remains suffering from the inevitable channel spreading.

In this paper, we formulate an off-grid sparse signal recovery (SSR) problem and propose to employ a sparse Bayesian learning (SBL) framework to estimate the original DD domain channel response for OTFS systems.
In particular, to avoid channel spreading and to fully reap the potential sparsity, we aim to estimate the original DD domain channel response directly rather than the effective DD domain channel, which is fundamentally different from \cite{ShenCEMassiveMIMO,ZhaoSBLOTFS}.
Besides, the proposed off-grid channel estimation scheme is essentially a semi-parametric estimation approach.
In particular, the exact Doppler and delay shifts can be estimated via their nearest grid indices in the recovered sparse vector while the corresponding off-grid components can be estimated via some parametric approach, such as the expectation-maximization (EM) algorithm.
There are three advantages of the proposed SBL-based channel estimation approaches. 
Firstly, the hyperprior structure of SBL 
\cite{YangZaiOffgridCE} is flexible such that the channel compactness and sparsity in the DD domain can be easily exploited.
Secondly, SBL can output the posterior distribution of the original DD domain channel response, where the mean denotes the channel estimate and the variance determines the channel estimation accuracy.
Thirdly, SBL avoids the use of sparse regularization parameters \cite{BabacanBCS} and it is not sensitive to the other initialization parameters, such as the number of paths and noise power.
As a result, the proposed channel estimation approach significantly outperforms the algorithm in \cite{LiPDMAOTFS}, which is the only work of compresses sensing-based channel estimation so far that can handle the fractional Doppler case.
The main contributions of this work are given as follows:
\begin{itemize}
	\item By characterizing the DD domain input-output relationship, we first reformulate the channel estimation of OTFS systems as a one-dimensional (1D) SSR problem, where we directly estimate the original DD domain response rather than the effective DD domain channel, e.g. \cite{RavitejaOTFSCE,KollengodeMIMOOTFSDetectionCE,WeiWindowOTFS}. Moreover, we establish a 1D off-grid compressed channel estimation model based on the first-order linear approximation, which enables us to improve the delay and Doppler shift estimation accuracy while maintaining a low computational complexity. The developed model serves as a generalized platform which is applicable to different pilot and data arrangements, arbitrary windows, as well as pulse shaping filters adopted at the OTFS transceivers.
	\item We first propose a 1D off-grid SBL-based channel estimation algorithm, which can estimate both the on-grid and off-grid components of the delay and Doppler shifts. In particular, the on-grid component is obtained by the entry indices with significant values in the recovered sparse vector. The corresponding off-grid components are treated as hyper-parameters in the SBL framework which can be estimated via an iterative algorithm based on the  expectation-maximization approach.
	\item To further reduce the computational complexity, we decouple the delay and Doppler shift estimation and develop a 2D off-grid compressed channel estimation model. A two-step algorithm is proposed to estimate the Doppler and delay shifts, respectively, where the first-step is realized via applying SBL to multiple snapshot cases and the second step is achieved by solving multiple 1D SSR problems in parallel. The computational complexity order of the proposed 2D off-grid method is proportional to the summation of the sizes of delay and Doppler grids, while the counterpart of the proposed 1D off-grid method is proportional to their product.
	\item Extensive simulations are conducted to evaluate and compare the performance of our proposed schemes. We demonstrate that our proposed 1D off-grid SBL-based scheme can achieve the most accurate channel estimates compared with the state-of-the-art schemes in the literature \cite{RavitejaOTFSCE,WeiWindowOTFS,ShenCEMassiveMIMO,LiPDMAOTFS,ZhaoSBLOTFS}. Also, the performance of our proposed 2D off-grid SBL-based scheme approaches closely to that of the 1D method, with a significantly reduced computational complexity.
\end{itemize}

{\textit{Notations:} Boldface capital and lower case letters are reserved for matrices and vectors, respectively; ${\left( \cdot \right)^{\mathrm{H}}}$ denotes the Hermitian transpose of a vector or a matrix; ${\left( \cdot \right)^{\mathrm{T}}}$ denotes the transpose of a vector or a matrix; ${\left( \cdot \right)^{-1}}$ denotes the inverse of a square matrix; $\mathbb{A}$ denotes the constellation set; $\mathbb{Z}^{+}$ denotes the set of all non-negative integers;
	$\mathbb{C}^{M\times N}$ denotes the set of all $M\times N$ matrices with complex entries;
	$\abs{\cdot}$ denotes the absolute value of a complex scalar or the cardinality of a set;
	$E_{X}\{\cdot\}$ denotes the expectation of the input with respect to (w.r.t.) the distribution of $X$;
	%
	%$\ast$ and $\circledast$ denote the convolution and circular convolution operations, respectively;
	$\left(\cdot\right)^*$ denotes the conjugate operation;
	$\otimes$ is the Kronecker product operator; $\odot$ is the point-wise product operator; $\left(\cdot\right)_N$ denotes the modulus operation w.r.t. $N$;
	$\diag\{\cdot\}$ returns a square diagonal matrix with the elements of input vector on the main diagonal;
	$\Re\{\cdot\}$ returns the real part of the input complex number;
	$\lfloor\cdot\rfloor$ is the floor function which returns the largest integer smaller than the input value;
	$\lceil\cdot\rceil$ is the ceiling function which returns the smallest integer larger than the input value;
	%$\left[x\right]^+ = \max\left\{0,x\right\}$;
	$\mathbf{I}_M$ denote the identity matrix of size $M \times M$; 
	$\{{\mathbf{X}}\}_n$ denotes the $n$-th column of the matrix $\mathbf{X}$ and $\{{\mathbf{X}}\}_{m,n}$ denotes the $\left[m,n\right]$-th entry of the matrix $\mathbf{X}$;
	$\{{\mathbf{x}}\}_n$ denotes the $n$-th entry of the vector $\mathbf{x}$ and $\{{\mathbf{x}}\}_{-n}$ denotes the vector without the $n$-th entry of $\mathbf{x}$.
	The big-O notation $\mathcal{O}\left(\cdot\right)$ asymptotically describes the order of computational complexity.
	The circularly symmetric complex Gaussian distribution with mean $\boldsymbol{\mu}$ and covariance matrix $\boldsymbol{\Sigma}$ is denoted by ${\cal CN}(\boldsymbol{\mu},\boldsymbol{\Sigma})$;
	the uniform distribution between $a$ and $b$ is denoted by $\mathcal{U}\left[a,b\right]$;
	$\sim$ stands for ``distributed as''; $\propto$ denotes ``proportional to''.
	The probability density function (PDF) for a random complex-valued vector ${\bf{x}} \sim {\cal CN}(\boldsymbol{\mu},\boldsymbol{\Sigma})$ is defined as $\mathcal{CN}\left( {{\bf{x}}\left| {{\boldsymbol{\mu }},{\bf{\Sigma }}} \right.} \right) = \frac{1}{{{\pi ^N}\left| {\bf{\Sigma }} \right|}}\exp \left\{ { - {{\left( {{\bf{x}} - {\boldsymbol{\mu }}} \right)}^{\rm{H}}}{{\bf{\Sigma }}^{ - 1}}\left( {{\bf{x}} - {\boldsymbol{\mu }}} \right)} \right\}$.
	The PDF for a random variable $x$ follows a Gamma distribution is defined as $\Gamma \left( {x\left| {a,b} \right.} \right) = \frac{{b^a}{x^{a - 1}}\exp \left\{ { - bx} \right\}}{{ {\Gamma \left( a \right)} }}$, where constants $a > 0$ and $b > 0$, and ${\Gamma \left( a \right)}$ is a Gamma function.}

\begin{figure}
	\centering
	\includegraphics[width=4.5in]{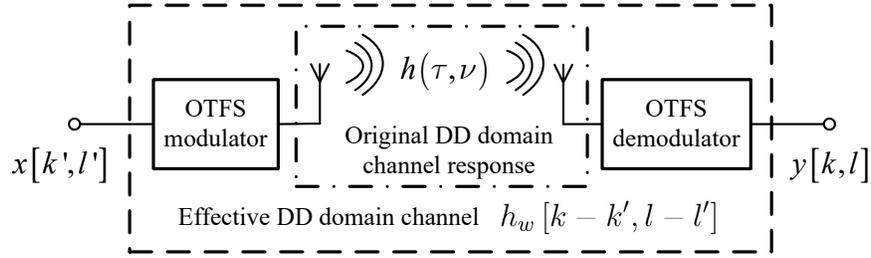}\vspace{-7mm}
	\caption{OTFS system model in the DD domain.}\vspace{-10mm}
	\label{OTFSSystemModel}%
\end{figure}

\vspace{-2mm}
\section{System Model}
For ease of presentation, as shown in Fig. \ref{OTFSSystemModel}, we directly present the DD domain input-output relationship of OTFS following the well-known derivations in the literature\cite{RavitejaOTFS,wei2020orthogonal,WeiWindowOTFS,LiPDMAOTFS}:
\vspace{-2mm}
\begin{equation}\label{IOOTFS}
y\left[ {k,l} \right] =  \sum\limits_{k' = 0}^{N - 1} {\sum\limits_{l' = 0}^{M - 1} {x\left[ {k',l'} \right]} } {h_w}\left[ {k - k',l - l'} \right] + z\left[k,l\right],\vspace{-2mm}
\end{equation}
where $x\left[ {k',l'} \right] \in \mathbb{A}$ is the modulated symbol arranged on the $\left[k',l'\right]$-th grid point in the DD domain and $y\left[ {k,l} \right] \in \mathbb{C}$ denotes the $\left[k,l\right]$-th sampled receiving signal in the DD domain.
Variable $z\left[k,l\right]$ denotes the additive noise whose distribution depends on the imposed receiver window \cite{WeiWindowOTFS}.
For example, adopting a rectangular window at the receiver, we have 
$z\left[k,l\right] \sim \mathcal{CN}\left(0,{\sigma}^2\right)$, where ${\sigma}^2$ denotes the unknown noise power\cite{WeiWindowOTFS}.
Integer $N$ denotes the number of time slots with a slot duration of $T$ for the OTFS system, resulting in one OTFS frame duration of $NT$.
Besides, integer $M$ represents the number of subcarriers with a subcarrier spacing of $\Delta_f$ for the OTFS system and the total system bandwidth is $M\Delta_f$.
Interested readers may refer to our previous work \cite{WeiWindowOTFS} for more derivation details of \eqref{IOOTFS}\footnote{Note that the input-output relationship in \eqref{IOOTFS} was derived for a single-antenna OTFS transceiver.
	As we mainly focus on handling the fractional Doppler and delay estimation issue of OTFS, we would like to adopt a single-input single-output system in this paper for obtaining important system design insights.
	The extension to multiple-input multiple-output systems \cite{ZhangMUltipleAntenna} will be considered in our future work.}.

In \eqref{IOOTFS}, ${h_w}\left[ {k - k',l - l'} \right]$ denotes the \textit{effective DD domain channel}, which is obtained by sampling the \textit{original DD domain channel response}.
In particular, the original DD domain channel response is given by
\vspace{-4mm}
\begin{equation}\label{OriginalDDDomainChannel}
	{h}\left(\nu,\tau\right) = \sum_{i=1}^{P} {h_i}  \delta\left(\nu - \nu_i\right) \delta\left(\tau - \tau_i\right),\vspace{-2mm}
\end{equation}
where $P$ represents the number of paths in the channel, ${h}_i$ is the channel coefficient of the $i$-th path, and $\delta\left(\cdot\right)$ is the Dirac delta function.
Variables ${\tau_i} \in \left(0, \tau_{\mathrm{max}}\right)$ and ${\nu_i} \in \left(-\nu_{\mathrm{max}}, \nu_{\mathrm{max}}\right)$ denote the delay and Doppler shifts of the $i$-th path, respectively.
In particular, constants $\nu_{\mathrm{max}}$ and $\tau_{\mathrm{max}}$ denote the maximum Doppler shift and the maximum delay shift in the propagation environments, which are related to the maximum relative velocity among the transceivers or scatters and the maximum propagation distance, respectively.
In this paper, we assume that the number of path $P$ is unknown while $\nu_{\mathrm{max}}$ and $\tau_{\mathrm{max}}$ are known in advance by our proposed channel estimation approaches.
In fact, the Doppler and delay shift boundary can be obtained empirically via long-term measurements, i.e., i.e., $\nu_{\mathrm{max}} = \max_{i} \{\left|\nu_i\right|\}$ and $\tau_{\mathrm{max}} = \max_{i} \{\tau_i\}$.
However, the number of paths for each channel realization mainly depends on the presenting scatters in the propagation environment, which is usually unpredictable.
As a result, the effective DD domain channel can be obtained by\cite{RavitejaOTFS,wei2020orthogonal,WeiWindowOTFS,LiPDMAOTFS}
\vspace{-2mm}
\begin{equation}\label{EffectiveDDDomainChannel}
{h_w}\left[ {k - k',l - l'} \right] = \sum_{i=1}^{P} {h_i} w(k-k'-k_{\nu_i}, l-l'-l_{\tau_i} ) {e^{ - j2\pi {\nu_i}l_{\tau_i}  }}
 = \sum_{i=1}^{P} {\tilde{h}_i} w(k-k'-k_{\nu_i}, l-l'-l_{\tau_i} ),\vspace{-2mm}
\end{equation}
where $k,k' \in \left\{0,\ldots,N-1\right\}$, $l,l' \in \left\{0,\ldots,M-1\right\}$, and $k_{\nu_i} = {\nu_i}{NT} \in \mathbb{R}$ and $l_{\tau_i} = {\tau_i}{M\Delta_f} \in \mathbb{R}$ represent the corresponding normalized version of $\nu_i$ and ${\tau_i}$, respectively.
To facilitate the channel estimation, the phase ${e^{ - j2\pi {\nu_i}l_{\tau_i}  }}$ is absorbed into the channel coefficient ${\tilde{h}_i} = {{h}_i}{e^{ - j2\pi {\nu_i}l_{\tau_i}  }}$ since the delay and Doppler variables are not separable in the exponential function.

In \eqref{EffectiveDDDomainChannel}, $w\left(k - k'-k_{\nu},l - l'-l_{\tau}\right)$ is the sampling function, which captures the joint effects of the Doppler and delay shift of the original DD domain channel response, the adopted pulse shaping filters and the TF domain window at the OTFS modulator and demodulator, as shown in Fig. \ref{OTFSSystemModel}, which can be obtained by\cite{RavitejaOTFS,WeiWindowOTFS}
\vspace{-2mm}
\begin{align}\label{WindowDDDomainRectPulse}
&w\left(k - k'-k_{\nu},l - l'-l_{\tau}\right) \notag\\[-1mm]
=& \frac{1}{{NM}}\sum\limits_{n = 0}^{N - 1} \sum\limits_{m = 0}^{M - 1} \sum\limits_{m' \ne m}^{M - 1} {A_{{g_{rx}}{g_{tx}}}}\left( { - {l_\tau }\frac{1}{{M\Delta f}},\left( {m - m'} \right)\Delta f - {k_\nu }\frac{1}{{NT}}} \right)\notag\\[-1mm]
 \times& U\left[ {n,m'} \right]V\left[ {n,m'} \right]{e^{ - j2\pi \frac{{n\left( {k - k' - {k_\nu }} \right)}}{N}}}{e^{ - j2\pi \frac{{\left( {ml - m'l' - m'{l_\tau }} \right)}}{M}}}    \notag\\[-1mm]
+&  {e^{ - j2\pi \frac{{k'}}{N}}}\frac{1}{{NM}}\sum\limits_{n = 1}^{N - 1} {\sum\limits_{m = 0}^{M - 1} {\sum\limits_{m' = 0}^{M - 1} {{A_{{g_{rx}}{g_{tx}}}}\left( {T - {l_\tau }\frac{1}{{M\Delta f}},\left( {m - m'} \right)\Delta f - {k_\nu }\frac{1}{{NT}}} \right)} } } \notag\\[-1mm]
\times& U\left[ {n - 1,m'} \right]V\left[ {n - 1,m'} \right]{e^{ - j2\pi \frac{{n\left( {k - k' - {k_\nu }} \right)}}{N}}}{e^{ - j2\pi \frac{{\left( {ml - m'l' - m'{l_\tau }} \right)}}{M}}},
\end{align}
\vspace{-10mm}\par\noindent
where $U\left[ {n,m} \right]$ and $V\left[ {n,m} \right]$ denote the adopted transmit window and receiver window in the $\left[ {n,m} \right]$-th TF domain point, respectively.
The ambiguity function ${{A_{{g_{rx}}{g_{tx}}}}\left(\tau,\nu \right)}$ for the adopted transmit pulse shaping filter ${g_{{\rm{tx}}}}\left( {t  } \right)$ and receiver pulse shaping filter ${g_{{\rm{rx}}}}\left( {t  } \right)$ can be defined by
\vspace{-2mm}
\begin{equation} \label{AmbiguityFunction}
{{A_{{g_{rx}}{g_{tx}}}}\left(\tau,\nu \right)} = \int_{t} {g_{{\rm{tx}}}}\left( {t} \right)g_{{\rm{rx}}}^ * \left( {t - \tau } \right){e^{ - j2\pi \nu \left( {t - \tau } \right)}}dt.\vspace{-2mm}
\end{equation}
Note that the considered system model in \eqref{IOOTFS}-\eqref{WindowDDDomainRectPulse} is a generalized framework, which are applicable to an arbitrary pulse shaping filter and an arbitrary TF domain window.
In \eqref{WindowDDDomainRectPulse}, we can observe that the sampling function has a circular structure with a period of $N$ and $M$ in the Doppler and delay domains, respectively.
As a result, the effective DD domain channel ${h_w}\left[ {k - k',l - l'} \right]$ also exhibits a circular structure with a period of $N$ and $M$ in the Doppler and delay domains, respectively, i.e., ${h_w}\left[ {k - k',l - l'} \right] = {h_w}\left[ {(k - k')_N,(l - l')_M} \right]$.

A commonly studied case in the OTFS literature is to apply an ideal pulse shaping filter\cite{RavitejaOTFS} to the transceiver, which satisfies the bi-orthogonal condition\cite{WeiWindowOTFS} as follows
\vspace{-2mm}
\begin{equation}\label{BIOFthogonal}
{{A_{{g_{rx}}{g_{tx}}}}\left( {\left( {n - n'} \right)T - {l_\tau }\frac{1}{{M\Delta f}} ,\left( {m - m'} \right)\Delta f - {k_\nu }\frac{1}{{NT}} } \right)} = q_{\tau_{\mathrm{max}}}\left(t-nT\right)q_{\nu_{\mathrm{max}}}\left(f-m\Delta f\right),\vspace{-2mm}
\end{equation}
with
\vspace{-4mm}
\begin{equation}
q_{\tau_{\mathrm{max}}}\hspace{-0.5mm}\left(t\hspace{-0.5mm}-\hspace{-0.5mm}nT\right)\hspace{-0.5mm} = \hspace{-0.5mm}\left\{ {\begin{array}{*{20}{c}}
	{\delta \left[ n \right]}&{\left| {t\hspace{-0.5mm}-\hspace{-0.5mm}nT} \right| \le \tau_{\mathrm{max}}},\\[-1mm]
	{q\left( t \right)}&{\mathrm{otherwise}},
	\end{array}} \right. \;\text{and}\;
q_{\nu_{\mathrm{max}}}\hspace{-0.5mm}\left(f\hspace{-0.5mm}-\hspace{-0.5mm}m\Delta f\right) \hspace{-0.5mm}=\hspace{-0.5mm} \left\{ {\begin{array}{*{20}{c}}
	{\delta \left[ m \right]}&{\left| {f\hspace{-0.5mm}-\hspace{-0.5mm}m\Delta f} \right| \le \nu_{\mathrm{max}}},\\[-1mm]
	{q\left( t \right)}&{\mathrm{otherwise}},
	\end{array}} \right. \vspace{-2mm}
\end{equation}
where $q\left( x \right)$ is an arbitrary function.
Furthermore, with adopting the rectangular TF domain window\cite{WeiWindowOTFS} at the transceiver, i.e., $U\left[ {n,m} \right] = V\left[ {n,m} \right] = 1$, $\forall m,n$, the sampling function in \eqref{WindowDDDomainRectPulse} can be simplified as 
\vspace{-2mm}
\begin{equation}\label{WindowDDDomain}
w\left(k - k'-k_{\nu},l - l'-l_{\tau}\right)
=  w_{\nu}\left(k - k'-k_{\nu}\right)w_{\tau}\left(l - l'-l_{\tau}\right),\vspace{-2mm}
\end{equation}
where the sampling function can be decomposed as the sampling functions in the Doppler and delay domains, respectively, which are given by
\vspace{-2mm}
\begin{align}\label{SamplingFunctionIdeal}
w_{\nu}\left(k - k'-k_{\nu}\right) &= \frac{1}{N}\left[ {{e^{ - j\left( {N - 1} \right)\pi \frac{{k - k'-k_{\nu}}}{N}}}\frac{{\sin \left( {\pi \left( k - k'-k_{\nu} \right)} \right)}}{{\sin \left( {\frac{{\pi \left( k - k'-k_{\nu} \right)}}{N}} \right)}}} \right]\;\text{and}\notag\\[-1mm]
w_{\tau}\left(l - l'-l_{\tau}\right) & = \frac{1}{M}\left[ {{e^{ - j\left( {M - 1} \right)\pi \frac{l - l'-l_{\tau} }{M}}}\frac{{\sin \left( {\pi \left( l - l'-l_{\tau} \right)} \right)}}{{\sin \left( {\frac{{\pi \left( l - l'-l_{\tau} \right)}}{M}} \right)}}} \right],
\end{align}
\vspace{-8mm}\par\noindent
respectively.

For notational simplicity, we focus on the case of ideal pulse shaping and rectangular window case in this paper. 
The designed channel estimation algorithm can be easily extended to the case of arbitrary pulse shaping filter and arbitrary window with straightforward modifications of the sampling functions following \eqref{WindowDDDomainRectPulse}.

\begin{figure}[t]
	\begin{minipage}{.47\textwidth}
		\centering\vspace{-3mm}
		\includegraphics[width=3.5in]{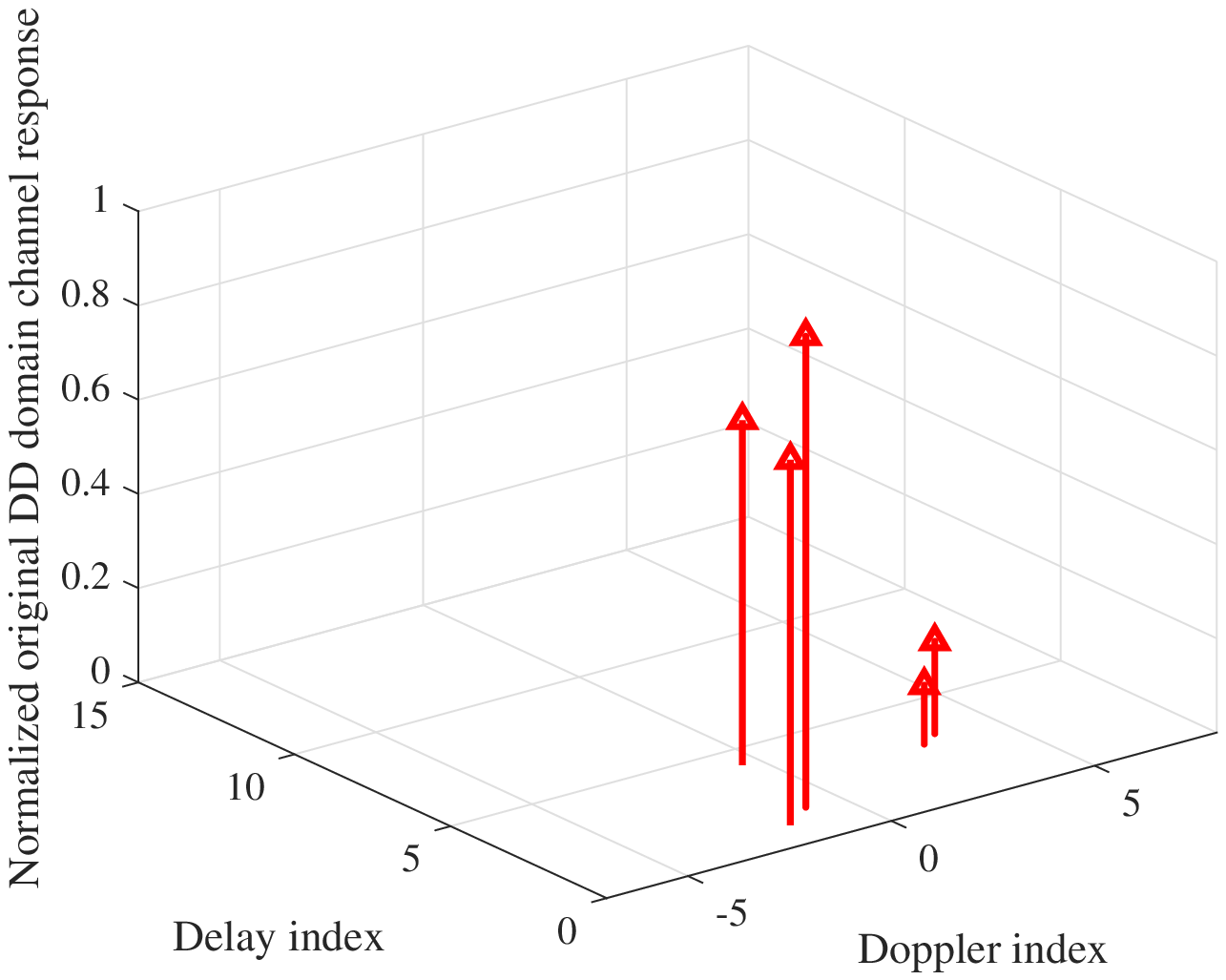}\vspace{-7mm}
		\caption{Normalized original DD domain channel response.}\vspace{-10mm}
		\label{OriginalDDDomainChannelResponse}
	\end{minipage}
	\hspace{2mm}
	\begin{minipage}{.47\textwidth}
		\centering
		\includegraphics[width=3.5in]{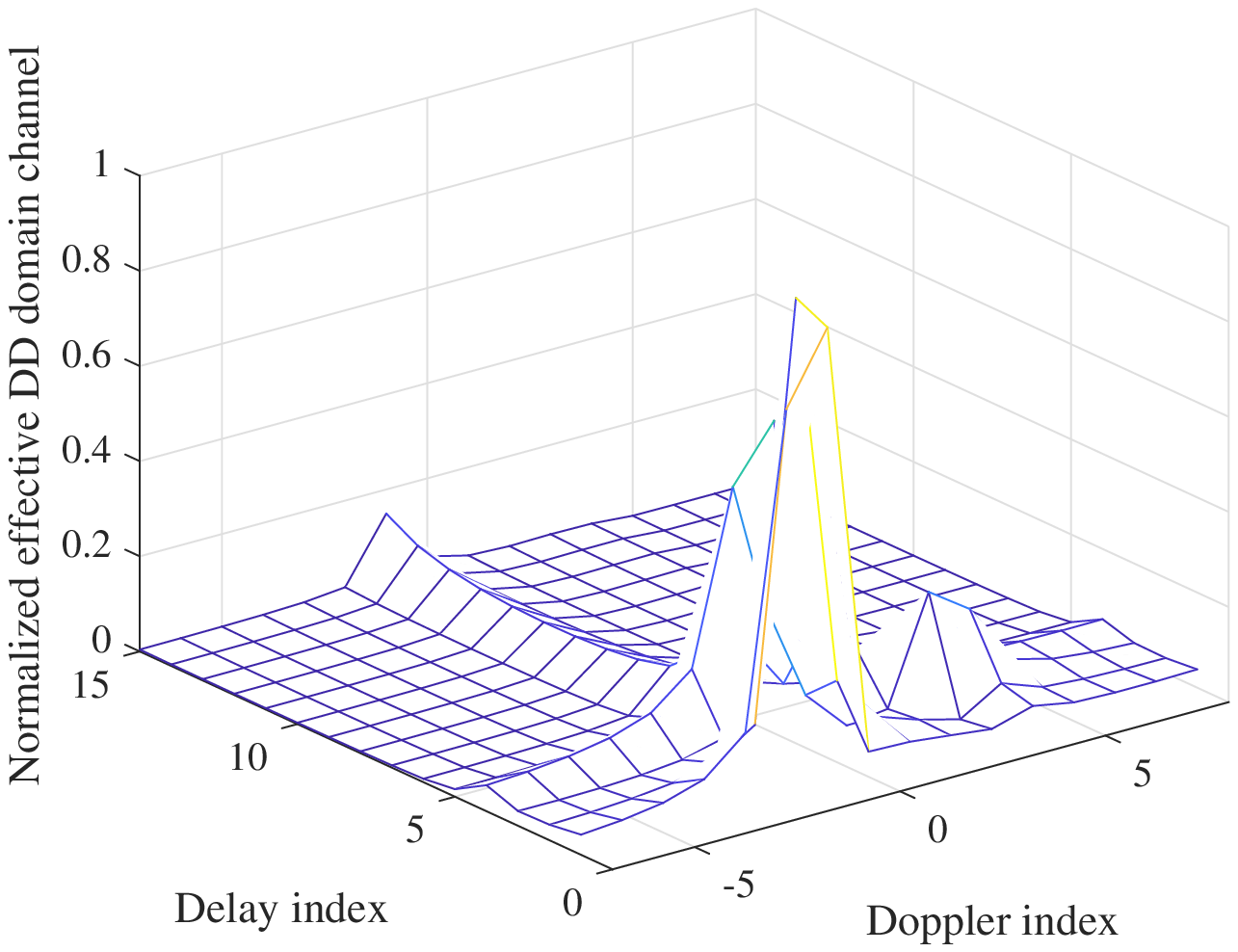}\vspace{-7mm}
		\caption{Normalized effective DD domain channel gain.}\vspace{-10mm}
		\label{EffectiveDDDomainChannelResponse}%
	\end{minipage}
\end{figure}

To visualize the difference between the original DD domain channel response and the effective DD domain channel, we plot a snapshot of the corresponding normalized channel gains in Fig. \ref{OriginalDDDomainChannelResponse} and Fig. \ref{EffectiveDDDomainChannelResponse}, respectively, where we assume $M = N=16$ and $P = 5$.
We can observe that the original DD domain channel response is indeed sparse.
However, due to the existence of fractional Doppler and delay, i.e., non-integer $k_{\nu}$ and $l_{\tau}$, even with ideal pulse shaping filters, the sampling functions in \eqref{SamplingFunctionIdeal} are non-zero for $k - k' \neq k_{\nu}$ and $l - l' \neq l_{\tau}$.
It causes channel spreading in the effective DD domain channel in \eqref{EffectiveDDDomainChannel}, which is far from sparse.
As a result, estimating the effective DD domain channel using compressed sensing techniques \cite{ShenCEMassiveMIMO,ZhaoSBLOTFS} may not lead to a satisfactory performance.
Therefore, in this paper, we propose to directly estimate the original DD domain channel response ${h}\left(\nu,\tau\right)$ and reconstruct the effective DD domain channel according to \eqref{EffectiveDDDomainChannel}, which is a key for realizing effective data detection.

\vspace{-2mm}
\section{1D Off-grid  Compressed Channel Estimation}\label{1DSSR}
In this section, with the DD domain input-output relationship in mind, we formulate the channel estimation problem as a 1D sparse signal recovery (SSR) problem.
%
%As discussed before, to avoid the channel spreading induced by the sampling function and to exploit the channel sparsity,  we aims to recovery the original DD domain channel response ${h}\left(\nu,\tau\right)$ 
\vspace{-2mm}
\subsection{Channel Estimation Problem Formulation}
To facilitate channel estimation, the system model in \eqref{IOOTFS} can rewritten as
\vspace{-2mm}
\begin{equation}\label{1DModel}
	{\bf{y = \Phi }}\left( {\mathbf{k}_{{\nu }}},{\mathbf{l}_{{\tau}}} \right){{\tilde{\mathbf{h}} + \mathbf{z}}},\vspace{-2mm}
\end{equation}
where $\mathbf{y} \in \mathbb{C}^{MN\times 1}$ is the vectorized form of $y\left[ {k,l} \right]$ and vector $\mathbf{z} \sim \mathcal{CN}\left(\mathbf{0},\sigma^2\mathbf{I}_{MN}\right) \in \mathbb{C}^{MN\times 1}$ is the vectorized form of $z\left[ {k,l} \right]$.
Vectors ${\mathbf{k}_{{\nu }}} = \left[k_{\nu_1}, \ldots, k_{\nu_{P}}\right]^{\mathrm{T}} \in \mathbb{R}^{P\times 1}$, ${\mathbf{l}_{{\tau}}} = \left[l_{\tau_1}, \ldots, l_{\tau_{P}}\right]^{\mathrm{T}} \in \mathbb{R}^{P\times 1}$, and $\tilde{\mathbf{h}} = \left[{\tilde{h}_1}, \ldots, {\tilde{h}_P}\right]^{\mathrm{T}} \in \mathbb{C}^{P\times 1}$ collect the normalized Doppler shifts, the normalized delay shifts, and the channel coefficients, respectively.
Matrix ${\bf{\Phi }}\left( {\mathbf{k}_{{\nu }}},{\mathbf{l}_{{\tau}}} \right) \in \mathbb{C}^{MN\times P}$ is the measurement matrix, which can be formulated as 
\vspace{-2mm}
\begin{equation}\label{MeasurmentMatrix}
	{\bf{\Phi }}\left( {\mathbf{k}_{{\nu }}},{\mathbf{l}_{{\tau}}} \right) = \left[\boldsymbol\phi\left( k_{\nu_1},l_{\tau_1} \right),\ldots,\boldsymbol\phi\left( k_{\nu_P},l_{\tau_P} \right)\right],\vspace{-2mm}
\end{equation}
where $\boldsymbol\phi\left( k_{\nu_i},l_{\tau_i} \right) \in \mathbb{C}^{MN \times 1}$ and its $\left(kM + l\right)$-th entry can be given by
\vspace{-2mm}
\begin{equation}
	\{\boldsymbol\phi\left( k_{\nu_i},l_{\tau_i} \right)\}_{kM + l} = \sum\limits_{k' = 0}^{N - 1} {\sum\limits_{l' = 0}^{M - 1} {x\left[ {k',l'} \right]w(k - k' - k_{\nu_i},l - l' - l_{\tau_i})} }.\vspace{-2mm}
\end{equation}

The channel estimation problem is to estimate the normalized Doppler shifts ${\mathbf{k}_{{\nu }}}$, the normalized delay shifts ${\mathbf{l}_{{\tau}}}$, and the channel coefficients $\tilde{\mathbf{h}}$ based on the function ${\bf{\Phi }}\left( {\mathbf{k}_{{\nu }}},{\mathbf{l}_{{\tau}}} \right)$ and the received signal $\bf{y}$.
One observation is that when the number of path $P$ is unknown, the length of $\tilde{\mathbf{h}}$ is not fixed and thus estimating $\tilde{\mathbf{h}}$ is a challenging task.
Besides, since the measurement matrix ${\bf{\Phi }}\left( {\mathbf{k}_{{\nu }}},{\mathbf{l}_{{\tau}}} \right)$ depends on the unknown variables ${\mathbf{k}_{{\nu }}}$ and ${\mathbf{l}_{{\tau}}}$, conventional methods, such as least square (LS) or 
minimum mean square error (MMSE) channel estimation are not applicable to the considered case.
Moreover, the measurement matrix ${\bf{\Phi }}\left( {\mathbf{k}_{{\nu }}},{\mathbf{l}_{{\tau}}} \right)$ also depends on the arrangement of data and pilot symbols in the DD domain $x\left[ {k',l'} \right]$.
For simplicity, we only exploit the pilot symbols in the measurement matrix ${\bf{\Phi }}\left( {\mathbf{k}_{{\nu }}},{\mathbf{l}_{{\tau}}} \right)$ for channel estimation.
In particular, the impacts of data symbols in the measurement matrix  ${\bf{\Phi }}\left( {\mathbf{k}_{{\nu }}},{\mathbf{l}_{{\tau}}} \right)$ are treated as the model error which is assumed to behave as independent Gaussian noise and can be absorbed into the measurement noise \cite{YangZaiOffgridCE}.
Thus, the received signal corresponding to the pilot can be obtained by 
\vspace{-2mm}
\begin{equation}\label{IOOTFSChannel}
y\left[ {k,l} \right] =  \sum\limits_{k' \in \mathcal{K}_{\mathrm{p}}} {\sum\limits_{l' \in \mathcal{L}_{\mathrm{p}}} {x\left[ {k',l'} \right]} } {h_w}\left[ {k - k',l - l'} \right] + \tilde z\left[k,l\right],\vspace{-2mm}
\end{equation}
and the measurement matrix can be rewritten as 
\vspace{-2mm}
\begin{equation}\label{MeasurmentMatrixII}
\{\boldsymbol\phi\left( k_{\nu_i},l_{\tau_i} \right)\}_{kM + l} = \sum\limits_{k' \in \mathcal{K}_{\mathrm{p}}} {\sum\limits_{l' \in \mathcal{L}_{\mathrm{p}}} {x\left[ {k',l'} \right]w(k - k' - k_{\nu_i},l - l' - l_{\tau_i})} },\vspace{-2mm}
\end{equation}
where $\mathcal{K}_{\mathrm{p}}$ and $\mathcal{L}_{\mathrm{p}}$ denote the index sets of pilot symbols in the Doppler and delay domains, respectively, and $\tilde z\left[k,l\right]$ denotes the new noise variable absorbing the interference of data symbols.
Note that the proposed model does not specify the pilot arrangement.
Specifically, when $\left|\mathcal{K}_{\mathrm{p}}\right| = \left|\mathcal{L}_{\mathrm{p}}\right| = 1$, the model in \eqref{MeasurmentMatrixII} degenerates to a single pilot impulse case in \cite{WeiWindowOTFS}. When $\left|\mathcal{K}_p\right| = \left|\mathcal{L}_p\right| > 1$, it is equivalent to multiple pulses or the pilot sequence cases in \cite{ShenCEMassiveMIMO}.
%
%Furthermore, the data-aided channel estimation method \cite{MaDataAidedCE} can be employed to further improve the channel estimation performance via substituting the estimated data symbols into the measurement matrix ${\bf{\Phi }}\left( {\mathbf{k}_{{\nu }}},{\mathbf{l}_{{\tau}}} \right)$.
%
For notational simplicity, we consider a single pulse case, i.e., $\mathcal{K}_{\mathrm{p}} = \{k_{\mathrm{p}}\}$ and $\mathcal{L}_{\mathrm{p}} = \{l_{\mathrm{p}}\}$, where a single pilot impulse is inserted in the $\left[k_{\mathrm{p}},l_{\mathrm{p}}\right]$-th DD grid point.

Due to the compactness of the DD domain channel, i.e., ${\tau_i} \in \left(0, \tau_{\mathrm{max}}\right)$ and ${\nu_i} \in \left(-\nu_{\mathrm{max}}, \nu_{\mathrm{max}}\right)$, only a fraction of the symbols in $\mathbf{y}$ are affected the single pilot impulse.
To reduce the computational complexity, we consider a truncated $\mathbf{y}_{\mathrm{T}}$ to estimate the channel, which is actually a \textit{sufficient statistic} for channel estimation of OTFS systems.
In particular, given the maximum Doppler shift $\nu_{\mathrm{max}}$ and the maximum delay shift $\tau_{\mathrm{max}}$, we only need to consider the received signal $y\left[k,l\right]$ in the range of $k_{\mathrm{p}} - k_{\mathrm{max}} \le k \le k_{\mathrm{p}} + k_{\mathrm{max}}$ and $l_{\mathrm{p}} \le l \le l_{\mathrm{p}} + k_{\mathrm{max}}$, where $k_{\mathrm{max}} = \left\lceil \nu_{\mathrm{max}}NT \right\rceil$ and $ l_{\mathrm{max}} = \left\lceil\tau_{\mathrm{max}}M\Delta_f \right\rceil$.
As a result, we have 
\vspace{-2mm}
\begin{equation}\label{1DModelTruncated}
\mathbf{y}_{\mathrm{T}} = \mathbf{\Phi}_{\mathrm{T}} \left( {\mathbf{k}_{{\nu }}},{\mathbf{l}_{{\tau}}} \right){{\tilde{\mathbf{h}} + \tilde{\mathbf{z}}_{\mathrm{T}}}},\vspace{-2mm}
\end{equation}
where $\mathbf{y}_{\mathrm{T}} \in \mathbb{C}^{M_{\mathrm{T}}N_{\mathrm{T}} \times 1}$ and $\tilde{\mathbf{z}}_{\mathrm{T}} \in \mathbb{C}^{M_{\mathrm{T}}N_{\mathrm{T}} \times 1}$ are obtained by $\{\mathbf{y}_{\mathrm{T}}\}_{kM_{\mathrm{T}} + l} = y\left[k,l\right]$ and $\{\tilde{\mathbf{z}}_{\mathrm{T}}\}_{kM_{\mathrm{T}} + l} = \tilde z\left[k,l\right]$, respectively, with $N_{\mathrm{T}} = 2k_{\mathrm{max}}+1$ and $M_{\mathrm{T}} = l_{\mathrm{max}}+1$.
The truncated measurement matrix is given by
\vspace{-2mm}
\begin{equation}
{\bf{\Phi }}_{\mathrm{T}}\left( {\mathbf{k}_{{\nu }}},{\mathbf{l}_{{\tau}}} \right) = \left[\boldsymbol\phi_{\mathrm{T}}\left( k_{\nu_1},l_{\tau_1} \right),\ldots,\boldsymbol\phi_{\mathrm{T}}\left( k_{\nu_P},l_{\tau_P} \right)\right] \in \mathbb{C}^{M_{\mathrm{T}}N_{\mathrm{T}}\times P},\vspace{-2mm}
\end{equation}
where $\boldsymbol\phi_{\mathrm{T}}\left( k_{\nu},l_{\tau} \right) \in \mathbb{C}^{M_{\mathrm{T}}N_{\mathrm{T}} \times 1}$ and it is defined by
\vspace{-2mm}
\begin{equation}
	\{\boldsymbol\phi_{\mathrm{T}}\left( k_{\nu_i},l_{\tau_i} \right)\}_{kM_{\mathrm{T}} + l} = { {x\left[ {k_\mathrm{p},l_\mathrm{p}} \right]w(k - k_\mathrm{p} - k_{\nu_i},l - l_\mathrm{p}  -  l_{\tau_i})} }, \vspace{-2mm}
\end{equation}
with $k_{\mathrm{p}} - k_{\mathrm{max}} \le k \le k_{\mathrm{p}} + k_{\mathrm{max}}$ and $l_{\mathrm{p}} \le l \le l_{\mathrm{p}} + k_{\mathrm{max}}$.
We can observe that the estimation model in \eqref{1DModelTruncated} is non-linear due to the coupling of unknown Doppler and delay shifts with the measurement matrix, which motivates us to propose the first-order linear approximation in next section.

%Then, restricting the range of $k,l$ and $k',l'$ to the guard space and reviewing the interference spreading from data to the guard space as noise, we can apply to OTFS scenario.
%
%Besides, we note that this model fits both the pilot pulse and pilot sequences scenario, where the pilot symbols is arranged in $x\left[ {k',l'} \right]$.
%
%How to adapt to the multiple antenna case? It is interesting.

%For single pulse case, we have $k' = k_{\mathrm{p}}$ and $l' = l_{\mathrm{p}}$ and $k_{\mathrm{p}} - k_{\mathrm{max}} \le k \le k_{\mathrm{p}} + k_{\mathrm{max}}$ and $l_{\mathrm{p}} \le l \le l_{\mathrm{p}} + k_{\mathrm{max}}$.

\vspace{-4mm}
\subsection{First-Order Linear Approximation}
Following the principle of compressed channel sensing\cite{BajwaCSC}, we formulate an SSR problem via a virtual representation based on a virtual sampling grid in the DD space.
Define a virtual sampling grid in the range of $\left(0, \tau_{\mathrm{max}}\right)$ and $\left(-\nu_{\mathrm{max}}, \nu_{\mathrm{max}}\right)$ in the DD domain, i.e., $\overline{\mathbf{k}}_{\nu} = \left[\overline k_{\nu_0},\ldots,\overline k_{\nu_{M_{\tau}N_{\nu}-1}}\right] \in \mathbb{R}^{M_{\tau}N_{\nu} \times 1}$ and $\overline{\mathbf{l}}_{\tau} = \left[\overline l_{\tau_0},\ldots,\overline l_{\tau_{M_{\tau}N_{\nu}-1}}\right] \in \mathbb{R}^{M_{\tau}N_{\nu} \times 1}$, where $N_{\nu}$ and $M_{\tau}$ denote the virtual sampling grid size in the Doppler and delay domains, respectively\footnote{Note that increasing the virtual Doppler and delay resolutions can improve the accuracy of the on-grid channel estimation schemes. However, the computational complexity will be prohibitively high. Therefore, our proposed off-grid channel estimation can mitigate the impact of coarse virtual sampling grid and thus can achieve a satisfactory performance with a much lower computational complexity. The impacts of virtual sampling resolution for both on-grid and off-grid methods will be investigated in Section VI through simulations.}.
Typically considering an equally spaced sampling grid with a virtual Doppler resolution $r_{\nu} = \frac{2\nu_{\mathrm{max}}}{N_{\nu}}$ and a virtual delay resolution $r_{\tau} = \frac{\tau_{\mathrm{max}}}{M_{\tau}}$, we have $\overline k_{\nu_{k''M_{\tau} + l''}} = k''r_{\nu} - \nu_{\mathrm{max}}$ and $\overline l_{\tau_{k''M_{\tau} + l''}} = l'' r_{\tau}$, $\forall k'' \in \left\{0,\ldots,N_{\nu}-1\right\}$ and $\forall l'' \in \left\{0,\ldots,M_{\tau}-1\right\}$.
To separate the estimation of the on-grid and off-grid components of Doppler and delay shifts based on the constructed virtual sampling grid, we can perform a linear approximation as follows:
\vspace{-2mm}
\begin{align}
	\boldsymbol\phi_{\mathrm{T}}\left( k_{\nu_i},l_{\tau_i} \right) &\approx \boldsymbol\phi_{\mathrm{T}}\left( \overline k^{i}_{\nu_{k''M_{\tau} + l''}},\overline l^{i}_{\tau_{k''M_{\tau} + l''}} \right) + \boldsymbol\phi'_{{\mathrm{T}},\nu}\left( \overline k^{i}_{\nu_{k''M_{\tau} + l''}},\overline l^{i}_{\tau_{k''M_{\tau} + l''}} \right) \left(k_{\nu_{i}} - \overline k^{i}_{\nu_{k''M_{\tau} + l''}}\right) \notag\\[-1mm]
	&+ \boldsymbol\phi'_{{\mathrm{T}},\tau}\left( \overline k^{i}_{\nu_{k''M_{\tau} + l''}},\overline l^{i}_{\tau_{k''M_{\tau} + l''}} \right) \left(l_{\tau_{i}} - \overline l^{i}_{\tau_{k''M_{\tau} + l''}}\right) \notag\\[-1mm]
	&= \boldsymbol\phi_{\mathrm{T}}\left( \overline k^{i}_{\nu_{k''M_{\tau} + l''}},\overline l^{i}_{\tau_{k''M_{\tau} + l''}} \right) + \boldsymbol\phi'_{{\mathrm{T}},\nu}\left( \overline k^{i}_{\nu_{k''M_{\tau} + l''}},\overline l^{i}_{\tau_{k''M_{\tau} + l''}} \right) {\kappa^{i} _{{\nu _{k''M_{\tau}+l''}}}} \notag\\[-1mm]
	&+ \boldsymbol\phi'_{{\mathrm{T}},\tau}\left( \overline k^{i}_{\nu_{k''M_{\tau} + l''}},\overline l^{i}_{\tau_{k''M_{\tau} + l''}} \right) {\iota^{i} _{{\tau_ {k''M_{\tau}+l''}}}},\label{1DVirtualSamplingModel}
\end{align}
\vspace{-10mm}\par\noindent
where the approximation is due to the exclusion of the higher order derivatives.
Variables $\overline k^{i}_{\nu_{k''M_{\tau} + l''}}$ and $\overline l^{i}_{\tau_{k''M_{\tau} + l''}}$ denote the unknown nearest grid points of $k_{\nu_{i}}$ and $l_{\tau_{i}}$, respectively.
Vectors $\boldsymbol\phi'_{{\mathrm{T}},\nu}\left( k_{\nu},l_{\tau}  \right) = \frac{{\partial \boldsymbol\phi_{\mathrm{T}}\left( k_{\nu},l_{\tau} \right)}}{{\partial {k_{{\nu }}}}} \in  \mathbb{C}^{M_{\mathrm{T}}N_{\mathrm{T}} \times 1}$ and $\boldsymbol\phi'_{{\mathrm{T}},\tau}\left(k_{\nu},l_{\tau}  \right) = \frac{{\partial \boldsymbol\phi_{\mathrm{T}}\left( k_{\nu},l_{\tau} \right)}}{{\partial {l_{{\tau }}}}}  \in \mathbb{C}^{M_{\mathrm{T}}N_{\mathrm{T}} \times 1}$ denote the first-order gradient of $\boldsymbol\phi_{\mathrm{T}}\left( k_{\nu},l_{\tau} \right)$ w.r.t. $k_{\nu}$ and $l_{\tau}$, respectively.
Also, the $\left(kM_{\tau} + l\right)$-th entries of $\boldsymbol\phi'_{{\mathrm{T}},\nu}\left( k_{\nu},l_{\tau}  \right)$ and $\boldsymbol\phi'_{{\mathrm{T}},\tau}\left(k_{\nu},l_{\tau}  \right)$ are given by
\vspace{-2mm}
\begin{align}
\{\boldsymbol\phi'_{{\mathrm{T}},\nu}\left( k_{\nu},l_{\tau}  \right)\}_{kM + l} &= { {x\left[ {k_\mathrm{p},l_\mathrm{p}} \right]w_{\tau}\left(l - l_\mathrm{p} -l_{\tau}\right)  w'_{\nu}\left(k - k_\mathrm{p} -k_{\nu}\right)} } \;\text{and}\notag\\[-1mm]
\{\boldsymbol\phi'_{{\mathrm{T}},\tau}\left(k_{\nu},l_{\tau}  \right)\}_{kM + l} &= { {x\left[ {k_\mathrm{p},l_\mathrm{p}} \right]w'_{\tau}\left(l - l_\mathrm{p} -l_{\tau}\right)  w_{\nu}\left(k - k_\mathrm{p} -k_{\nu}\right)} },
\end{align}
\vspace{-10mm}\par\noindent
respectively, where 
\vspace{-2mm}
\begin{align}
w'_{\nu}\left(k - k_\mathrm{p} -{k_{{\nu }}}\right) & = \frac{{\partial w_{\nu}\left(k - k_\mathrm{p} -k_{\nu}\right) }}{{\partial {k_{{\nu }}}}} =\frac{1}{N}\sum\limits_{n = 0}^{N - 1} {\frac{{j2\pi n}}{N}{e^{ - j2\pi nT\frac{{\left( {k - k_\mathrm{p} - {k_\nu }} \right)}}{{NT}}}}} \;\text{and}\notag\\[-1mm]
w'_{\tau}\left(l - l_\mathrm{p} -l_{\tau}\right) & = \frac{{\partial w_{\tau}\left(l - l_\mathrm{p}-l_{\tau}\right) }}{{\partial {l_{\tau}}}} =-\frac{1}{M}\sum\limits_{m = 0}^{M - 1} {\frac{{j2\pi m}}{M}{e^{ j2\pi m\Delta_f\frac{{\left( {l - l_\mathrm{p} - {l_\tau }} \right)}}{{M\Delta_f}}}}}.
\end{align}
\vspace{-8mm}\par\noindent
Variables ${\kappa^{i} _{{\nu _{k''M_{\tau}+l''}}}}$ and ${\iota^{i} _{{\tau_ {k''M_{\tau}+l''}}}}$ represent the off-grid components of the Doppler and delay shifts of the $i$-th path, respectively, and they are defined as:
\vspace{-1mm}
\begin{equation}\label{FractionalBounds}
	{\kappa^{i} _{{\nu _{k''M_{\tau}+l''}}}} \hspace{-1mm}=\hspace{-1mm} k_{\nu_{i}} - \overline k^{i}_{\nu_{k''M_{\tau} + l''}} \in  \left[-\frac{1}{2}r_{\nu},\frac{1}{2}r_{\nu}\right]\;\text{and}\;
	{\iota^{i} _{{\tau_ {k''M_{\tau}+l''}}}} \hspace{-1mm} = \hspace{-1mm} l_{\tau_{i}} - \overline l^{i}_{\tau_{k''M_{\tau} + l''}} \in  \left[-\frac{1}{2}r_{\tau},\frac{1}{2}r_{\tau}\right],\vspace{-2mm}
\end{equation}
respectively.
Now, substituting \eqref{1DVirtualSamplingModel} into the system model in \eqref{1DModelTruncated}, we have
\vspace{-2mm}
\begin{align}\label{1DApproxModel}
\mathbf{y}_{\mathrm{T}} &\approx  \sum_{i=1}^{P} \left[\boldsymbol\phi_{\mathrm{T}}\left( \overline k^{i}_{\nu_{k''M_{\tau} + l''}},\overline l^{i}_{\tau_{k''M_{\tau} + l''}} \right) + \boldsymbol\phi'_{{\mathrm{T}},\nu}\left( \overline k^{i}_{\nu_{k''M_{\tau} + l''}},\overline l^{i}_{\tau_{k''M_{\tau} + l''}} \right) {\kappa^{i} _{{\nu _{k''M_{\tau}+l''}}}} \right.\notag\\[-1mm]
&\left. + \boldsymbol\phi'_{{\mathrm{T}},\tau}\left( \overline k^{i}_{\nu_{k''M_{\tau} + l''}},\overline l^{i}_{\tau_{k''M_{\tau} + l''}} \right) {\iota^{i} _{{\tau_ {k''M_{\tau}+l''}}}}\right]\tilde{h}_i + \tilde{\mathbf{z}}_{\mathrm{T}}.
\end{align}
\vspace{-8mm}\par\noindent
We can observe that the Doppler and delay shifts can be estimated via estimating the on-grid components, i.e., the nearest grid point $\left( \overline k^{i}_{\nu_{k''M_{\tau} + l''}},\overline l^{i}_{\tau_{k''M_{\tau} + l''}} \right)$, and the off-grid components, i.e., $\left({\kappa^{i} _{{\nu _{k''M_{\tau}+l''}}}},{\iota^{i} _{{\tau_ {k''M_{\tau}+l''}}}}\right)$, separately.
After obtaining the on-grid and off-grid components, we can recover the Doppler and delay shift as $k_{\nu_{i}} = \overline k^{i}_{\nu_{k''M_{\tau} + l''}} + {\kappa^{i} _{{\nu _{k''M_{\tau}+l''}}}}$ and $l_{\tau_i} = \overline l^{i}_{\tau_{k''M_{\tau} + l''}} + {\iota^{i} _{{\tau_ {k''M_{\tau}+l''}}}}$, respectively.
This separation enables the possibility to recast the channel estimation as a 1D off-grid SSR problem. 

\vspace{-2mm}
\subsection{1D Off-grid SSR Model}
In \eqref{1DApproxModel}, the nearest grid point $\left( \overline k^{i}_{\nu_{k''M_{\tau} + l''}},\overline l^{i}_{\tau_{k''M_{\tau} + l''}} \right)$, the corresponding off-grid component $\left({\kappa^{i} _{{\nu _{k''M_{\tau}+l''}}}},{\iota^{i} _{{\tau_ {k''M_{\tau}+l''}}}}\right)$, and the channel coefficient $\tilde{h}_i$, $\forall i$, are unknown and remained to be estimated.
In other words, the goal is to estimate $\left\{\overline k^{i}_{\nu_{k''M_{\tau} + l''}},\overline l^{i}_{\tau_{k''M_{\tau} + l''}},{\kappa^{i} _{{\nu _{k''M_{\tau}+l''}}}},{\iota^{i} _{{\tau_ {k''M_{\tau}+l''}}}}, \tilde{h}_i\right\}$ given $\bf{y}$ and the functions ${\boldsymbol{\phi }}_{\mathrm{T}}\left( {{k}}_{\nu},{{l}}_{{\tau}} \right)$, $\boldsymbol\phi'_{{\mathrm{T}},\nu}\left( k_{\nu},l_{\tau}  \right)$ and $\boldsymbol\phi'_{{\mathrm{T}},\tau}\left(k_{\nu},l_{\tau}  \right)$.
As $P \ll M_{\tau}N_{\nu}$ usually holds, we recast \eqref{1DApproxModel} as an SSR problem based on the virtual sampling grid:
\vspace{-2mm}
\begin{equation}\label{SSRDModel}
\mathbf{y}_{\mathrm{T}} = \mathbf{\overline \Phi} _\mathrm{T}\left( {\boldsymbol{\kappa}_{{\nu }}},{\boldsymbol{\iota}_{{\tau}}} \right)\overline{\mathbf{h}} + \overline{\mathbf{z}}_{\mathrm{T}},\vspace{-2mm}
\end{equation}
where the linear approximation error in \eqref{1DApproxModel} is also absorbed into the noise term.
The new measurement matrix can be formulated as 
\vspace{-2mm}
\begin{equation}
	{\bf \overline \Phi}_\mathrm{T} \left( {\boldsymbol{\kappa}_{{\nu }}},{\boldsymbol{\iota}_{{\tau}}} \right) = { \bf \Phi}_\mathrm{T} + { \bf \Phi}_{\mathrm{T},\nu} \diag\left({\boldsymbol{\kappa}_{{\nu }}}\right) + { \bf \Phi}_{\mathrm{T},\tau} \diag\left({\boldsymbol{\iota}_{{\tau}}}\right)\in  \mathbb{C}^{M_{\mathrm{T}}N_{\mathrm{T}} \times M_{\tau}N_{\nu}},\vspace{-2mm}
\end{equation}
where ${ \bf \Phi}_\mathrm{T} \in  \mathbb{C}^{M_{\mathrm{T}}N_{\mathrm{T}} \times M_{\tau}N_{\nu}}$,  ${ \bf \Phi}_{\mathrm{T},\nu} \in  \mathbb{C}^{M_{\mathrm{T}}N_{\mathrm{T}} \times M_{\tau}N_{\nu}}$, ${ \bf \Phi}_{\mathrm{T},\tau} \in  \mathbb{C}^{M_{\mathrm{T}}N_{\mathrm{T}} \times M_{\tau}N_{\nu}}$,  $\boldsymbol{\kappa}_{{\nu }} \in  \mathbb{R}^{M_{\tau}N_{\nu} \times 1}$, and ${\boldsymbol{\iota}_{{\tau}}}\in  \mathbb{R}^{M_{\tau}N_{\nu} \times 1}$ are defined based on the virtual sampling grid $\overline{\mathbf{k}}_{\nu}$ and $\overline{\mathbf{l}}_{\tau}$, and they are given by
\vspace{-2mm}
\begin{align}
	{ \bf \Phi}_\mathrm{T} & = \left[\boldsymbol\phi_\mathrm{T}\left(\overline k_{\nu_{0}},\overline l_{\tau_{0}}\right),\ldots,\boldsymbol\phi_\mathrm{T}\left(\overline k_{\nu_{k''M_{\tau} + l''}},\overline l_{\tau_{k''M_{\tau} + l''}}\right),\ldots,\boldsymbol\phi_\mathrm{T}\left(\overline k_{\nu_{M_{\tau}N_{\nu}-1}}, \overline l_{\tau_{M_{\tau}N_{\nu}-1}}\right)\right],\\[-1mm]
	{ \bf \Phi}_{\mathrm{T},\nu} & = \left[\boldsymbol\phi'_{\mathrm{T},\nu}\left( \overline k_{\nu_{0}},\overline l_{\tau_{0}}  \right),\ldots,\boldsymbol\phi'_{\mathrm{T},\nu}\left(\overline k_{\nu_{k''M_{\tau} + l''}},\overline l_{\tau_{k''M_{\tau} + l''}}\right),\ldots,\boldsymbol\phi'_{\mathrm{T},\nu}\left( \overline k_{\nu_{M_{\tau}N_{\nu}-1}},\overline l_{\tau_{M_{\tau}N_{\nu}-1}}  \right)\right],\notag\\[-1mm]
	{ \bf \Phi}_{\mathrm{T},\tau} & = \left[\boldsymbol\phi'_{\mathrm{T},\tau}\left(\overline k_{\nu_{0}},\overline l_{\tau_{0}}  \right),\ldots,\boldsymbol\phi'_{\mathrm{T},\tau}\left(\overline k_{\nu_{k''M_{\tau} + l''}},\overline l_{\tau_{k''M_{\tau} + l''}}\right),\ldots,\boldsymbol\phi'_{\mathrm{T},\tau}\left(\overline k_{\nu_{M_{\tau}N_{\nu}-1}},\overline l_{\tau_{M_{\tau}N_{\nu}-1}}  \right)\right],\notag\\[-1mm]
	\boldsymbol{\kappa}_{{\nu }} &= \left[{\kappa _{{\nu _{0}}}},\ldots,{\kappa _{{\nu _{k''M_{\tau} + l''}}}},\ldots,{\kappa _{{\nu _{M_{\tau}N_{\nu}-1}}}}\right]^{\mathrm{T}},\;\text{and}\;
	{\boldsymbol{\iota}_{{\tau}}} = \left[{\iota _{{\tau _{0}}}},\ldots,{\iota _{{\tau _{k''M_{\tau} + l''}}}},\ldots,{\iota _{{\tau _{M_{\tau}N_{\nu}-1}}}}\right]^{\mathrm{T}} , \notag
\end{align}
\vspace{-8mm}\par\noindent
respectively.
The new noise vector is defined as $\overline{\mathbf{z}}_{\mathrm{T}} \sim \mathcal{CN}\left(\mathbf{0},\overline{\sigma^2}\mathbf{I}_{M_{\mathrm{T}}N_{\mathrm{T}}}
\right)  \in \mathbb{C}^{M_{\mathrm{T}}N_{\mathrm{T}}\times 1}$, where $\overline{\sigma^2}$ is the unknown noise power including the background noise power, the interference power due to the unknown data symbols, as well as the modeling error induced by our proposed first-order approximation in \eqref{1DApproxModel}.

In \eqref{SSRDModel}, the unknown vectors $\overline{\bf{h}}$, $\boldsymbol{\kappa}_{{\nu }}$, and ${\boldsymbol{\iota}_{{\tau}}}$ are remained to be estimated.
We can observe that when $k'' = \overline k^{i}_{\nu_{k''M_{\tau} + l''}}$ and $l'' = \overline l^{i}_{\tau_{k''M_{\tau} + l''}}$, $\forall i$, these unknown vectors can be defined as $\{\overline{\bf{h}}\}_{k''M_{\tau} + l''} = {\tilde h_i}$, ${\kappa _{{\nu_{{k''M_{\tau} + l''}}}}} = k_{\nu_{i}} - \overline k^{i}_{\nu_{k''M_{\tau} + l''}}$, and ${\iota _{{\tau_{{k''M_{\tau} + l''}} }}} = l_{\tau_{i}} - \overline l^{i}_{\tau_{k''M_{\tau} + l''}}$.
The remain entries in $\overline{\bf{h}}$, $\boldsymbol{\kappa}_{{\nu }}$, and ${\boldsymbol{\iota}_{{\tau}}}$ are all zeros.
In other words, $\overline{\bf{h}}$, $\boldsymbol{\kappa}_{{\nu }}$, ${\boldsymbol{\iota}_{{\tau}}}$ are all $P$-sparse channel vectors with the same support, i.e., joint sparsity \cite{YangZaiOffgridCE}.
The on-grid components are determined by the indices of the non-zero entries in $\overline{\bf{h}}$, while the off-grid components can be given by the estimated parameters $\boldsymbol{\kappa}_{{\nu }}$ and ${\boldsymbol{\iota}_{{\tau}}}$, as detailed in the next section.
We can observe that the proposed model degenerates to the case of on-grid sparse channel estimation \cite{ZhaoSBLOTFS} when forcing the off-grid components as zeros, i.e., ${\boldsymbol{\kappa}_{{\nu }}} = {\boldsymbol{\iota}_{{\tau}}} = \mathbf{0}$.
%
%\section{Off-grid 2D Compressed Channel Estimation}

\vspace{-2mm}
\section{1D Off-grid SBL-based Channel Estimation Algorithm}\label{1DCEOffgrid}
In this section, we propose the off-grid SBL algorithm for estimating the unknown sparse vectors in the 1D SSR problem in \eqref{SSRDModel}.

\vspace{-2mm}
\subsection{Hierarchical Hyperprior Distribution}
\begin{figure}
	\centering\vspace{-3mm}
	\includegraphics[width=3.3in]{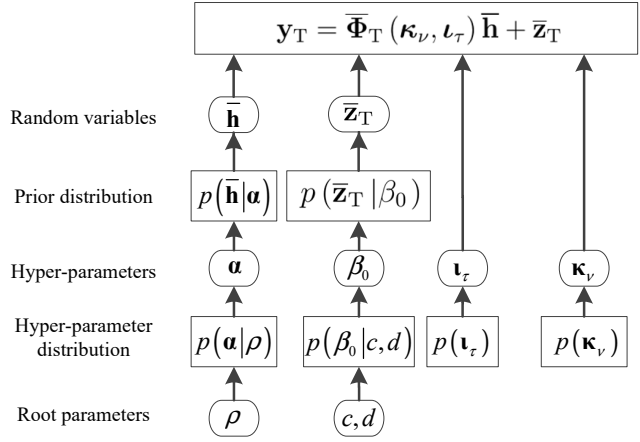}\vspace{-7mm}
	\caption{A hierarchical hyperprior distribution for the proposed off-grid SBL channel estimation.}\vspace{-10mm}
	\label{SBLStructure}%
\end{figure}

To exploit the sparsity of the original DD domain channel response, we formulate a hierarchical hyperprior distribution following the SBL framework \cite{YangZaiOffgridCE}, as illustrated in Fig. \ref{SBLStructure}.
In particular, the prior distribution of unknown channel vector $\overline {\bf{h}}$ is assumed to follow
\vspace{-2mm}
\begin{equation}\label{PriorH}
p\left( {\overline {\bf{h}} \left| {\boldsymbol{\alpha }} \right.} \right) = \mathcal{CN}\left( {\overline {\bf{h}} \left| {{\bf{0}},{\bf{\Lambda }}} \right.} \right),\vspace{-2mm}
\end{equation}
where the covariance matrix ${\bf{\Lambda }} = \diag\left( {\boldsymbol{\alpha }}  \right)$ is a diagonal matrix and ${\boldsymbol{\alpha }} = \left[{{\alpha _0}, \ldots ,{\alpha _{k''M_{\tau} + l''}}, \ldots {\alpha _{M_{\tau}N_{\nu} - 1}}}\right]^{\mathrm{T}} \in \mathbb{R}^{M_{\tau}N_{\nu} \times 1}$ with ${\alpha _{k''M_{\tau} + l''}}\ge0$ is the hyper-parameter to model the sparsity of $\overline {\bf{h}}$.
In particular, when ${\alpha _{k''M_{\tau} + l''}} = 0$, the corresponding channel coefficient $\{\overline{\bf{h}}\}_{k''M_{\tau} + l''} = 0$. Otherwise, the corresponding channel coefficient is non-zero.
To avoid the impact of the initialization of ${\boldsymbol{\alpha }}$ on the channel estimation performance, we further assume that the hyper-parameters ${\boldsymbol{\alpha }}$ are independent and follow the Gamma distribution \cite{YangZaiOffgridCE}: 
\vspace{-2mm}
\begin{equation}\label{PriorALPHA}
p\left( {{\boldsymbol{\alpha }}\left| \rho  \right.} \right) = \mathop \prod \limits_{k'' = 0}^{N_{\nu} - 1} \mathop \prod \limits_{l'' = 0}^{M_{\tau} - 1} \Gamma \left( {{\alpha _{k''M_{\tau} + l''}}\left| {1,\rho } \right.} \right),\vspace{-2mm}
\end{equation}
where $\rho  > 0$ is a fixed root parameter for ${\boldsymbol{\alpha }}$.
The hyperprior formulation defined in \eqref{PriorH} and \eqref{PriorALPHA} results in a Laplace prior, see (17) in \cite{BabacanBCS}, and hence a sparse estimation of $\overline {\bf{h}}$ is preferred.

Moreover, we assume that the noise vector $\overline {\bf{z}}_{\mathrm{T}}$ follows a complex Gaussian distribution
\vspace{-2mm}
\begin{equation}
	p\left( {\overline {\bf{z}}_{\mathrm{T}} \left| {{\beta _0}} \right.} \right) = \mathcal{CN}\left( {\overline {\bf{z}}_{\mathrm{T}} \left| {{\bf{0}},\beta _0^{ - 1}{{\bf{I}}_{M_{\mathrm{T}}N_{\mathrm{T}}}}} \right.} \right),\vspace{-2mm}
\end{equation}
where $\beta _0$ denotes the inverse of the unknown noise variance $\overline{\sigma^2}$ and it further follows:
\vspace{-2mm}
\begin{equation}
p\left( {{\beta _0}\left| {c,d} \right.} \right) = \Gamma \left( {{\beta _0}\left| {c,d} \right.} \right),\vspace{-2mm}
\end{equation}
with $c,d > 0$ are fixed root parameters.
Furthermore, unknown off-grid Doppler and delay variables follow a uniform distribution within their bounds in \eqref{FractionalBounds}, i.e.,
\vspace{-2mm}
\begin{equation}
{\kappa _{{\nu _{k''M_{\tau} + l''}}}} \sim \mathcal{U}\left[ { - \frac{1}{2}r_{\nu},\frac{1}{2}r_{\nu}} \right] \;\text{and}\;
{\iota _{{\tau _{k''M_{\tau} + l''}}}} \sim \mathcal{U}\left[ { - \frac{1}{2}r_{\tau},\frac{1}{2}}r_{\tau} \right],\forall k'',l''. \vspace{-2mm}
\end{equation}
As shown in Fig. \ref{SBLStructure}, the off-grid Doppler and delay $\left( {\boldsymbol{\kappa}_{{\nu }}},{\boldsymbol{\iota}_{{\tau}}} \right)$ are modeled as hyper-parameters, which will be updated concurrently with the posterior distribution of channel vector $\overline {\bf{h}}$ in our proposed algorithm in next section.

\vspace{-2mm}
\subsection{Off-grid SBL-based Channel Estimation Algorithm}\label{1DSSRAlgorithm}
According to the Bayesian principle, the optimal estimation of the unknown variables in \eqref{SSRDModel} can be obtained from the posterior distribution $p\left( {\overline {\bf{h}} ,{\boldsymbol{\alpha }},{{\boldsymbol{\kappa }}_\nu },{{\boldsymbol{\iota }}_\tau },{\beta _0}}\left| {\bf{y}}_{\mathrm{T}}\right. \right)$, which includes the full information of both their prior distribution and the measurement $\mathbf{y}_{\mathrm{T}}$.
However, the posterior distribution $p\left( {\overline {\bf{h}} ,{\boldsymbol{\alpha }},{{\boldsymbol{\kappa }}_\nu },{{\boldsymbol{\iota }}_\tau },{\beta _0}}\left| {\bf{y}}_{\mathrm{T}}\right. \right)$ cannot be expressed explicitly due to the non-linearly coupled unknown variables in the measurement matrix $\overline {\bf{\Phi }}_{\mathrm{T}} \left( {{{\boldsymbol{\kappa }}_\nu },{{\boldsymbol{\iota }}_\tau }} \right)$ in \eqref{SSRDModel}.
The unknown hyper-parameters ${\boldsymbol{\alpha }}$ and ${{\beta _0}}$, which determine the prior distribution of $\overline {\bf{h}}$ and the likelihood function $p\left( { {\bf{y}}_{\mathrm{T}}\left|\overline {\bf{h}} \right.}\right)$, respectively, are also the main obstacles to derive the posterior distribution.
Following the Bayes' rule, the posterior distribution can be decomposed as 
\vspace{-2mm}
\begin{equation}
	p\left( {\overline {\bf{h}} ,{\boldsymbol{\alpha }},{{\boldsymbol{\kappa }}_\nu },{{\boldsymbol{\iota }}_\tau },{\beta _0}}\left| {\bf{y}}_{\mathrm{T}}\right. \right) = p\left( \overline {\bf{h}}\left| {\bf{y}}_{\mathrm{T}},{\boldsymbol{\alpha }},{{\boldsymbol{\kappa }}_\nu },{{\boldsymbol{\iota }}_\tau },{\beta _0}\right. \right)p\left({\boldsymbol{\alpha }}, {{{\boldsymbol{\kappa }}_\nu },{{\boldsymbol{\iota }}_\tau },{\beta _0}}\left| {\bf{y}}_{\mathrm{T}}\right. \right).\vspace{-2mm}\label{PosteriorDistributionFull}
\end{equation}
Inspired by this decomposition, we prefer to derive the conditional posterior distribution of $\overline {\bf{h}}$ for given hyper-parameters $\left({\boldsymbol{\alpha }}, {{\boldsymbol{\kappa }}_\nu },{{\boldsymbol{\iota }}_\tau },{\beta _0}\right)$, i.e., $p\left( {\overline {\bf{h}} \left| {{\bf{y}}_{\mathrm{T}};{\boldsymbol{\alpha }},{{\boldsymbol{\kappa }}_\nu },{{\boldsymbol{\iota }}_\tau },{\beta _0}} \right.} \right)$, and then update the hyper-parameters $\left({\boldsymbol{\alpha }}, {{\boldsymbol{\kappa }}_\nu },{{\boldsymbol{\iota }}_\tau },{\beta _0}\right)$ to maximize the posterior distribution $p\left({\boldsymbol{\alpha }}, {{{\boldsymbol{\kappa }}_\nu },{{\boldsymbol{\iota }}_\tau },{\beta _0}}\left| {\bf{y}}_{\mathrm{T}}\right. \right)$.
In turn, the updated point estimates of the hyper-parameters can be further adopted to update the conditional posterior distribution.
Such an alternative inference calculation will be proceeded repeatedly until convergence.

In the $t$-th iteration, given the hyper-parameters $\left({{{\boldsymbol{\alpha }}^{\left( t \right)}},{\boldsymbol{\kappa }}_\nu ^{\left( t \right)},{\boldsymbol{\iota }}_\tau ^{\left( t \right)},\beta _0^{\left( t \right)}}\right)$, the conditional posterior distribution of $\overline {\bf{h}}$ is given by
\vspace{-4mm}
\begin{align}\label{PosteriorDistribution}
&p\left( {\overline {\bf{h}} \left| {{\bf{y}}_{\mathrm{T}};{{\boldsymbol{\alpha }}^{\left( t \right)}},{{\boldsymbol{\kappa }}_\nu^{\left( t \right)} },{{\boldsymbol{\iota }}_\tau^{\left( t \right)} },{\beta _0^{\left( t \right)}}} \right.} \right)  = \frac{{p\left( {{\bf{y}}_{\mathrm{T}},\overline {\bf{h}} ;{{\boldsymbol{\alpha }}^{\left( t \right)}},{{\boldsymbol{\kappa }}_\nu^{\left( t \right)} },{{\boldsymbol{\iota }}_\tau^{\left( t \right)} },{\beta _0^{\left( t \right)}}} \right)}}{{p\left( {{\bf{y}}_{\mathrm{T}}; {{\boldsymbol{\alpha }}^{\left( t \right)}},{{{\boldsymbol{\kappa }}_\nu^{\left( t \right)} },{{\boldsymbol{\iota }}_\tau^{\left( t \right)} },{\beta _0^{\left( t \right)}}} } \right)}}\notag\\[-1mm]
=&\frac{{p\left( {{\bf{y}}_{\mathrm{T}}\left| {\overline {\bf{h}} ;{{\boldsymbol{\kappa }}_\nu^{\left( t \right)} },{{\boldsymbol{\iota }}_\tau^{\left( t \right)} },{\beta _0^{\left( t \right)}}} \right.} \right)p\left( {\overline {\bf{h}} ; {\boldsymbol{\alpha }^{\left( t \right)}} } \right)}}{{p\left( {{\bf{y}}_{\mathrm{T}}; {{\boldsymbol{\alpha }}^{\left( t \right)}},{{{\boldsymbol{\kappa }}_\nu^{\left( t \right)} },{{\boldsymbol{\iota }}_\tau^{\left( t \right)} },{\beta _0^{\left( t \right)}}} } \right)}} \propto {p\left( {{\bf{y}}_{\mathrm{T}}\left| {\overline {\bf{h}} ;{{\boldsymbol{\kappa }}_\nu^{\left( t \right)} },{{\boldsymbol{\iota }}_\tau^{\left( t \right)} },{\beta _0^{\left( t \right)}}} \right.} \right)p\left( {\overline {\bf{h}} ; {\boldsymbol{\alpha }^{\left( t \right)}} } \right)},
%& \propto \frac{1}{{{\pi ^{MN}}\beta _0^{ - MN}}}\exp \left\{ { - {\beta _0}{{\left( {{\bf{y}}_{\mathrm{T}}_{\mathrm{T}} - \overline {\bf{\Phi }} \left( {{{\bf{\kappa }}_\nu },{{\bf{\iota }}_\tau }} \right)\overline {\bf{h}} } \right)}^{\rm{H}}}\left( {{\bf{y}}_{\mathrm{T}} - \overline {\bf{\Phi }} \left( {{{\bf{\kappa }}_\nu },{{\bf{\iota }}_\tau }} \right)\overline {\bf{h}} } \right)} \right\}\frac{1}{{{\pi ^{MN}}\mathop \prod \limits_{k'' = 0}^{N - 1} \mathop \prod \limits_{l'' = 0}^{M - 1} {\alpha _{k''M + l''}}}}\exp \left\{ { - {{\overline {\bf{h}} }^{\rm{H}}}{{\bf{\Lambda }}^{ - 1}}\overline {\bf{h}} } \right\}
\end{align}
\vspace{-6mm}\par\noindent
where the prior distribution $p\left( {\overline {\bf{h}} ; {\boldsymbol{\alpha }^{\left( t \right)}} } \right)$ is given by \eqref{PriorH} and $p\left( {{\bf{y}}_{\mathrm{T}}\left| {\overline {\bf{h}} ;{{\boldsymbol{\kappa }}_\nu^{\left( t \right)} },{{\boldsymbol{\iota }}_\tau^{\left( t \right)} },{\beta _0^{\left( t \right)}}} \right.} \right)$ denotes the likelihood function for given hyper-parameters $\left({{{\boldsymbol{\kappa }}_\nu^{\left( t \right)} },{{\boldsymbol{\iota }}_\tau^{\left( t \right)} },{\beta _0^{\left( t \right)}}}\right)$ and it is given by
\vspace{-2mm}
\begin{equation}\label{Likelihood}
	p\left( {{\bf{y}}_{\mathrm{T}}\left| {\overline {\bf{h}} ;{{\boldsymbol{\kappa }}_\nu^{\left( t \right)} },{{\boldsymbol{\iota }}_\tau^{\left( t \right)} },{\beta _0^{\left( t \right)}}} \right.} \right) = \mathcal{CN}\left( {{\bf{y}}_{\mathrm{T}}\left| {\overline {\bf{\Phi }}_{\mathrm{T}} \left( {{{\boldsymbol{\kappa }}_\nu^{\left( t \right)} },{{\boldsymbol{\iota }}_\tau^{\left( t \right)} }} \right) \overline {\bf{h}} ,\left(\beta _0^{\left( t \right)}\right)^{ - 1}{{\bf{I}}_{{M_{\mathrm{T}}N_{\mathrm{T}}}}}} \right.} \right).\vspace{-2mm}
\end{equation}
Note that ${p\left( {{\bf{y}}_{\mathrm{T}}; {{\boldsymbol{\alpha }}^{\left( t \right)}},{{{\boldsymbol{\kappa }}_\nu^{\left( t \right)} },{{\boldsymbol{\iota }}_\tau^{\left( t \right)} },{\beta _0^{\left( t \right)}}} } \right)} $ in \eqref{Likelihood} is a normalized term, which is omitted to facilitate the derivation.
In fact, ${p\left( {{\bf{y}}_{\mathrm{T}}; {{\boldsymbol{\alpha }}^{\left( t \right)}},{{{\boldsymbol{\kappa }}_\nu^{\left( t \right)} },{{\boldsymbol{\iota }}_\tau^{\left( t \right)} },{\beta _0^{\left( t \right)}}} } \right)}$ can be interpreted as the model evidence, which measures the possibility of the observations ${\bf{y}}_{\mathrm{T}}$ given the hyper-parameters. 

Combining \eqref{PriorH}, \eqref{PosteriorDistribution}, and \eqref{Likelihood}, the conditional posterior distribution of $\overline {\bf{h}}$  can be obtained by
\vspace{-5mm}
\begin{equation}
p\left( {\overline {\bf{h}} \left| {{\bf{y}}_{\mathrm{T}};{{\boldsymbol{\alpha }}^{\left( t \right)}},{{\boldsymbol{\kappa }}_\nu^{\left( t \right)} },{{\boldsymbol{\iota }}_\tau^{\left( t \right)} },{\beta _0^{\left( t \right)}}} \right.} \right) = \mathcal{CN}\left(\overline {\bf{h}} \left|\boldsymbol{\mu}_{\overline {\bf{h}}}^{\left( t \right)},\boldsymbol{\Sigma }_{\overline {\bf{h}}}^{\left( t \right)} \right.\right)\vspace{-2mm}
\end{equation}
and the conditional posterior mean and covariance matrix in the $t$-th iteration are given by\footnote{The estimation of the posterior mean and covariance matrix in \eqref{PosteriorMean} and \eqref{PosteriorcOV} are the optimal solution of the inverse problem in \eqref{SSRDModel} for given $\left({{{\boldsymbol{\alpha }}^{\left( t \right)}},{\boldsymbol{\kappa }}_\nu ^{\left( t \right)},{\boldsymbol{\iota }}_\tau ^{\left( t \right)},\beta _0^{\left( t \right)}}\right)$.
Note that some low-complexity approximate inference methods\cite{ThomasSBL}, such as variational Bayesian or message passing algorithm, can achieve a suboptimal solution, which are out of the scope of this paper.}
\vspace{-2mm}
\begin{align}
	\boldsymbol{\mu}_{\overline {\bf{h}}}^{\left( t \right)} & = {\beta _0^{\left( t \right)}}{{\bf{\Sigma }}_{\overline {\bf{h}} }^{\left( t \right)}}{\overline {\bf{\Phi }}_{\mathrm{T}} ^{\rm{H}}}\left( {{{\boldsymbol{\kappa }}_\nu^{\left( t \right)} },{{\boldsymbol{\iota }}_\tau^{\left( t \right)} }} \right){\bf{y}}_{\mathrm{T}}\;\text{and}\label{PosteriorMean}\\[-1mm]
	\boldsymbol{\Sigma }_{\overline {\bf{h}}}^{\left( t \right)} & = {\left( {{\beta _0}{{\overline {\bf{\Phi }}_{\mathrm{T}}\left( {{{\boldsymbol{\kappa }}_\nu^{\left( t \right)} },{{\boldsymbol{\iota }}_\tau^{\left( t \right)} }} \right) }^{\rm{H}}}\overline {\bf{\Phi }}_{\mathrm{T}}\left( {{{\boldsymbol{\kappa }}_\nu^{\left( t \right)} },{{\boldsymbol{\iota }}_\tau^{\left( t \right)} }} \right) + {\left({\bf{\Lambda }}^{\left( t \right)}\right)^{ - 1}}} \right)^{ - 1}} \label{PosteriorcOV} \\[-1mm]
	& = {\boldsymbol{\Lambda }}^{\left( t \right)} - {\boldsymbol{\Lambda }}^{\left( t \right)}\overline {\boldsymbol{\Phi }}_{\mathrm{T}} {\left( {{{\boldsymbol{\kappa }}_\nu^{\left( t \right)} },{{\boldsymbol{\iota }}_\tau^{\left( t \right)} }} \right)^{\rm{H}}}\label{PosteriorcOVII}\\[-1mm]
	&\times {\left( {\left(\beta _0^{\left( t \right)}\right)^{ - 1}{{\bf{I}}_{M_{\mathrm{T}}N_{\mathrm{T}}}} + \overline {\boldsymbol{\Phi }}_{\mathrm{T}} \left( {{{\boldsymbol{\kappa }}_\nu^{\left( t \right)} },{{\boldsymbol{\iota }}_\tau^{\left( t \right)} }} \right){\boldsymbol{\Lambda }}^{\left( t \right)}\overline {\boldsymbol{\Phi }}_{\mathrm{T}} {{\left( {{{\boldsymbol{\kappa }}_\nu^{\left( t \right)} },{{\boldsymbol{\iota }}_\tau^{\left( t \right)} }} \right)}^{\rm{H}}}} \right)^{ - 1}}\overline {\boldsymbol{\Phi }}_{\mathrm{T}} \left( {{{\boldsymbol{\kappa }}_\nu^{\left( t \right)} },{{\boldsymbol{\iota }}_\tau^{\left( t \right)} }} \right){\boldsymbol{\Lambda }}^{\left( t \right)}, \notag
\end{align}
\vspace{-10mm}\par\noindent
respectively.
Note that the covariance matrix in \eqref{PosteriorcOVII} is obtained by the matrix inversion lemma, which can significantly reduce the computational complexity of the matrix inverse operation in \eqref{PosteriorcOV} as usually $M_{\tau} \gg M_{\mathrm{T}}$ and $N_{\nu} \gg N_{\mathrm{T}}$.

Then, given the conditional posterior distribution $p\left( {\overline {\bf{h}} \left| {{\bf{y}}_{\mathrm{T}};{{\boldsymbol{\alpha }}^{\left( t \right)}},{{\boldsymbol{\kappa }}_\nu^{\left( t \right)} },{{\boldsymbol{\iota }}_\tau^{\left( t \right)} },{\beta _0^{\left( t \right)}}} \right.} \right)$ in the $t$-th iteration, the hyper-parameters $\left({\boldsymbol{\alpha }}, {{\boldsymbol{\kappa }}_\nu },{{\boldsymbol{\iota }}_\tau },{\beta _0}\right)$ can be updated by maximizing the posterior distribution $p\left({\boldsymbol{\alpha }}, {{{\boldsymbol{\kappa }}_\nu },{{\boldsymbol{\iota }}_\tau },{\beta _0}}\left| {\bf{y}}_{\mathrm{T}}\right. \right)$.
Note that $p\left({\boldsymbol{\alpha }}, {{{\boldsymbol{\kappa }}_\nu },{{\boldsymbol{\iota }}_\tau },{\beta _0}}\left| {\bf{y}}_{\mathrm{T}}\right. \right) \propto p\left( {{\bf{y}}_{\mathrm{T}} ,{\boldsymbol{\alpha }}, {{\boldsymbol{\kappa }}_\nu },{{\boldsymbol{\iota }}_\tau },{\beta _0}} \right)$ as the observation ${\bf{y}}_{\mathrm{T}}$ is independent of the hyper-parameters.
An expectation-maximization (EM) algorithm\cite{MoonSPM} is employed for updating the hyper-parameters in the $\left(t+1\right)$-th iteration.
In particular, the expectation step is w.r.t. the joint distribution function $p\left( {{\bf{y}}_{\mathrm{T}},\overline {\bf{h}} ,{\boldsymbol{\alpha }}, {{\boldsymbol{\kappa }}_\nu },{{\boldsymbol{\iota }}_\tau },{\beta _0}} \right)$ via marginalizing over the conditional posterior distribution of $\overline {\bf{h}}$.
According to the Bayes' rule, the joint distribution $p\left( {{\bf{y}}_{\mathrm{T}},\overline {\bf{h}} ,{\boldsymbol{\alpha }},{{\boldsymbol{\kappa }}_\nu },{{\boldsymbol{\iota }}_\tau },{\beta _0}} \right)$ can be decomposed as
\vspace{-2mm}
\begin{equation}\label{JointDistribution}
p\left( {{\bf{y}}_{\mathrm{T}},\overline {\bf{h}} ,{\boldsymbol{\alpha }},{{\boldsymbol{\kappa }}_\nu },{{\boldsymbol{\iota }}_\tau },{\beta _0}} \right) = p\left( {{\bf{y}}_{\mathrm{T}}\left| {\overline {\bf{h}} ,{{\boldsymbol{\kappa }}_\nu },{{\boldsymbol{\iota }}_\tau },{\beta _0}} \right.} \right)p\left( {\overline {\bf{h}} \left| {\boldsymbol{\alpha }} \right.} \right)p\left( {\boldsymbol{\alpha }} \right)p\left( {{{\boldsymbol{\kappa }}_\nu }} \right)p\left( {{{\boldsymbol{\iota }}_\tau }} \right)p\left( {{\beta _0}} \right).\vspace{-2mm}
\end{equation}

Now, the expectation and maximization steps can be defined as follows:
\begin{itemize}
	\item Expectation step: In this step, we aim to obtain an objective function of the hyper-parameters $\left({\boldsymbol{\alpha }}, {{\boldsymbol{\kappa }}_\nu },{{\boldsymbol{\iota }}_\tau },{\beta _0}\right)$ via averaging the joint distribution in \eqref{JointDistribution} over the conditional posterior distribution of $\overline {\bf{h}}$, i.e.,
	\vspace{-2mm}
	\begin{equation}\label{ExpectationStep}
	\hspace{-5mm}Q\left( {{\boldsymbol{\alpha }},{{\boldsymbol{\kappa }}_\nu },{{\boldsymbol{\iota }}_\tau },{\beta _0}\left| {{{\boldsymbol{\alpha }}^{\left( t \right)}},{\boldsymbol{\kappa }}_\nu ^{\left( t \right)},{\boldsymbol{\iota }}_\tau ^{\left( t \right)},\beta _0^{\left( t \right)}} \right.} \right) = {E_{\overline {\bf{h}} \left| {{\bf{y}}_{\mathrm{T}};{{\boldsymbol{\alpha }}^{\left( t \right)}},{\boldsymbol{\kappa }}_\nu ^{\left( t \right)},{\boldsymbol{\iota }}_\tau ^{\left( t \right)},\beta _0^{\left( t \right)}} \right.}}\left\{ {\ln \left[ {p\left( {{\bf{y}}_{\mathrm{T}},\overline {\bf{h}} ,{\boldsymbol{\alpha }},{{\boldsymbol{\kappa }}_\nu },{{\boldsymbol{\iota }}_\tau },{\beta _0}} \right)} \right]} \right\}.\vspace{-2mm}
	\end{equation} 
	
	\item Maximization step: In this step, we optimize each hyper-parameter alternatively to maximize the obtained objective function in the expectation step , i.e., 
	\vspace{-2mm}
	\begin{align} 
	{{\boldsymbol{\alpha }}^{\left( t + 1\right)}} &= \arg \maxo Q\left( {{\boldsymbol{\alpha }}\left| {{{\boldsymbol{\alpha }}^{\left( t \right)}},{\boldsymbol{\kappa }}_\nu ^{\left( t \right)},{\boldsymbol{\iota }}_\tau ^{\left( t \right)},\beta _0^{\left( t \right)}} \right.} \right), \label{MaximizationStepI}\\[-1mm]
	{\boldsymbol{\kappa }}_\nu ^{\left( t+1 \right)} &= \arg \maxo Q\left( {{{\boldsymbol{\kappa }}_\nu }\left| {{{\boldsymbol{\alpha }}^{\left( t \right)}},{\boldsymbol{\kappa }}_\nu ^{\left( t \right)},{\boldsymbol{\iota }}_\tau ^{\left( t \right)},\beta _0^{\left( t \right)}} \right.} \right), \label{MaximizationStepII}\\[-1mm]
	{\boldsymbol{\iota }}_\tau ^{\left( t +1 \right)} &= \arg \maxo Q\left( {{{\boldsymbol{\iota }}_\tau }\left| {{{\boldsymbol{\alpha }}^{\left( t \right)}},{\boldsymbol{\kappa }}_\nu ^{\left( t \right)},{\boldsymbol{\iota }}_\tau ^{\left( t \right)},\beta _0^{\left( t \right)}} \right.} \right),\;\text{and} \label{MaximizationStepIII}\\[-1mm]
	\beta _0^{\left( t+1 \right)} &= \arg \maxo Q\left( {{\beta _0}\left| {{{\boldsymbol{\alpha }}^{\left( t \right)}},{\boldsymbol{\kappa }}_\nu ^{\left( t \right)},{\boldsymbol{\iota }}_\tau ^{\left( t \right)},\beta _0^{\left( t \right)}} \right.} \right). \label{MaximizationStepIV}
	\end{align}
	\vspace{-10mm}\par\noindent
\end{itemize}

Following definitions in \eqref{ExpectationStep}-\eqref{MaximizationStepIV}, the closed-form updating rule for each hyper-parameter will be derived as follows.
{In particular, the updating rule for the hyper-parameter $\left({\boldsymbol{\alpha }},{\beta _0}\right)$ are stated in the following proposition.}

{\begin{proposition}\label{Propo1}
The hyper-parameter $\left({\boldsymbol{\alpha }},{\beta _0}\right)$ can be updated by
\vspace{-2mm}
\begin{align}
\alpha _{k''M_{\tau} + l''}^{\left( {t + 1} \right)} &= \frac{{\sqrt {1 + 4\rho\left\{ {{\left| {\{{{\boldsymbol{\mu }}^{\left(t\right)}_{\overline {\bf{h}} }}\}_{k''M_{\tau} + l''}} \right|}^2} + \{{{\boldsymbol{\Sigma }}^{\left(t\right)}_{\overline {\bf{h}} }}\}_{k''M_{\tau} + l'',k''M_{\tau} + l''} \right\}}  - 1}}{{2\rho }}\;\text{and}\label{HyperParameterI}\\[-1mm]
{\beta _0}^{\left( {t + 1} \right)} &= \frac{{c - 1 + M_{\mathrm{T}}N_{\mathrm{T}}}}{{d + {A_{{\beta _0}}}}},\label{HyperParameterIV}
\end{align}
\vspace{-10mm}\par\noindent
respectively, where 
\vspace{-4mm}
\begin{equation}\label{Abeta0}
	{A_{{\beta _0}}} \hspace{-1mm}= \hspace{-1mm} {\left\| {{\bf{y}}_{\mathrm{T}}  \hspace{-0.5mm}- \hspace{-0.5mm} \overline {\boldsymbol{\Phi }}_{\mathrm{T}}  \hspace{-0.5mm}\left( {{\boldsymbol{\kappa }}_\nu ^{\left( t \right)},{\boldsymbol{\iota }}_\tau ^{\left( t \right)}} \right) \hspace{-0.5mm}{\boldsymbol{\mu }}_{\overline {\bf{h}} }^{\left(t\right)}} \right\|^2}{\rm{  \hspace{-0.5mm}+ \hspace{-0.5mm} }}\left( \hspace{-0.5mm}\beta^{\left(t\right)} _0 \hspace{-0.5mm}\right) ^{ - 1} \hspace{-0.5mm}\mathop \sum \limits_{k'' = 0}^{N_\nu - 1} \mathop \sum \limits_{l'' = 0}^{M_\tau - 1}  \hspace{-0.5mm}\left( \hspace{-0.5mm} {1  \hspace{-0.5mm}- \hspace{-0.5mm}  \left(\alpha _{k''M_\tau + l''}^{\left( t \right)}\right)^{ - 1}{{ \hspace{-0.5mm}\left\{ {{\bf{\Sigma }}_{\overline {\bf{h}} }^{\left( t \right)}} \right\}}_{k'',l''}}}  \hspace{-0.5mm}\right).\vspace{-2mm}
\end{equation}
\end{proposition}}

\begin{proof}
Please refer to Appendix of Proposition \ref{Propo1}.
\end{proof}

To update off-grid components of the Doppler shift, i.e., ${{\boldsymbol{\kappa }}_\nu }$, the objective function in \eqref{MaximizationStepII} can be obtained by
\vspace{-2mm}
\begin{align}
	& Q\left( {{{\boldsymbol{\kappa }}_\nu }\left| {{{\boldsymbol{\alpha }}^{\left( t \right)}},{\boldsymbol{\kappa }}_\nu ^{\left( t \right)},{\boldsymbol{\iota }}_\tau ^{\left( t \right)},\beta _0^{\left( t \right)}} \right.} \right) = {E_{\overline {\bf{h}} \left| {{\bf{y}}_{\mathrm{T}};{{\boldsymbol{\alpha }}^{\left( t \right)}},{\boldsymbol{\kappa }}_\nu ^{\left( t \right)},{\boldsymbol{\iota }}_\tau ^{\left( t \right)},\beta _0^{\left( t \right)}} \right.}}\left\{ {\ln \left[ {p\left( {{\bf{y}}_{\mathrm{T}}\left| {\overline {\bf{h}} ,{{\boldsymbol{\kappa }}_\nu },{{\boldsymbol{\iota }}_\tau^{\left( t \right)} },{\beta _0^{\left( t \right)}}} \right.} \right)p\left( {{{\boldsymbol{\kappa }}_\nu }} \right)} \right]} \right\} \notag\\[-1mm]
	= & {E_{\overline {\bf{h}} \left| {{\bf{y}}_{\mathrm{T}};{{\boldsymbol{\alpha }}^{\left( t \right)}},{\boldsymbol{\kappa }}_\nu ^{\left( t \right)},{\boldsymbol{\iota }}_\tau ^{\left( t \right)},\beta _0^{\left( t \right)}} \right.}}\left\{ {\ln \left[ {{\cal C}{\cal N}\left( {{\bf{y}}_{\mathrm{T}}\left| {\overline {\boldsymbol{\Phi }}_{\mathrm{T}} \left( {{{\boldsymbol{\kappa }}_\nu },{{\boldsymbol{\iota }}_\tau^{\left( t \right)} }} \right)\overline {\bf{h}} ,\left(\beta _0^{\left( t \right)}\right)^{ - 1}{{\bf{I}}_{M_{\mathrm{T}}N_{\mathrm{T}}}}} \right.} \right)} \right]} \right\}.\label{ObjectiveFunctionNu}
\end{align}
\vspace{-8mm}\par\noindent
It can be seen that maximizing the objective function in \eqref{ObjectiveFunctionNu} is equivalent to
\vspace{-2mm}
\begin{equation}
	 \mathop {\mino }\limits_{{\boldsymbol{\kappa }}_\nu} {E_{\overline {\bf{h}} \left| {{\bf{y}}_{\mathrm{T}};{{\boldsymbol{\alpha }}^{\left( t \right)}},{\boldsymbol{\kappa }}_\nu ^{\left( t \right)},{\boldsymbol{\iota }}_\tau ^{\left( t \right)},\beta _0^{\left( t \right)}} \right.}}\hspace{-0.5mm}\left\{ {{{\left\| {{\bf{y}}_{\mathrm{T}} \hspace{-0.5mm}-\hspace{-0.5mm} \overline {\boldsymbol{\Phi }}_{\mathrm{T}} \left( {{{\boldsymbol{\kappa }}_\nu },{\boldsymbol{\iota }}_\tau^{\left( t \right)}} \right)\overline {\bf{h}} } \right\|}^2}} \right\}\hspace{-0.5mm}
	\Leftrightarrow\hspace{-0.5mm}   \mathop {\mino }\limits_{{\boldsymbol{\kappa }}_\nu} \left( {{\boldsymbol{\kappa }}_\nu ^{\rm{T}}{{\bf{A}}_\nu }{{\boldsymbol{\kappa }}_\nu } \hspace{-0.5mm}-\hspace{-0.5mm} 2{\bf{b}_\nu^{\rm{T}}}{{\boldsymbol{\kappa }}_\nu }} \right),\vspace{-2mm}
\end{equation}
where ${{\bf{A}}_\nu } \in \mathbb{R}^{M_{\tau}N_{\nu} \times M_{\tau}N_{\nu}}$ and ${{\bf{b}}_{{\nu }}} \in \mathbb{R}^{M_{\tau}N_{\nu} \times 1}$ are given by
\vspace{-2mm}
\begin{align}
{{\bf{A}}_\nu } &= \Re \left\{{\boldsymbol{\Phi }}_{{\mathrm{T}},\nu} ^{\rm{H}}{{\boldsymbol{\Phi }}_{{\mathrm{T}},\nu} } \odot \left( { \left({\boldsymbol{\mu }}_{\overline {\bf{h}} }^{\left(t\right)}\right) ^*\left({\boldsymbol{\mu }}_{\overline {\bf{h}} }^{\left(t\right)}\right) ^{\rm{T}} + \left({\boldsymbol{\Sigma }}_{\overline {\bf{h}} }^{\left(t\right)}\right)^{\rm{T}}} \right)\right\} \;\text{and}\\[-1mm]
{{\bf{b}}_{{\nu }}} &= \Re \left\{ \diag\left( {{\boldsymbol{\mu }}_{\overline {\bf{h}} }^{\left(t\right)}} \right){\boldsymbol{\Phi }}_{{\mathrm{T}},\nu} ^{\rm{T}}{{\bf{y}}_{\mathrm{T}}^*} \hspace{-0.5mm} - \hspace{-0.5mm}\diag\left( {\left( {{\boldsymbol{\mu }}_{\overline {\bf{h}} }^{\left(t\right)}  \left({\boldsymbol{\mu }}_{\overline {\bf{h}} }^{\left(t\right)}\right)^{\rm{H}} \hspace{-0.5mm}+ \hspace{-0.5mm}{\boldsymbol{\Sigma }}_{\overline {\bf{h}} }^{\left(t\right)}} \right){{\left( {{\boldsymbol{\Phi }}_{{{\mathrm{T}}}} \hspace{-0.5mm}+\hspace{-0.5mm} {{\boldsymbol{\Phi }}_{{\mathrm{T}},\tau} }{\diag}\left( {{{\boldsymbol{\iota }}_\tau^{\left(t\right)} }} \right)} \right)}^{\rm{H}}}{{\boldsymbol{\Phi }}_{{\mathrm{T}},\nu} }} \right) \right\}, \notag
\end{align}
\vspace{-9mm}\par\noindent
respectively.

Since ${{\boldsymbol{\kappa }}_\nu }$ is jointly sparse with $\overline {\bf{h}}$ sharing the same support, to reduce the complexity for optimizing ${{\boldsymbol{\kappa }}_\nu }$, we can truncate ${{\bf{A}}_\nu }$, ${{\boldsymbol{\kappa }}_\nu }$, and ${{\bf{b}}_{\boldsymbol{\nu }}}$ according to ${{\boldsymbol{\alpha }}^{\left( t \right)}}$.
However, the number of paths $P$, i.e., the number of non-zero entries in $\overline {\bf{h}}$ is unknown.
As typically a sparse vector with a length of $M_{\tau}N_{\nu}$ and at most $\frac{M_{\mathrm{T}}N_{\mathrm{T}}}{\ln\left(M_{\tau}N_{\nu}\right)}$ non-zero entries can be recovered based on ${M_{\mathrm{T}}N_{\mathrm{T}}}$ measurements by compressed sensing technique\cite{CandesCS}, we set $\hat P = \left\lfloor {\frac{M_{\mathrm{T}}N_{\mathrm{T}}}{\ln\left(M_{\tau}N_{\nu}\right)}} \right\rfloor$ and truncate ${{\bf{A}}_\nu }$, ${{\boldsymbol{\kappa }}_\nu }$, and ${{\bf{b}}_{\boldsymbol{\nu }}}$ accordingly.
In particular, if ${\alpha _{k''M_{\tau} + l''}^{\left(t\right)}}$ is one of the largest $\hat P$ entries of ${{\boldsymbol{\alpha }}^{\left( t \right)}}$, the $\left[{k''M_{\tau} + l''},{k''M_{\tau} + l''}\right]$-th entry of ${{\bf{A}}_\nu }$ and the $\left({k''M_{\tau} + l''}\right)$-th entry of ${{\bf{b}}_{{\nu }}}$ will be selected, resulting in ${{\bf{A}}^{\mathrm{T}}_\nu } \in \mathbb{R}^{\hat P \times \hat P}$ and ${{\bf{b}}^{\mathrm{T}}_{{\nu }}} \in \mathbb{R}^{\hat P \times 1}$, respectively.
The index $k''M_{\tau} + l''$ will be included in the truncation index set $\mathcal{S}^{\left(t\right)}_{\nu}$ with $\left|\mathcal{S}^{\left(t\right)}_{\nu}\right| = \hat P$.
Also, the $\left({k''M_{\tau} + l''}\right)$-th entry of ${{\boldsymbol{\kappa }}_\nu }$ will be selected, resulting a truncated version of unknown hyper-parameter, i.e., ${{\boldsymbol{\kappa }}^{\mathrm{T}}_\nu } \in \mathbb{R}^{\hat P \times 1}$.
When ${{\bf{A}}^{\mathrm{T}}_\nu }$ is invertible, ${{\boldsymbol{\kappa }}^{\mathrm{T}}_\nu }$ can be updated by
\vspace{-4mm}
\begin{equation}
\left({{\boldsymbol{\kappa }}^{\mathrm{T}}_\nu }\right)^{\left(t+1\right)} = \left({{\bf{A}}^{\mathrm{T}}_\nu }\right)^{-1} {{\bf{b}}^{\mathrm{T}}_{{\nu }}}.\vspace{-3mm}
\end{equation}
Otherwise, we perform an element-wise update for ${{\boldsymbol{\kappa }}^{\mathrm{T}}_\nu }$, i.e., 
\vspace{-2mm}
\begin{equation}
\{\left({{\boldsymbol{\kappa }}^{\mathrm{T}}_\nu }\right)^{\left(t+1\right)}\}_n =  \frac{\{{\bf{b}}^{\mathrm{T}}_{{\nu }}\}_n - \{\{{{\bf{A}}^{\mathrm{T}}_\nu }\}_n\}_{-n}^{\mathrm{T}} \{{{\boldsymbol{\kappa }}^{\mathrm{T}}_\nu }\}_{-n}}{\{{{\bf{A}}^{\mathrm{T}}_\nu }\}_{n,n}}, \forall n \in \{1, \ldots, \hat P\}.\vspace{-2mm}
\end{equation}
Note that ${\kappa _{{\nu_{{k''M_{\tau} + l''}}}}} \in \left[-\frac{1}{2}r_{\nu},\frac{1}{2}r_{\nu}\right]$ and thus we have
\vspace{-2mm}
\begin{equation}\label{HyperParameterII}
{\{ \overline {\left({{\boldsymbol{\kappa }}^{\mathrm{T}}_\nu }\right)^{\left(t+1\right)}} \} _n} = \left\{ {\begin{array}{*{20}{c}}
	{{{\{ \left({{\boldsymbol{\kappa }}^{\mathrm{T}}_\nu }\right)^{\left(t+1\right)}\} }_n}},&{{\rm{if }}\;{{\{ \left({{\boldsymbol{\kappa }}^{\mathrm{T}}_\nu }\right)^{\left(t+1\right)}\} }_n} \in \left[ { - \frac{1}{2}r_{\nu},\frac{1}{2}r_{\nu}} \right]},\\[-1mm]
	{ - \frac{1}{2}r_{\nu}},&{{\rm{if }}\;{{\{ \left({{\boldsymbol{\kappa }}^{\mathrm{T}}_\nu }\right)^{\left(t+1\right)}\} }_n} <  - \frac{1}{2}r_{\nu}},\\[-1mm]
	{\frac{1}{2}r_{\nu}},&{{\rm{otherwise}}}.
	\end{array}} \right.\vspace{-2mm}
\end{equation}
Now, given the estimated $\overline {\left({{\boldsymbol{\kappa }}^{\mathrm{T}}_\nu }\right)^{\left(t+1\right)}}$, we can recover $\overline {\left({{\boldsymbol{\kappa }}_\nu }\right)^{\left(t+1\right)}}$ based on the truncation index set $\mathcal{S}^{\left(t\right)}_{\nu}$.
Note that due to the estimation error of ${{\boldsymbol{\alpha }}^{\left( t \right)}}$, the truncation above may introduce performance loss in estimating the off-grid Doppler shift.
However, when the iterations is close to convergence, the estimation of ${{\boldsymbol{\alpha }}^{\left( t \right)}}$ is accurate and thus the performance loss due to the proposed truncation is marginal.

Similarly, the off-grid components of the delay shift, i.e., ${{\boldsymbol{\iota }}_\tau }$, can be updated by
\vspace{-2mm}
\begin{equation}
\mathop {\mino }\limits_{{\boldsymbol{\iota }}_\tau} {E_{\overline {\bf{h}} \left| {{\bf{y}}_{\mathrm{T}};{{\boldsymbol{\alpha }}^{\left( t \right)}},{\boldsymbol{\kappa }}_\nu ^{\left( t \right)},{\boldsymbol{\iota }}_\tau ^{\left( t \right)},\beta _0^{\left( t \right)}} \right.}}\left\{ {{{\left\| {{\bf{y}}_{\mathrm{T}} - \overline {\boldsymbol{\Phi }}_{\mathrm{T}} \left( {{{\boldsymbol{\kappa }}_\nu^{\left( t \right)} },{\boldsymbol{\iota }}_\tau} \right)\overline {\bf{h}} } \right\|}^2}} \right\}
\Leftrightarrow  \mathop {\mino }\limits_{{\boldsymbol{\iota }}_\tau} \left( {{\boldsymbol{\iota }}_\tau ^{\rm{T}}{{\bf{A}}_\tau }{\boldsymbol{\iota }}_\tau ^{} - 2{\bf{b}}_\tau ^{\rm{T}}{\boldsymbol{\iota }}_\tau ^{}} \right),\vspace{-3mm}
\end{equation}
where ${{\bf{A}}_\tau } \in \mathbb{R}^{M_{\tau}N_{\nu} \times M_{\tau}N_{\nu}}$ and ${{\bf{b}}_{{\tau }}} \in \mathbb{R}^{M_{\tau}N_{\nu} \times 1}$ are given by
\vspace{-2mm}
\begin{align}
{{\bf{A}}_\tau } &= \Re \left\{{\boldsymbol{\Phi }}_{{\mathrm{T}},\tau} ^{\rm{H}}{{\boldsymbol{\Phi }}_{{\mathrm{T}},\tau} } \odot \left( { \left({\boldsymbol{\mu }}_{\overline {\bf{h}} }^{\left(t\right)}\right) ^*\left({\boldsymbol{\mu }}_{\overline {\bf{h}} }^{\left(t\right)}\right) ^{\rm{T}} + \left({\boldsymbol{\Sigma }}_{\overline {\bf{h}} }^{\left(t\right)}\right)^{\rm{T}}} \right)\right\} \;\text{and}\\[-1mm]
{{\bf{b}}_{{\tau }}} &= \Re \left\{ \diag\left( {{\boldsymbol{\mu }}_{\overline {\bf{h}} }^{\left(t\right)}} \right){\boldsymbol{\Phi }}_{{\mathrm{T}},\tau} ^{\rm{T}}{{\bf{y}}_{\mathrm{T}}^*} \hspace{-0.5mm} - \hspace{-0.5mm}\diag\left( {\left( {{\boldsymbol{\mu }}_{\overline {\bf{h}} }^{\left(t\right)}  \left({\boldsymbol{\mu }}_{\overline {\bf{h}} }^{\left(t\right)}\right)^{\rm{H}}\hspace{-0.5mm} +\hspace{-0.5mm} {\boldsymbol{\Sigma }}_{\overline {\bf{h}} }^{\left(t\right)}} \right){{\left( {{\boldsymbol{\Phi }}_{\mathrm{T}} \hspace{-0.5mm}+\hspace{-0.5mm} {{\boldsymbol{\Phi }}_{\mathrm{T},\nu}}\diag\left( {{{\boldsymbol{\kappa }}_\nu^{\left( t \right)} }} \right)} \right)}^{\rm{H}}}{{\boldsymbol{\Phi }}_{{\mathrm{T}},\tau} }} \right) \right\},  \notag
\end{align}
\vspace{-8mm}\par\noindent
respectively.
Similarly, we can truncate ${{\bf{A}}_\tau }$, ${{\mathbf{b}}_{{\tau }}}$, and ${{\boldsymbol{\iota }}_\tau }$ according to ${{\boldsymbol{\alpha }}^{\left( t \right)}}$ as ${{\bf{A}}^{\mathrm{T}}_\tau } \in \mathbb{R}^{\hat P \times \hat  P}$, ${{\bf{b}}^{\mathrm{T}}_{{\tau }}} \in \mathbb{R}^{\hat P \times 1}$, and ${{\boldsymbol{\iota }}^{\mathrm{T}}_\tau } \in \mathbb{R}^{\hat P \times 1}$ respectively, and obtain a truncation index set $\mathcal{S}^{\left(t\right)}_{\tau}$.
When ${{\bf{A}}^{\mathrm{T}}_\tau }$ is invertible, ${{\boldsymbol{\iota }}^{\mathrm{T}}_\tau }$ can be updated by
\vspace{-4mm}
\begin{equation}
\left({{\boldsymbol{\iota }}^{\mathrm{T}}_\tau }\right)^{\left(t+1\right)}  = \left({{\bf{A}}^{\mathrm{T}}_\tau }\right)^{-1} {{\bf{b}}^{\mathrm{T}}_{{\tau }}}.\vspace{-2mm}
\end{equation}
Otherwise, an element-wise update for ${{\boldsymbol{\iota }}^{\mathrm{T}}_\tau }$ can be obtained by
\vspace{-2mm}
\begin{equation}
\{{{\boldsymbol{\iota }}^{\mathrm{T}}_\tau }\}_n =  \frac{\{{\bf{b}}^{\mathrm{T}}_{{\tau }}\}_n - \{\{{\bf{A}}^{\mathrm{T}}_\tau \}_n\}_{-n}^{\mathrm{T}} \{{{\boldsymbol{\kappa }}^{\mathrm{T}}_\tau }\}_{-n}}{\{{{\bf{A}}^{\mathrm{T}}_\tau }\}_{n,n}}, \forall n \in \{1, \ldots, \hat P\}.\vspace{-2mm}
\end{equation}
Since ${\iota _{{\tau_{{k''M + l''}}}}} \in \left[-\frac{1}{2}r_{\tau},\frac{1}{2}r_{\tau}\right]$, we have
\vspace{-2mm}
\begin{equation}\label{HyperParameterIII}
{\{ \overline{\left({{\boldsymbol{\iota }}^{\mathrm{T}}_\tau }\right)^{\left(t+1\right)}} \} _n} = \left\{ {\begin{array}{*{20}{c}}
	{{{\{ \left({{\boldsymbol{\iota }}^{\mathrm{T}}_\tau }\right)^{\left(t+1\right)}\} }_n}},&{{\rm{if }}{{\{ \left({{\boldsymbol{\iota }}^{\mathrm{T}}_\tau }\right)^{\left(t+1\right)}\} }_n} \in \left[ { - \frac{1}{2}r_{\tau},\frac{1}{2}r_{\tau}} \right]},\\[-1mm]
	{ - \frac{1}{2}r_{\tau}},&{{\rm{if }}{{\{ \left({{\boldsymbol{\iota }}^{\mathrm{T}}_\tau }\right)^{\left(t+1\right)}\} }_n} <  - \frac{1}{2}r_{\tau}},\\[-1mm]
	{\frac{1}{2}r_{\tau}},&{{\rm{otherwise}}}.
	\end{array}} \right.\vspace{-2mm}
\end{equation}
Now, we can recover $\overline{\left({{\boldsymbol{\iota }}_\tau }\right)^{\left(t+1\right)}}$ based on the estimated $\overline{\left({{\boldsymbol{\iota }}^{\mathrm{T}}_\tau }\right)^{\left(t+1\right)}}$ and the truncation index set $\mathcal{S}^{\left(t\right)}_{\tau}$.

\begin{table}
	\vspace*{-5mm}
	\begin{algorithm} [H]                    % enter the algorithm environment
		\caption{1D Off-grid SBL-based Channel Estimation Algorithm}     % give the algorithm a caption
		\label{alg1}                             % and a label for \ref{} commands later in the document
		\begin{algorithmic} [1]
			\footnotesize          % enter the algorithmic environment
			\STATE \textbf{Initialization}\\
			Initialize the convergence tolerance $\epsilon$, the maximum number of iterations $T_\mathrm{max}$, the iteration counter $t = 1$, the root parameters $\rho$, $c$ and $d$, and the noise variance $\overline{\sigma^2} = \frac{\left\|\mathbf{y}_{\mathrm{T}}\right\|^2}{100 M_{\mathrm{T}}N_{\mathrm{T}}}$.
			Initialize the hyper-parameters ${{\boldsymbol{\alpha }}^{\left( t \right)}} = \left|\mathbf{\overline \Phi} _\mathrm{T}^{\mathrm{H}}\left( {\boldsymbol{\kappa}_{{\nu }}},{\boldsymbol{\iota}_{{\tau}}} \right)\mathbf{y}_{\mathrm{T}}\right|$, ${\boldsymbol{\kappa }}_\nu ^{\left( t \right)} = {\boldsymbol{\iota }}_\tau ^{\left( t \right)} = \mathbf{0}$, and $\beta _0^{\left( t \right)} = \frac{1}{\overline{\sigma^2}}$.
			
			\REPEAT
			\STATE Update the conditional posterior mean $\boldsymbol{\mu}_{\overline {\bf{h}}}^{\left( t \right)}$ and  covariance matrix $\boldsymbol{\Sigma }_{\overline {\bf{h}}}^{\left( t \right)}$ according to \eqref{PosteriorMean} and \eqref{PosteriorcOV}, respectively
			\STATE Update the hyper-parameters $\left({{{\boldsymbol{\alpha }}^{\left( t+1 \right)}},{\boldsymbol{\kappa }}_\nu ^{\left( t+1 \right)},{\boldsymbol{\iota }}_\tau ^{\left( t+1 \right)},\beta _0^{\left( t+1 \right)}}\right)$ according to \eqref{HyperParameterI}, \eqref{HyperParameterII}, \eqref{HyperParameterIII}, and \eqref{HyperParameterIV}, respectively
			\STATE $t = t + 1$ 
			\UNTIL
			$t = T_\mathrm{max}$ or $\frac{\left\| {{\boldsymbol{\alpha }}^{\left( t+1 \right)}} - {{\boldsymbol{\alpha }}^{\left( t \right)}} \right\|_2}{\left\| {{\boldsymbol{\alpha }}^{\left( t \right)}}  \right\|_2} \le \epsilon$
			\STATE Return the converged solution $\hat{\overline {\bf{h}}} = \boldsymbol{\mu}_{\overline {\bf{h}}}^{\left( t \right)}$, $\hat{\mathbf{k}}_{{\nu }} = \overline{\mathbf{k}}_{\nu} + {\boldsymbol{\kappa }}_\nu ^{\left( t \right)}$, and $\hat{\mathbf{l}}_{{\tau}} = \overline{\mathbf{l}}_{\tau} + {\boldsymbol{\iota }}_\tau ^{\left( t \right)}$.
		\end{algorithmic}
	\end{algorithm}\vspace*{-15mm}
\end{table}

Now, the proposed 1D off-grid SBL-based channel estimation algorithm for OTFS can be summarized in \textbf{Algorithm} \ref{alg1}, where the hyper-parameters are updated alternatively according to \eqref{HyperParameterI}, \eqref{HyperParameterII}, \eqref{HyperParameterIII}, and \eqref{HyperParameterIV}, with concurrently updating the conditional posterior distribution of $\overline {\bf{h}}$ according to \eqref{PosteriorMean} and \eqref{PosteriorcOV}.
When the maximum number of iterations is reached or the changes of hyper-parameter ${\boldsymbol{\alpha }}$ are smaller than a predefined convergence tolerance $\epsilon$, the algorithm terminates and outputs the estimated channel coefficient $\hat{\overline{\bf{h}}}$, the estimated normalized Doppler shift $\hat{\mathbf{k}}_{{\nu }}$, and the estimated normalized delay shift $\hat{\mathbf{l}}_{{\tau}}$.
By the virtue of the EM algorithm\cite{MoonSPM}, the proposed algorithm is guaranteed to converge to the stationary point of $p\left({\boldsymbol{\alpha }}, {{{\boldsymbol{\kappa }}_\nu },{{\boldsymbol{\iota }}_\tau },{\beta _0}}\left| {\bf{y}}_{\mathrm{T}}\right. \right)$ since it is non-decreasing over iterations.
The computational complexity of the proposed algorithm is dominated by computing the conditional posterior covariance matrix in \eqref{PosteriorcOV}, which has an order of $\mathcal{O}\left(M^2_{\tau}N^2_{\nu} M_{\mathrm{T}}N_{\mathrm{T}}\right)$ \cite{YangZaiOffgridCE}.
We can observe that the 1D off-grid SBL-based channel estimation algorithm suffers from high computational complexity with a large virtual sampling grid size, i.e., $M_{\tau} \to \infty$ and $N_{\nu} \to \infty$.
This motivates us to propose the 2D off-grid SBL-based channel estimation as detailed in the next section, which enjoys a lower computational complexity.

\begin{Remark}
	The above proposed framework treats the unknown data symbols in the measurement matrix in \eqref{SSRDModel} as the impact of noise.
	To mitigate the impacts of the unknown data symbols, we can further proposed an iterative data-aided channel estimation scheme.
	The key idea of data-aided channel estimation \cite{MaDataAidedCE} is that the estimated data symbols can be viewed as pilots, which can in turn further improve the channel estimation performance.
	In particular, we first perform the proposed off-grid 1D channel estimation, i.e., \textbf{Algorithm} \ref{alg1}, based on \eqref{SSRDModel} assuming only pilot symbol available in its measurement matrix.
	Then, we reconstruct the effective DD domain channel through \eqref{EffectiveDDDomainChannel} and perform data detection to estimate the data symbols $\hat x\left[k,l\right]$ based on \eqref{IOOTFS}.
	Substituting the estimated symbols $\hat x\left[k,l\right]$ into the measurement matrix in \eqref{SSRDModel}, we can perform our proposed \textbf{Algorithm} \ref{alg1} again to update the channel estimation.
	Such an iteration procedure proceeds until the estimated channel does not change.
	The performance of the proposed data-aided channel estimation will be evaluated in the Section VI.
\end{Remark}

\vspace{-2mm}
\section{2D Off-grid SBL-based Channel Estimation}
In this section, to reduce the computational complexity of the SBL-based channel estimation, we propose a 2D off-grid SSR model and a 2D off-grid SBL-based channel estimation algorithm.
Note that all the previous works for compressed channel sensing of OTFS \cite{ShenCEMassiveMIMO,ZhaoSBLOTFS,LiPDMAOTFS} are based on 1D SSR model, which are fundamentally different from our proposed 2D model.
\vspace{-2mm}
\subsection{{2D Off-grid SSR Model}}
When the pilot sequence is separable in Doppler and delay domains, i.e., $\mathbf{X} = \mathbf{x}_{\nu}  \otimes \mathbf{x}_{\tau}^{\mathrm{H}}$, where $\mathbf{X} \in \mathbb{C}^{\left|\mathcal{K}_{\mathrm{p}}\right|\times \left|\mathcal{L}_{\mathrm{p}}\right|}$ collects the pilot symbols $x\left[ {k',l'} \right]$, $\mathbf{x}_{\nu} \in \mathbb{C}^{\left|\mathcal{K}_{\mathrm{p}}\right|\times 1}$ collects the pilot symbols $x_{\nu}\left[ {k'} \right]$ in the Doppler domain, and $\mathbf{x}_{\tau} \in \mathbb{C}^{\left|\mathcal{L}_{\mathrm{p}}\right|\times 1}$ collects the pilot symbols $x_{\tau}\left[ {l'} \right]$ in the delay domain, $\forall k' \in \mathcal{K}_{\mathrm{p}}$ and $\forall l' \in \mathcal{L}_{\mathrm{p}}$, the system model in \eqref{IOOTFSChannel} can be rewritten as 
\vspace{-2mm}
\begin{align}\label{IOOTFS2D}
\hspace{-20mm}y\left[ {k,l} \right] &=  \sum\limits_{k' \in \mathcal{K}_{\mathrm{p}}} {\sum\limits_{l' \in \mathcal{L}_{\mathrm{p}}} x_{\nu}\left[k'\right]x_{\tau}\left[l'\right] } {h_w}\left[ {k - k',l - l'} \right] + z\left[k,l\right] \notag\\[-1mm]
& = \sum_{i=1}^{P} {\tilde{h}_i} \sum\limits_{k' \in \mathcal{K}_{\mathrm{p}}} x_{\nu}\left[k'\right] w_\nu(k-k'-k_{\nu_i}) \sum\limits_{l' \in \mathcal{L}_{\mathrm{p}}} x_{\tau}\left[l'\right] w_\tau(l-l'-l_{\tau_i}) + z\left[k,l\right] \notag\\[-1mm]
& \mathop {\rm{ = }}\limits^{\left( a \right)}  \sum_{i=1}^{P} {\tilde{h}_i} x_{\nu}\left[k_p\right] w_\nu(k-k_p-k_{\nu_i}) x_{\tau}\left[l_p\right] w_\tau(l-l_p-l_{\tau_i}) + z\left[k,l\right].
\end{align}
\vspace{-8mm}\par\noindent
Equality $(a)$ is obtained with adopting a single pilot impulse, i.e., $\left|\mathcal{K}_{\mathrm{p}}\right| = \left|\mathcal{L}_{\mathrm{p}}\right| = 1$.
Note that for the case of a single pilot impulse, the pilot sequence is separable as $x\left[ {k_\mathrm{p},l_\mathrm{p}} \right] = x_{\nu}\left[k_p\right]  x_{\tau}\left[l_p\right]$ always holds.

From \eqref{IOOTFS2D}, we can observe that it is possible to decouple the estimation of the normalized Doppler shift $k_{\nu_i}$ and the normalized delay shift $l_{\tau_i}$, which is different from the 1D model in Section \ref{1DSSR}.
Benefiting from this decoupling, it is possible to reduce the computational complexity based on the proposed 2D SSR model as follows.
To this end, we define the virtual sampling grid for Doppler and delay domains individually, which is different from that in \eqref{1DVirtualSamplingModel}.
In particular, the Doppler virtual sampling grid\footnote{For notational consistency, in this section, we reuse some notations of Section III.} is defined in the range of $\left(-\nu_{\mathrm{max}}, \nu_{\mathrm{max}}\right)$ with a virtual Doppler resolution $r_{\nu} = \frac{2\nu_{\mathrm{max}}}{N_{\nu}}$, resulting in $\overline{\mathbf{k}}_{\nu} = \left[\overline k_{\nu_0},\ldots,\overline k_{\nu_{N_{\nu}-1}}\right] \in \mathbb{R}^{N_{\nu} \times 1}$ and $\overline k_{\nu_{k''}} = k''r_{\nu} - \nu_{\mathrm{max}}$.
On the other hand, the delay virtual sampling grid is defined in the range of $\left(0, \tau_{\mathrm{max}}\right)$ with a virtual delay resolution $r_{\tau} = \frac{\tau_{\mathrm{max}}}{M_{\tau}}$, resulting in $\overline{\mathbf{l}}_{\tau} = \left[\overline l_{\tau_0},\ldots,\overline l_{\tau_{M_{\tau}-1}}\right] \in \mathbb{R}^{M_{\tau}\times 1}$ and $\overline l_{\tau_{l''}} = l'' r_{\tau}$.
Now, similar to \eqref{SSRDModel}, we can recast the system model in \eqref{IOOTFS2D} as a 2D SSR problem based on the defined virtual sampling grids and the first-order approximation:
\vspace{-2mm}
\begin{equation}\label{IOOTFS2DApprox}
\mathbf{Y}_{\mathrm{T}} = \overline{\mathbf{\Phi}}_{\mathrm{L}} \left(\boldsymbol{\kappa}_\nu\right) \overline{\mathbf{H}} \overline{\mathbf{\Phi}}_{\mathrm{R}}^{\mathrm{T}} \left(\boldsymbol{\iota}_\tau \right) + \overline{\mathbf{Z}}_{\mathrm{T}},\vspace{-2mm}
\end{equation}
where $\mathbf{Y}_{\mathrm{T}} \in \mathbb{C}^{N_{\mathrm{T}} \times M_{\mathrm{T}}}$ collects the  received signal $y\left[k,l\right]$ in the range of $k_{\mathrm{p}} - k_{\mathrm{max}} \le k \le k_{\mathrm{p}} + k_{\mathrm{max}}$ and $l_{\mathrm{p}} \le l \le l_{\mathrm{p}} + k_{\mathrm{max}}$ thanks to the compactness of the DD domain channel.
Also, matrix $\overline{\mathbf{Z}}_{\mathrm{T}} \in \mathbb{C}^{N_{\mathrm{T}} \times M_{\mathrm{T}}}$ collects the new noise $\overline{z}\left[k,l\right] \sim \mathcal{CN}\left(\mathbf{0},\overline{\sigma^2}\right)$, where $\overline{\sigma^2}$ is the unknown noise power taking into account the measurement noise, approximation error, as well as the interference from unknown data symbols.
The unknown channel matrix $\overline{\mathbf{H}} \in \mathbb{C}^{N_{\nu} \times M_{\tau}}$ collects the channel coefficients $\tilde{h}_i$, $\forall i$, i.e., $\{\overline{\mathbf{H}}\}_{\overline k^{i}_{\nu_{k''}},\overline l^{i}_{\tau_{l''}}} = \tilde{h}_i$, where
$\overline k^{i}_{\nu_{k''}}$ and $\overline l^{i}_{\tau_{l''}}$ denote the unknown nearest points of $k_{\nu_{i}}$ and $l_{\tau_{i}}$, respectively.
In other words, there are at most $P$ non-zero entries in the sparse unknown channel matrix $\overline{\mathbf{H}}$.

In \eqref{IOOTFS2DApprox}, the left measurement matrix $\overline{\mathbf{\Phi}}_{\mathrm{L}}\left(\boldsymbol{\kappa}_\nu \right) \in \mathbb{C}^{N_{\mathrm{T}} \times N_{\nu}}$ and the right measurement matrix $\overline{\mathbf{\Phi}}_{\mathrm{R}}\left(\boldsymbol{\iota}_\tau \right) \in \mathbb{C}^{M_{\mathrm{T}} \times M_{\tau}}$ are defined as
\vspace{-2mm}
\begin{equation}
	\overline{\mathbf{\Phi}}_{\mathrm{L}}\left(\boldsymbol{\kappa}_\nu \right) = \mathbf{\Phi}_\mathrm{L} + \mathbf{\Phi}_{\mathrm{L},\nu} \diag \left(\boldsymbol{\kappa}_\nu\right)\;\text{and}\;\overline{\mathbf{\Phi}}_{\mathrm{R}}\left(\boldsymbol{\iota}_\tau \right) = \mathbf{\Phi}_\mathrm{R} + \mathbf{\Phi}_{\mathrm{R},\tau} \diag \left(\boldsymbol{\iota}_\tau\right),\vspace{-2mm}
\end{equation}
respectively, where $\boldsymbol{\kappa}_\nu = \left[\kappa_{\nu_0},\ldots,\kappa_{\nu_{N_{\nu}-1}}\right] \in \mathbb{R}^{N_{\nu} \times 1}$ and $\boldsymbol{\iota}_\tau = \left[ \iota_{\tau_0},\ldots, \iota_{\tau_{M_{\tau}-1}}\right] \in \mathbb{R}^{M_{\tau} \times 1}$.
And the matrices $\mathbf{\Phi}_\mathrm{L}, \mathbf{\Phi}_{\mathrm{L},\nu} \in \mathbb{C}^{N_{\mathrm{T}}\times N_{\nu}}$ and $\mathbf{\Phi}_\mathrm{R},\mathbf{\Phi}_{\mathrm{R},\tau} \in \mathbb{C}^{M_{\mathrm{T}}\times M_{\tau}}$ are defined as $\{\mathbf{\Phi}_{\mathrm{L}}\}_{k,k''}  = x_{\nu}\left[k_{\mathrm{p}}\right] w_\nu(k-k_{\mathrm{p}}-\overline k_{\nu_{k''}})$, $\{\mathbf{\Phi}_{\mathrm{L,\nu}}\}_{k,k''}  = x_{\nu}\left[k_{\mathrm{p}}\right] w'_\nu(k-k_{\mathrm{p}}-\overline k_{\nu_{k''}})$, $\{\mathbf{\Phi}_{\mathrm{R}}\}_{l,l''}  = x_{\tau}\left[l_{\mathrm{p}}\right] w_\tau(l-l_{\mathrm{p}}-\overline l_{\tau_{l''}})$, and
$\{\mathbf{\Phi}_{\mathrm{R,\tau}}\}_{l,l''}  = x_{\tau}\left[l_{\mathrm{p}}\right] w'_\tau(l-l_{\mathrm{p}}-\overline l_{\tau_{l''}})$, respectively, $\forall k'' \in \left\{0,\ldots,N_{\nu}-1\right\}$ and $\forall l'' \in \left\{0,\ldots,M_{\tau}-1\right\}$.

\vspace{-2mm}
\subsection{2D Off-grid SBL-based Channel Estimation Algorithm}
Based on the built 2D off-grid SSR model in \eqref{IOOTFS2DApprox}, we propose a two-step SBL-based channel estimation algorithm in this section, where the first step estimates the composite matrix $\mathbf{D} = \overline{\mathbf{H}} \overline{\mathbf{\Phi}}_{\mathrm{R}}^{\mathrm{T}}\left(\boldsymbol{\iota}_\tau \right)$ and the second step aims to recovery $\overline{\mathbf{H}}$ based on the estimated $\mathbf{D}$.

\subsubsection{First Step}
Defining a composite matrix $\mathbf{D} = \overline{\mathbf{H}} \overline{\mathbf{\Phi}}_{\mathrm{R}}^{\mathrm{T}}\left(\boldsymbol{\iota}_\tau \right) = \left[\mathbf{d}_{\mathrm{c},0},\ldots,\mathbf{d}_{\mathrm{c},M_{\mathrm{T}}-1}\right] \in \mathbb{C}^{N_{\nu} \times M_{\mathrm{T}}}$, the system model in \eqref{IOOTFS2DApprox} can be rewritten as
\vspace{-2mm}
\begin{equation}\label{MultiplesnapshotSBL}
\mathbf{Y}_{\mathrm{T}} = \overline{\mathbf{\Phi}}_{\mathrm{L}} \left(\boldsymbol{\kappa}_\nu\right) \mathbf{D} + \overline{\mathbf{Z}}_{\mathrm{T}},\vspace{-2mm}
\end{equation}
where $\mathbf{D}$ is a row-sparse channel matrix sharing the same row-sparsity with $\overline{\mathbf{H}}$, i.e., the sparsity in the Doppler domain, $\mathbf{d}_{\mathrm{c},l} \in \mathbb{C}^{N_{\nu} \times 1}$ denotes the $l$-th column of $\mathbf{D}$, and all columns $\mathbf{d}_{\mathrm{c},l}$ are jointly sparse with the same support, $\forall l \in \{0,\ldots,M_{\mathrm{T}}-1\}$.
To recovery $\mathbf{D}$, we can develop an SBL algorithm for a multiple snapshots case assuming a row sparsity in $\mathbf{D}$.
In particular, the unknown matrix $\mathbf{D}$ and the hyper-parameters $\left({\boldsymbol{\alpha }}_{\nu}, {{\boldsymbol{\kappa }}_\nu },{\beta _0}\right)$ are updated iteratively, where ${\boldsymbol{\alpha }}_{\nu} = \left[{\alpha _{\nu_0}, \ldots ,{\alpha _{\nu_{k''}}}, \ldots {\alpha _{\nu_{N_{\nu} - 1}}}}\right] \in \mathbb{R}^{N_{\nu} \times 1}$ with ${\alpha _{\nu_{k''}}} > 0$ models the row-sparsity of $\mathbf{D}$.

Assuming independent columns of $\mathbf{D}$, the prior distribution of the unknown matrix $\mathbf{D}$ assumes to follow
\vspace{-2mm}
\begin{equation}
	 p\left( { {\bf{D}} \left| {\boldsymbol{\alpha }}_{\nu} \right.} \right) = \prod_{l=1}^{M_{\mathrm{T}}}\mathcal{CN}\left( { {\mathbf{d}_{\mathrm{c},l}} \left| {{\bf{0}},{\bf{\Lambda }}} \right.} \right),\vspace{-2mm}
\end{equation}
where ${\bf{\Lambda }} = \diag\left( {\boldsymbol{\alpha }}_{\nu}  \right)$ and $p\left( {{\boldsymbol{\alpha }}_{\nu}\left| \rho  \right.} \right) = \prod \limits_{k'' = 0}^{N_{\nu} - 1} \mathop \Gamma \left( {{\alpha _{\nu_{k''}}}\left| {1,\rho } \right.} \right)$, with $\rho  > 0$.
And the unknown off-grid components of the Doppler shift in the measurement matrix of \eqref{MultiplesnapshotSBL} follows ${\kappa _{{\nu _{k''}}}} \sim \mathcal{U}\left[ { - \frac{1}{2}r_{\nu},\frac{1}{2}r_{\nu}} \right]$, $\forall k'' \in \{0,\ldots,N_{\nu}-1\}$.
The new noise matrix follows
\vspace{-2mm}
\begin{equation}
	 p\left( { \overline{\mathbf{Z}}_{\mathrm{T}} \left| {{\beta _0}} \right.} \right) = \prod_{l=0}^{M_{\mathrm{T}}-1} \mathcal{CN}\left( { {\overline{\mathbf{z}}_l} \left| {{\bf{0}},\beta _0^{ - 1}{{\bf{I}}_{N_{\mathrm{T}}}}} \right.} \right),\vspace{-2mm}
\end{equation}
where $\overline{\mathbf{Z}}_{\mathrm{T}} = \left[{\overline{\mathbf{z}}_0},\ldots,{\overline{\mathbf{z}}_{M_{\mathrm{T}}-1}}\right]$, $\overline{\mathbf{z}}_l \in \mathbb{C}^{N_{\mathrm{T}} \times 1}$, and $p\left( {{\beta _0}\left| {c,d} \right.} \right) = \Gamma \left( {{\beta _0}\left| {c,d} \right.} \right)$, with $c,d > 0$.
Now, the likelihood function for given $\left({{{\boldsymbol{\kappa }}_\nu },{\beta _0}}\right)$ for \eqref{MultiplesnapshotSBL} is given by
\vspace{-2mm}
\begin{equation}
	 p\left( {{\bf{Y}}_{\mathrm{T}}\left| { {\bf{D}} ;{{\boldsymbol{\kappa }}_\nu },{\beta _0}} \right.} \right) = \prod_{l=0}^{M_{\mathrm{T}}-1} \mathcal{CN}\left( {{\mathbf{y}_l}\left| {\overline{\mathbf{\Phi}}_{\mathrm{L}} \left(\boldsymbol{\kappa}_\nu\right) \mathbf{d}_{\mathrm{c},l} ,\beta _0^{ - 1}{{\bf{I}}_{N_{\mathrm{T}}}}} \right.} \right),\vspace{-2mm}
\end{equation}
and the joint distribution can be decomposed by
\vspace{-2mm}
\begin{equation}
	 p\left( {{\bf{Y}}_{\mathrm{T}}, {\bf{D}} ,{\boldsymbol{\alpha }}_{\nu},{{\boldsymbol{\kappa }}_\nu },{\beta _0}} \right) = p\left( {{\bf{Y}}_{\mathrm{T}}\left| { {\bf{D}} ,{{\boldsymbol{\kappa }}_\nu },{\beta _0}} \right.} \right)p\left( { {\bf{D}} \left| {\boldsymbol{\alpha }}_{\nu} \right.} \right)p\left( {\boldsymbol{\alpha }}_{\nu} \right)p\left( {{{\boldsymbol{\kappa }}_\nu }} \right)p\left( {{\beta _0}} \right),\vspace{-2mm}
\end{equation}
where ${\mathbf{y}_l} \in \mathbb{C}^{N_{\mathrm{T}}\times 1}$ is the $l$-th column of ${\bf{Y}}_{\mathrm{T}}$.
	 
Now, given the hyper-parameters $\left({\boldsymbol{\alpha }}_{\nu}^{\left(t\right)}, {{\boldsymbol{\kappa }}^{\left(t\right)}_\nu },{\beta ^{\left(t\right)}_0}\right)$ in the $t$-th iteration, the conditional posterior distribution of $ {\bf{D}}$ can be obtained by
\vspace{-2mm}
\begin{equation}
	 p\left( {\bf{D}}\left| {\bf{Y}}_{\mathrm{T}}; {{\boldsymbol{\kappa }}^{\left(t\right)}_\nu },{\beta^{\left(t\right)} _0}\right. \right) = \prod_{l=0}^{M_{\mathrm{T}}-1} \mathcal{CN}\left(\mathbf{d}_{\mathrm{c},l} \left|\boldsymbol{\mu}^{\left(t\right)}_{\mathbf{d}_{\mathrm{c},l}},\boldsymbol{\Sigma }^{\left(t\right)}_{\mathbf{d}_{\mathrm{c}}} \right.\right),\vspace{-2mm}
\end{equation}
where the conditional posterior mean for $\mathbf{d}_{\mathrm{c},l}$, $\forall \in \{0,\ldots,M_{\mathrm{T}}-1\}$, and their common covariance matrix are given by
\vspace{-2mm}
	 \begin{align}
	 \boldsymbol{\mu}^{\left(t\right)}_{\mathbf{d}_{\mathrm{c},l}} & = {\beta^{\left(t\right)} _0}\boldsymbol{\Sigma }^{\left(t\right)}_{\mathbf{d}_{\mathrm{c}}}\overline{\mathbf{\Phi}}_{\mathrm{L}} \left(\boldsymbol{\kappa}^{\left(t\right)}_\nu\right) {\mathbf{y}_l}\;\text{and}\label{PosteriorMeanDD}\\[-1mm]
	 \boldsymbol{\Sigma }^{\left(t\right)}_{\mathbf{d}_{\mathrm{c}}} & = {\left( {{\beta^{\left(t\right)} _0}{{\overline{\mathbf{\Phi}}_{\mathrm{L}}^{\rm{H}} \left(\boldsymbol{\kappa}^{\left(t\right)}_\nu\right) }}\overline{\mathbf{\Phi}}_{\mathrm{L}} \left(\boldsymbol{\kappa}^{\left(t\right)}_\nu\right) + {\left({\bf{\Lambda }}^{\left(t\right)}\right)^{ - 1}}} \right)^{ - 1}} \notag\\[-1mm]
	 & = {\boldsymbol{\Lambda }}^{\left(t\right)} - {\boldsymbol{\Lambda }}^{\left(t\right)}\overline{\mathbf{\Phi}}_{\mathrm{L}}^{\mathrm{H}} \left(\boldsymbol{\kappa}^{\left(t\right)}_\nu\right){\left( {\left(\beta ^{\left(t\right)}_0\right)^{ - 1}{{\bf{I}}_{{N_{\mathrm{T}}}}} + \overline{\mathbf{\Phi}}_{\mathrm{L}} \left(\boldsymbol{\kappa}^{\left(t\right)}_\nu\right){\boldsymbol{\Lambda }} \overline{\mathbf{\Phi}}_{\mathrm{L}}^{\mathrm{H}} \left(\boldsymbol{\kappa}^{\left(t\right)}_\nu\right)} \right)^{ - 1}}\overline{\mathbf{\Phi}}_{\mathrm{L}} \left(\boldsymbol{\kappa}^{\left(t\right)}_\nu\right){\boldsymbol{\Lambda }}^{\left(t\right)},\label{PosteriorCovMatrixDD}
	 \end{align}
	 \vspace{-10mm}\par\noindent
respectively.

Similar to Proposition \ref{Propo1}, following the EM principle \cite{MoonSPM}, the hyper-parameters $\left({\boldsymbol{\alpha }}, {\beta _0}\right)$ can be updated by the following rules
\vspace{-2mm}
\begin{align}
	\alpha _{k''}^{\left( {t + 1} \right)} &= \frac{{\sqrt {1 + 4\frac{\rho}{M_{\mathrm{T}}} \frac{1}{M_{\mathrm{T}}}\sum_{l = 0}^{M_{\mathrm{T}}-1}\{{{{\left| \{\boldsymbol{\mu}^{\left(t\right)}_{\mathbf{d}_{\mathrm{c},l}}\}_{k''} \right|}}^2} + \{\boldsymbol{\Sigma }^{\left(t\right)}_{\mathbf{d}_{\mathrm{c}}}\}_{k'',k''}\}}  - 1}}{{2\frac{\rho}{M_{\mathrm{T}}} }}\; \text{and}\label{ALPHAUPDATEDD}\\[-1mm]
	\beta_0^{\left( {t + 1} \right)} &= \frac{{c - 1 + M_{\mathrm{T}}N_{\mathrm{T}}}}{{d + \sum_{l=0}^{M_{\mathrm{T}}-1}{A^l_{{\beta _0}}}}},\label{betaUPDATEDD}
\end{align}
\vspace{-6mm}\par\noindent
respectively, and ${A^l_{{\beta _0}}} = {\left\| {{\bf{y}}_l - \overline{\mathbf{\Phi}}_{\mathrm{L}}^{\mathrm{H}} \left(\boldsymbol{\kappa}^{\left(t\right)}_\nu\right)\boldsymbol{\mu}^{\left(t\right)}_{\mathbf{d}_{\mathrm{c},l}} } \right\|^2}{\rm{ + }}\left(\beta ^{\left(t\right)}_0\right)^{ - 1}\mathop \sum \limits_{k'' = 0}^{N_\nu - 1} \left( {1 - \left(\alpha _{\nu_{k''}}\right)^{ - 1}\{\boldsymbol{\Sigma }^{\left(t\right)}_{\mathbf{d}_{\mathrm{c}}}\}_{k'',k''}} \right)$.

Also, the hyper-parameter ${{\boldsymbol{\kappa }}_\nu }$ can be updated by
\vspace{-4mm}
\begin{equation}
{{\boldsymbol{\kappa }}_\nu } = \left({{\bf{A}}_\nu }\right)^{-1} {{\bf{b}}^{}_{{\nu }}},\vspace{-2mm}
\end{equation}
when ${{\bf{A}}_\nu } = \frac{1}{{M_{\mathrm{T}}}}\sum_{l=1}^{{{M_{\mathrm{T}}}}}\Re \left\{\mathbf{\Phi}_{\mathrm{L},\nu} ^{\rm{H}}\mathbf{\Phi}_{\mathrm{L},\nu} \odot \left( {\left(\boldsymbol{\mu}^{\left(t\right)}_{\mathbf{d}_{\mathrm{c},l}}\right)^*\left(\boldsymbol{\mu}^{\left(t\right)}_{\mathbf{d}_{\mathrm{c},l}}\right)^{\rm{T}} + \left(\boldsymbol{\Sigma }^{\left(t\right)}_{\mathbf{d}_{\mathrm{c}}}\right)^{\rm{T}}} \right)\right\}\in \mathbb{R}^{N_{\nu} \times N_{\nu}}$ is invertible and ${{\bf{b}}_{{\nu }}}$ is given by
\vspace{-2mm}
\begin{equation}
	 {{\bf{b}}_{{\nu }}} = \frac{1}{{M_{\mathrm{T}}}}\sum_{l=1}^{{{M_{\mathrm{T}}}}} \Re \left\{ {\diag\left( \boldsymbol{\mu}^{\left(t\right)}_{\mathbf{d}_{\mathrm{c},l}} \right)\mathbf{\Phi}_{\mathrm{L},\nu} ^{\rm{T}}{{\mathbf{y}}^*_l} - \diag\left( {\left( {\boldsymbol{\mu}^{\left(t\right)}_{\mathbf{d}_{\mathrm{c},l}} \left(\boldsymbol{\mu}^{\left(t\right)}_{\mathbf{d}_{\mathrm{c},l}}\right)^{\mathrm{H}} + \boldsymbol{\Sigma }^{\left(t\right)}_{\mathbf{d}_{\mathrm{c}}}} \right){\mathbf{\Phi}_{\mathrm{L},\nu}}} \right)} \right\} \in \mathbb{R}^{N_{\nu} \times 1}. \vspace{-2mm}
\end{equation}
When ${{\bf{A}}_\nu }$ is not invertible, we perform an element-wise update for ${{\boldsymbol{\kappa }}^{}_\nu }$, i.e.,
\vspace{-2mm}
\begin{equation}
\{{{\boldsymbol{\kappa }}^{\left(t+1\right)}_\nu }\}_n =  \frac{\{{\bf{b}}^{}_{{\nu }}\}_n - \{\{{\bf{A}}^{}_\nu \}_n\}_{-n}^{\mathrm{T}} \{{{\boldsymbol{\kappa }}^{}_\nu }\}_{-n}}{\{{{\bf{A}}^{}_\nu }\}_{n,n}}.\vspace{-2mm}
\end{equation}
Note that ${\kappa _{{\nu_{{k''}}}}} \in \left[-\frac{1}{2}r_{\nu},\frac{1}{2}r_{\nu}\right]$ and thus we have
\vspace{-2mm}
\begin{equation}\label{HyperparameterDD}
	{\{ \overline{{\boldsymbol{\kappa }}^{\left(t+1\right)}_\nu} \} _n} = \left\{ {\begin{array}{*{20}{c}}
	 {{{\{ {\boldsymbol{\kappa }}_\nu ^{\left(t+1\right)}\} }_n}},&{{\rm{if }}\;{{\{ {\boldsymbol{\kappa }}_\nu ^{\left(t+1\right)}\} }_n} \in \left[ { - \frac{1}{2}r_{\nu},\frac{1}{2}r_{\nu}} \right]},\\[-1mm]
	 { - \frac{1}{2}r_{\nu}},&{{\rm{if }}\;{{\{ {\boldsymbol{\kappa }}_\nu ^{\left(t+1\right)}\} }_n} <  - \frac{1}{2}r_{\nu}},\\[-1mm]
	 {\frac{1}{2}r_{\nu}},&{{\rm{otherwise}}}.
	 \end{array}} \right.\vspace{-2mm}
\end{equation}

Now, the conditional posterior distribution of the unknown matrix $\mathbf{D}$ and the values of hyper-parameters $\left({\boldsymbol{\alpha }}, {{\boldsymbol{\kappa }}_\nu },{\beta _0}\right)$ can be updated according to \eqref{PosteriorMeanDD}, \eqref{PosteriorCovMatrixDD}, \eqref{ALPHAUPDATEDD}, \eqref{betaUPDATEDD}, and \eqref{HyperparameterDD} iteratively, as detailed in line $1-8$ in \textbf{Algorithm} \ref{alg2} as follows.
After convergence, we can obtain $ \hat{\bf{D}} = \left[\boldsymbol{\mu}^{\left(t\right)}_{\mathbf{d}_{\mathrm{c},0}},\ldots,\boldsymbol{\mu}^{\left(t\right)}_{\mathbf{d}_{\mathrm{c},M_{\mathrm{T}}-1}} \right]$ and the Doppler shift $\hat{\mathbf{k}}_{{\nu }} = \overline{\mathbf{k}}_{\nu} + {\boldsymbol{\kappa }}_\nu ^{\left( t \right)}$, where $\hat{\bf{D}}$ will be used to estimate the unknown channel matrix and the delay shift in the second step.
	
\subsubsection{Second Step} 
The estimated composite matrix $\hat{\bf{D}}$ can be rewritten as $\hat{\bf{D}} = \left[\hat{\mathbf{d}}_{\mathrm{r},0},\ldots,\hat{\mathbf{d}}_{\mathrm{r},N_{\nu}-1}\right]^{\mathrm{T}}$, where $\left(\hat{\mathbf{d}}_{\mathrm{r},k}\right)^{\mathrm{T}} \in \mathbb{C}^{1 \times {M_{\mathrm{T}}}}$ denotes the $k$-th row vector of $\hat{\bf{D}}$.
Besides, based on the definition of ${\bf{D}}$ in \eqref{MultiplesnapshotSBL}, we can assume $\hat{\bf{D}} = \overline{\mathbf{H}} \overline{\mathbf{\Phi}}_{\mathrm{R}}^{\mathrm{T}}\left(\boldsymbol{\iota}_\tau \right)$.
Therefore, defining $\overline{\mathbf{H}} = \left[\overline{\mathbf{h}}_0,\ldots,\overline{\mathbf{h}}_{N_{\nu}-1}\right]^{\mathrm{T}}$, where $\overline{\mathbf{h}}^{\mathrm{T}}_k \in \mathbb{C}^{1 \times {M_{\tau}}}$ denotes the $k$-th row vector of $\overline{\mathbf{H}}$, we can obtain ${N_{\nu}}$ 1D SSR problems:
\vspace{-4mm}
\begin{equation}\label{SecondStepSSR}
\hat{\mathbf{d}}_{\mathrm{r},k} = \overline{\mathbf{\Phi}}_{\mathrm{R}}\left(\boldsymbol{\iota}_\tau \right) \overline{\mathbf{h}}_k, \forall k \in \{0,\ldots,{N_{\nu}}-1\},\vspace{-2mm}
\end{equation}
which can be solved in parallel using \textbf{Algorithm} \ref{alg1}.
The details for recovering $\overline{\mathbf{h}}_k$ are omitted here to concise our presentation.

\begin{table}
	\vspace*{-5mm}
	\begin{algorithm} [H]                    % enter the algorithm environment
		\caption{2D Off-grid SBL-based Channel Estimation Algorithm}     % give the algorithm a caption
		\label{alg2}                             % and a label for \ref{} commands later in the document
		\begin{algorithmic} [1]
			\footnotesize          % enter the algorithmic environment
			\STATE \textbf{Initialization}\\
			Initialize the convergence tolerance $\epsilon$, the maximum number of iterations $T_\mathrm{max}$, the iteration counter $t = 1$, the root parameters $\rho$, $c$, and $d$, and the noise variance $\overline{\sigma^2} = \frac{\left\|\mathbf{Y}_{\mathrm{T}}\right\|^2}{100 M_{\mathrm{T}}N_{\mathrm{T}}}$.
			Initialize the hyper-parameters ${\alpha _{\nu_{k''}}^{\left( t \right)}} = \frac{1}{M_{\mathrm{T}}} \sum_{l=0}^{M_{\mathrm{T}}-1}\left\{\left|\{\overline{\mathbf{\Phi}}^{\mathrm{H}}_{\mathrm{L}} \left(\boldsymbol{\kappa}_\nu\right)\mathbf{Y}_{\mathrm{T}}\}_{k'',l}\right|\right\}$, ${\boldsymbol{\kappa }}_\nu ^{\left( t \right)} = \mathbf{0}$, and $\beta _0^{\left( t \right)} = \frac{1}{\overline{\sigma^2}}$.
			
			\REPEAT[\textbf{First Step}] 
			\STATE Update the conditional posterior mean $\boldsymbol{\mu}^{\left(t\right)}_{\mathbf{d}_{\mathrm{c},l}}$ and  covariance matrix $\boldsymbol{\Sigma }^{\left(t\right)}_{\mathbf{d}_{\mathrm{c}}}$ according to \eqref{PosteriorMeanDD} and \eqref{PosteriorCovMatrixDD}, respectively
			\STATE Update the hyper-parameters $\left({{{\boldsymbol{\alpha }}_{\nu}^{\left( t+1 \right)}},{\boldsymbol{\kappa }}_\nu ^{\left( t+1 \right)},\beta _0^{\left( t+1 \right)}}\right)$ according to \eqref{ALPHAUPDATEDD}, \eqref{HyperparameterDD}, and \eqref{betaUPDATEDD}, respectively
			\STATE $t = t + 1$
			\UNTIL
			$t = T_\mathrm{max}$ or $\frac{\left\| {{\boldsymbol{\alpha }}_{\nu}^{\left( t+1 \right)}} - {{\boldsymbol{\alpha }}_{\nu}^{\left( t \right)}} \right\|_2}{\left\| {{\boldsymbol{\alpha }}_{\nu}^{\left( t \right)}}  \right\|_2} \le \epsilon$
			\STATE Output the converged solution $\hat{\bf{D}} = \left[\boldsymbol{\mu}^{\left(t\right)}_{\mathbf{d}_{\mathrm{c},0}},\ldots,\boldsymbol{\mu}^{\left(t\right)}_{\mathbf{d}_{\mathrm{c},M_{\mathrm{T}}-1}} \right] = \left[\hat{\mathbf{d}}_{\mathrm{r},0},\ldots,\hat{\mathbf{d}}_{\mathrm{r},N_{\nu}-1}\right]^{\mathrm{T}}$ and ${\hat{\mathbf{k}}_{{\nu }}} = \overline{\mathbf{k}}_{\nu} + {\boldsymbol{\kappa }}_\nu ^{\left( t \right)}$.
			\STATE {\textbf{Second Step}:} Recover $\hat{\overline {\bf{h}}}_k$ and $\hat{\mathbf{l}}_{{\tau}}$ according to \textbf{Algorithm} \ref{alg1}, based on the model in \eqref{SecondStepSSR}, $\forall k$.
		\end{algorithmic}
	\end{algorithm}\vspace*{-15mm}
\end{table}

Now, the 2D off-grid SBL-based channel estimation algorithm can be summarized in \textbf{Algorithm} \ref{alg2}.
The convergence of the first step can be guaranteed by the virtue of the EM algorithm\cite{MoonSPM}, where 
$p\left({\boldsymbol{\alpha }}, {{{\boldsymbol{\kappa }}_\nu },{\beta _0}}\left| {\bf{Y}}_{\mathrm{T}}\right. \right)$ will be non-decreasing for each iteration.
The computational complexity of the proposed algorithm is dominated by channel inversion operations in both the first step and second step, which has an order of $\mathcal{O}\left(N^2_{\nu} N_{\mathrm{T}} + M^2_{\tau}M_{\mathrm{T}}\right)$ \cite{YangZaiOffgridCE}.
We can observe that the computational complexity of \textbf{Algorithm} \ref{alg2} is much lower compared with \textbf{Algorithm} \ref{alg1}.
In particular, benefiting from the decoupling of Doppler and delay shift estimation of our proposed 2D method, its computational complexity order is proportional to the summation of the sizes of delay and Doppler grids, i.e., $\mathcal{O}\left(N^2_{\nu} N_{\mathrm{T}} + M^2_{\tau}M_{\mathrm{T}}\right)$, while the counterpart of the proposed 1D off-grid method is proportional to their product, i.e., $\mathcal{O}\left(N^2_{\nu} M^2_{\tau}N_{\mathrm{T}} M_{\mathrm{T}}\right)$.

\vspace{-2mm}
\section{Simulations}
In this section, we evaluate the channel estimation performance of our proposed 1D and 2D off-grid SBL-based channel estimation algorithms.
Unless specified otherwise, the system parameters used in the our simulations are given as follows.
The DD domain grid size is $M = N = 32$, the carrier frequency is $3$ GHz, the subcarrier spacing is $\Delta_f = 15$ kHz, and the slot duration is $T = \frac{1}{\Delta_f}$.
The total number of path is set as $P = 5$, the normalized maximum Doppler shift and maximum delay shift are given by $k_{\mathrm{max}} = 3$ and $l_{\mathrm{max}} = 4$, respectively.
The Doppler and delay shifts are randomly and uniformly generated within $\left[-\nu_{\mathrm
max}, \nu_{\mathrm
max}\right]$ and $\left[0, \tau_{\mathrm
max}\right]$, respectively.
Note that the adopted maximum Doppler and delay shift correspond
to the maximum relative velocity of $506.25$ km/h among the transceivers or scatters and the maximum propagation distance difference of $2.5$ km between propagating paths, respectively.
The channel coefficients $h_i$ are generated according to the distribution $h_{i}\sim \mathcal{CN}(0,q^{l_{\tau_i}})$, where $q^{l_{\tau_i}}$ follows a normalized exponential power delay profile\footnote{Here, we assume that the path loss and shadowing have been compensated following the literature studying OTFS \cite{RavitejaOTFS,YuanOTFS}.}, i.e., $q^{l_{\tau_i}}=\frac{\exp(-0.1 l_{\tau_i})}{\sum_i \exp(-0.1 l_{\tau_i})}$.
A single pilot impulse is inserted at $\left[k_p,l_p\right] = \left[\frac{M}{2}+1,\frac{N}{2}+1\right]$ in the DD domain and a normalized constellation is assumed for data symbols, i.e., $E\left\{ {{{\left| {x\left[ {k,l} \right]} \right|}^2}} \right\} = 1$.
The power of the pilot symbol is 30 dB higher than that of the data symbols, i.e., $10 \log_{10} \frac{\left|x_p\right|^2}{E\left\{ {{{\left| {x\left[ {k,l} \right]} \right|}^2}} \right\}} = 30$ dB.
The system signal-to-noise ratio (SNR) is defined as $\mathrm{SNR} = \frac{1}{N_0}$ and is set in the range of $\left[0,30\right]$ dB.
The virtual Doppler and delay sampling resolutions are set in the range $r_{\nu}, r_{\tau} \in \left[0.2,0.9\right]$.
Following \cite{RavitejaOTFSCE,WeiWindowOTFS}, the DD domain grids in the range ${k_p} - 2{k_{\max }} \le k \le {k_p} + 2{k_{\max }}$ and ${l_p} - {l_{\mathrm{max}} } \le l \le {l_p} + {l_{\mathrm{max}} }$, with $ k \neq {k_p}$ and $ l \neq {l_p}$, remain null serving as guard space to mitigate the interference from the data symbols to the pilot symbol.
{In our proposed \textbf{Algorithms} \ref{alg1} and \ref{alg2}, we set $\epsilon = 10^{-3}$, $\rho = 10^{-2}$, and $c = d = 10^{-4}$ \cite{YangZaiOffgridCE}.}
%^
For comparison, we evaluate the performance of the traditional impulse-based channel estimation method \cite{RavitejaOTFSCE}, our previously proposed channel estimation exploiting DC window in the TF domain\cite{WeiWindowOTFS}, and NOMP-based channel estimation in \cite{LiPDMAOTFS}, which are denoted as ``Traditional impulse'', ``Traditional impulse + DC window'', ``Off-grid NOMP'', respectively.
Besides, to demonstrate the performance gain of our proposed off-grid methods, the channel estimation performance for on-grid schemes with ${\mathbf{k}_{{\nu }}} = {\mathbf{l}_{{\tau}}} =\mathbf{0}$ based on OMP\cite{ShenCEMassiveMIMO}, 1D SBL\cite{ZhaoSBLOTFS}, and our proposed 2D SBL are also evaluated, which are denoted as ``On-grid OMP'', ``1D on-grid SBL'', ``2D on-grid SBL'', respectively.
Note that the additional steps introduced by our proposed off-grid methods is the hyper-parameter update compared with the on-grid methods, which can be efficiently computed.
However, the computational complexities of both the on-grid and off-grid methods are dominated by the matrix inversion operation and thus they are in the same order.

Note that the traditional channel estimation methods of OTFS, such as ``Traditional impulse'' and ``Traditional impulse + DC window'', are mainly to estimate the effective DD domain channel rather than the original DD domain channel response as our proposed scheme in this paper.
Therefore, for our proposed schemes, we reconstruct the effective DD domain channel ${\hat{h}_w}\left[ {k - k',l - l'} \right]$ based on the estimated original DD domain channel response according to \eqref{EffectiveDDDomainChannel} and define the normalized mean square error (NMSE) of channel estimation as
\vspace{-1mm}
\begin{equation}
	\mathrm{NMSE} = \frac{\sum_{k - k' = 0}^{N-1}\sum_{l-l' = 0}^{M-1} \left|{h_w}\left[ {k - k',l - l'} \right] - {\hat{h}_w}\left[ {k - k',l - l'} \right]\right|^2}{\sum_{k - k' = 0}^{N-1}\sum_{l-l' = 0}^{M-1} \left|{h_w}\left[ {k - k',l - l'} \right] \right|^2},\vspace{-1mm}
\end{equation}
where the summation is over $k - k' \in \left[0,N-1\right]$ and $l-l' \in \left[0,M-1\right]$ as ${h_w}\left[ {k - k',l - l'} \right] = {h_w}\left[ {(k - k')_N,(l - l')_M} \right]$.
All the simulation results are averaged over $1,000$ OTFS frames.

\begin{figure}[t]
	\begin{minipage}{.47\textwidth}
		\centering\vspace{-3mm}
		\includegraphics[width=3.5in]{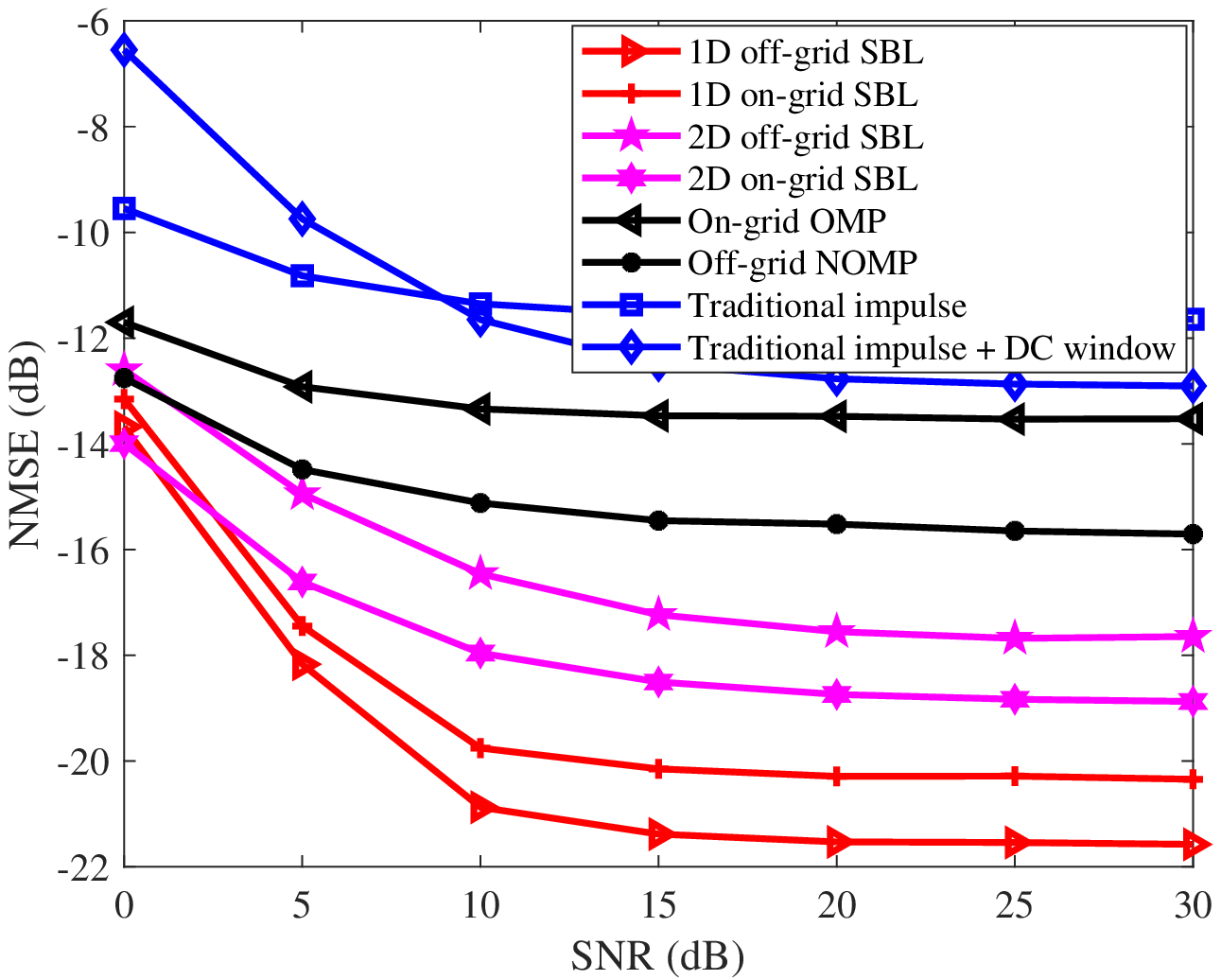}\vspace{-7mm}
		\caption{NMSE of effective channel estimation versus SNR with a high virtual sampling resolution $r_{\nu} = r_{\tau} = 0.5$.}\vspace{-10mm}
		\label{NMSE_WithGuard_HighResolution}
	\end{minipage}
	\hspace{2mm}
	\begin{minipage}{.47\textwidth}
		\centering\vspace{-3mm}
		\includegraphics[width=3.5in]{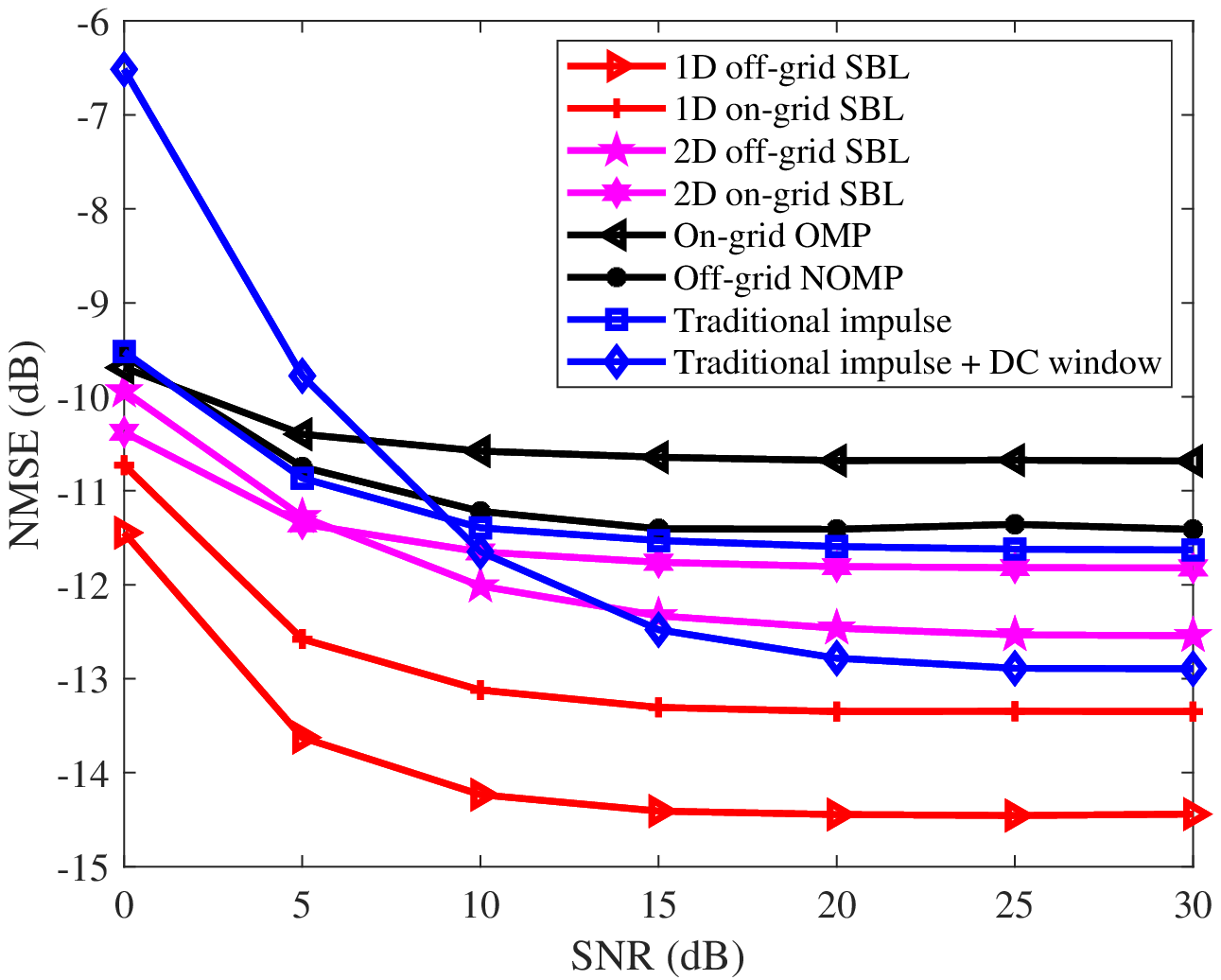}\vspace{-7mm}
		\caption{NMSE of effective channel estimation versus SNR with a low virtual sampling resolution $r_{\nu} = r_{\tau} = 0.8$.}\vspace{-10mm}
		\label{NMSE_WithGuard_LowerResolution}%
	\end{minipage}
\end{figure}

\vspace{-2mm}
\subsection{NMSE versus SNR with Different Virtual Sampling Resolutions}
Fig. \ref{NMSE_WithGuard_HighResolution} and Fig. \ref{NMSE_WithGuard_LowerResolution} illustrate the performance of our proposed channel estimation schemes with high and low virtual sampling resolutions, respectively.
In the case of a high virtual sampling resolution with $r_{\nu} = r_{\tau} = 0.5$, we can observe that our proposed 1D off-grid SBL-based channel estimation achieves the lowest NMSE and the performance gain over the on-grid method is about 1.5 dB in the high SNR regime.
Moreover, the proposed 2D off-grid/on-grid SBL-based schemes achieve a coarser channel estimation than the proposed 1D methods due to potential estimation error propagation from the first step to the second step in \textbf{Algorithm} \ref{alg2}.
However, owing to the decoupling of our proposed 2D model, the computational complexity is significantly reduced.
Note that the 2D on-grid SBL-based scheme outperforms the 2D off-grid SBL-based scheme.
This is because the non-linear coupling of the hyper-parameters $\left({{\boldsymbol{\kappa }}_\nu },\boldsymbol{\iota}_\tau\right)$ in the measurement matrices in our proposed 2D off-grid model in \eqref{IOOTFS2DApprox} may reduce the gain offered by estimating the off-grid components for a high virtual sampling resolution.
Additionally, the OMP-based methods, including both ``On-grid OMP'' and ``Off-grid NOMP'', achieve a better NMSE performance than the impulse-based methods, including both ``Traditional impulse'' and ``Traditional impulse + DC window'', since the former methods can exploit the channel sparsity in the DD domain.
However, the OMP-based methods perform worse than our proposed SBL-based methods as OMP is heuristic and is generally far from the optimal solution.
In fact, our proposed SBL-based methods are derived from the maximum a posteriori probability estimate in \eqref{PosteriorDistributionFull} which can effectively capture the main properties of the channel for estimation.
Although the NOMP method \cite{LiPDMAOTFS} can refine the delay and Doppler shifts estimation through Newton's method and achieves a 2 dB NMSE gain compared with the on-grid OMP method \cite{ShenCEMassiveMIMO}, its NMSE is still substantially higher than that of our proposed SBL-based schemes owing to the superiority of the SBL framework.
Note that our previous work of DC window-based channel estimation \cite{WeiWindowOTFS} can achieve a better performance than the traditional impulse method in the high SNR regime thanks to the improved channel sparsity obtained by windowing.
However, the work in \cite{WeiWindowOTFS} still suffers from a high NMSE compared with the compressed sensing-based channel estimation approaches.
In fact, the work in \cite{WeiWindowOTFS} estimates the effective DD domain channel instead of the original DD domain channel response, where the sparsity of the former channel is sacrificed compared with the latter one due to inevitable channel spreading.

For the case of a low virtual sampling resolution with $r_{\nu} = r_{\tau} = 0.8$, we can observe a similar performance trend as Fig. \ref{NMSE_WithGuard_HighResolution} for the compressed sensing-based channel estimation schemes.
The only difference is that for our proposed 2D model, the off-grid method outperforms the on-grid method since the former can estimate the off-grid components of delay and Doppler shifts and thus can achieve a high channel estimation accuracy.
We can observe that the OMP-based methods are even worse than the impulse-based approaches as the former suffer from the low virtual sampling resolution.
However, even with a low virtual sampling resolution, our proposed 1D off-grid SBL-based channel estimation can still achieve the best performance.

\begin{figure}[t]
	\begin{minipage}{.47\textwidth}
		\centering\vspace{-3mm}
		\includegraphics[width=3.5in]{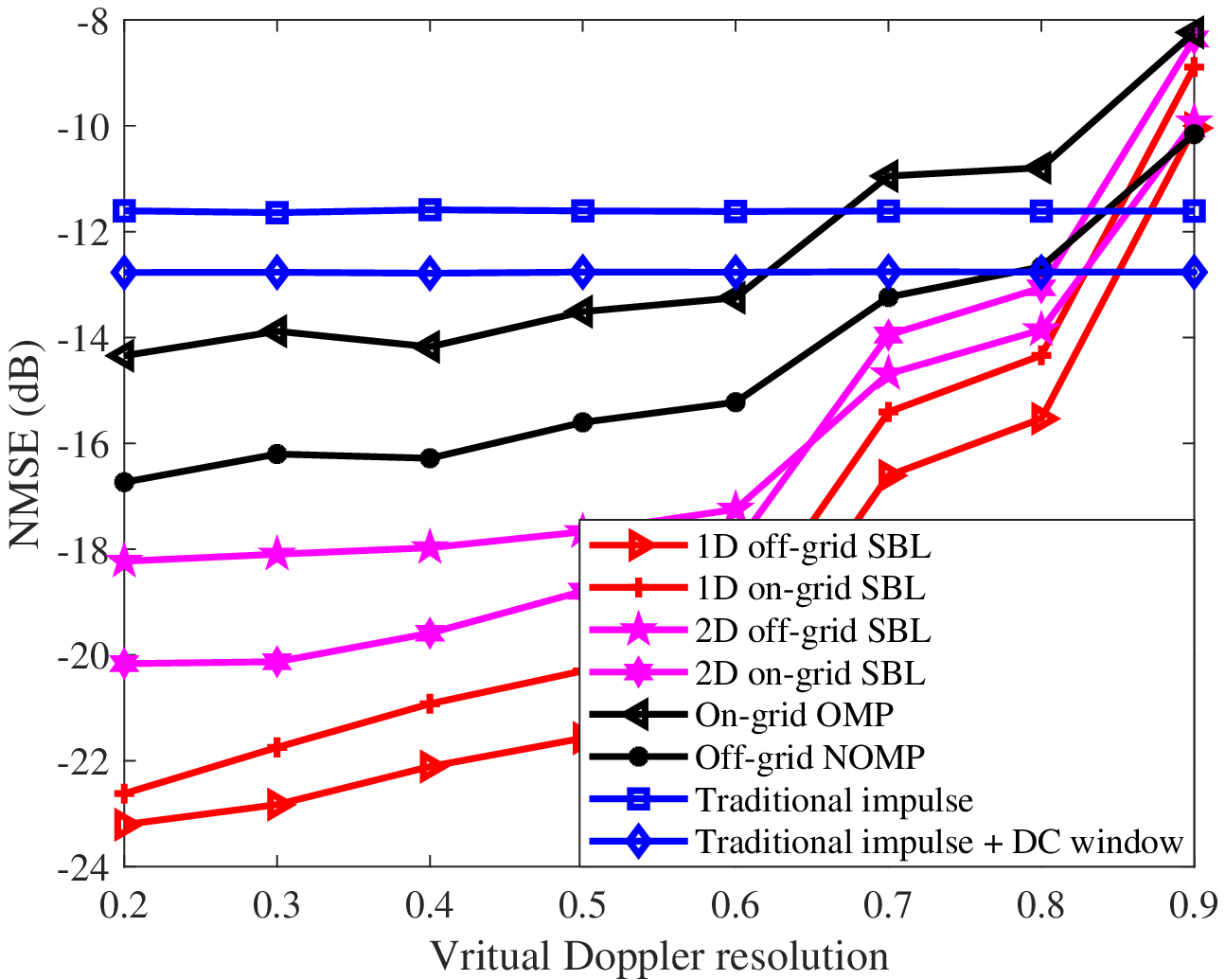}\vspace{-7mm}
		\caption{NMSE of effective channel estimation versus the virtual resolution.}\vspace{-10mm}
		\label{NMSE_WithGuard_Resolution}
	\end{minipage}
	\hspace{2mm}
	\begin{minipage}{.47\textwidth}
		\centering\vspace{-3mm}
		\includegraphics[width=3.5in]{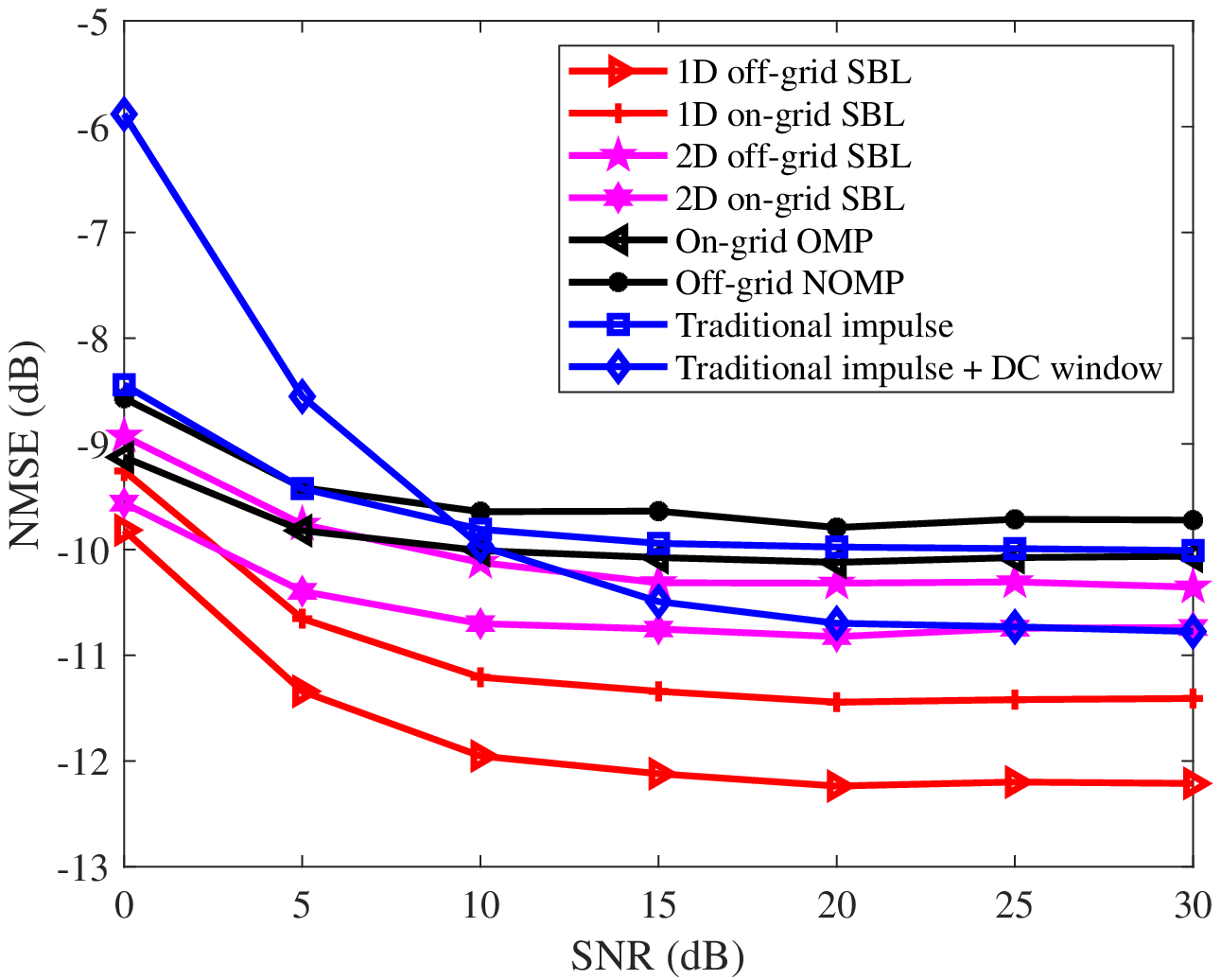}\vspace{-7mm}
		\caption{NMSE of effective channel estimation versus SNR without guard space for $r_{\nu} = r_{\tau} = 0.8$.}\vspace{-10mm}
		\label{NMSE_WithoutGuard}%
	\end{minipage}
\end{figure}

\vspace{-2mm}
\subsection{NMSE versus the Virtual Sampling Resolution}
To investigate the impact of the virtual sampling resolution, Fig. \ref{NMSE_WithGuard_Resolution} illustrates the channel estimation performance versus the virtual sampling resolution in the Doppler domain with a fixed virtual delay sampling resolution $r_{\tau} = 0.5$.
We can observe that the NMSE of the compressed sensing-based channel estimation increases with decreasing the virtual Doppler resolution from $r_{\nu} = 0.2$ to $r_{\nu} = 0.9$.
In contrast, the NMSE of the impulse-based channel estimation methods is invariant since it estimates the effective DD domain channel which does not depend on the virtual sampling resolution.
However, the computational complexity of the compressed sensing-based approaches increases significantly with  the increasing virtual sampling resolution. 
For our proposed 2D model, the on-grid approach outperforms the off-grid method with a high virtual Doppler resolution while the former scheme is worse than the latter one with a low virtual Doppler resolution.
In fact, the 2D on-grid SBL-based scheme becomes more accurate for a high virtual sampling resolution and enjoys a simpler estimation model with fewer unknown parameters.
With reducing the virtual sampling resolution, i.e., increasing $r_{\nu}$, the NMSE of the proposed 2D on-grid SBL-based scheme becomes higher since all the estimated delay and Doppler shifts are constrained on the coarse DD domain grid.
In contrast, our proposed off-grid approach can compensate the loss of the low virtual sampling resolution 
through estimating the off-grid components as hyper-parameters and thus enjoys a more accurate channel estimation.

\vspace{-2mm}
\subsection{NMSE versus SNR without Guard Space}
In practice, guard spaces are introduced to mitigate the interference from data symbols to the pilot symbol.
However, the insertion causes a significantly high signaling overhead for OTFS.
In particular, nulling the DD domain space in the range of ${k_p} - 2{k_{\max }} \le k \le {k_p} + 2{k_{\max }}$ and ${l_p} - {l_{\mathrm{max}} } \le l \le {l_p} + {l_{\mathrm{max}} }$ results in $\left(4{k_{\max }}+1\right)\left(2{l_{\mathrm{max}} } + 1\right)$ overhead symbols out of $MN$ symbols.
Therefore, it is interesting to investigate the robustness of our proposed channel estimation schemes against the interference via evaluating the channel estimation performance without the use of guard space.
Fig. \ref{NMSE_WithoutGuard} demonstrates the NMSE of channel estimation without guard space for  $r_{\nu} = r_{\tau} = 0.8$.
We can observe that our proposed 1D off-grid SBL-based channel estimation can achieve the lowest NMSE and possesses about 1 dB gain compared to that of 1D on-grid SBL-based channel estimation, which is consistent to previous results.
Besides, it is worth to note that the on-grid methods outperform the off-grid methods for both OMP-based and 2D SBL-based channel estimation schemes.
This is because the on-grid methods are based on a linear SSR model while the off-grid methods are based on a non-linear model, e.g. \eqref{IOOTFS2DApprox}.
As a result, the on-grid methods are more robust against the interference than that of the off-grid methods.
Also, comparing Fig. \ref{NMSE_WithGuard_LowerResolution} and Fig. \ref{NMSE_WithoutGuard}, it can be seen that there is about $1 \sim 2$ dB loss in channel estimation NMSE for all the evaluated schemes due to the higher amount of interference from data symbols without guard space.
This motivates us to further employ the proposed data-aided channel estimation scheme in next subsection.

\begin{figure}
	\centering\vspace{-3mm}
	\includegraphics[width=4in]{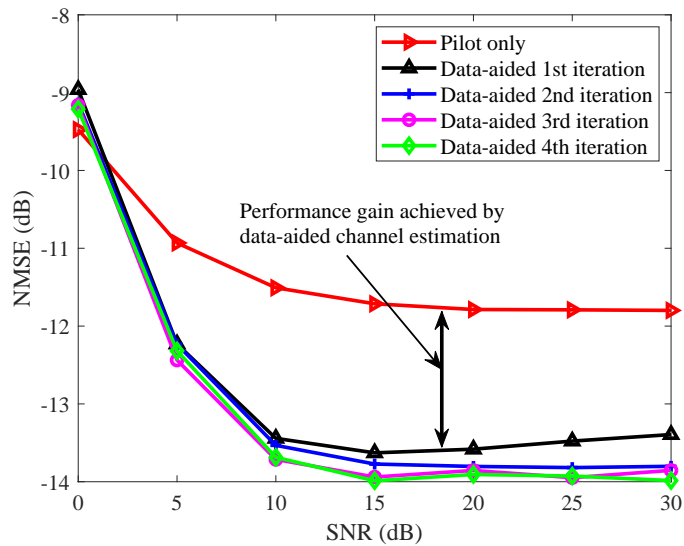}\vspace{-7mm}
	\caption{NMSE of effective channel estimation versus SNR with data-aided channel estimation.}\vspace{-10mm}
	\label{NMSE_WithoutGuard_DataAided}%
\end{figure}

\vspace{-2mm}
\subsection{Data-Aided Channel Estimation}
Fig. \ref{NMSE_WithoutGuard_DataAided} illustrates the NMSE performance for different iterations in our proposed 1D off-grid SBL-based channel estimation adopting the data-aided philosophy\cite{MaDataAidedCE}.
We adopt binary phase shift keying (BPSK) modulation for $x\left[k,l\right]$ and the sum-product algorithm proposed in our previous work \cite{WeiWindowOTFS} for data detection.
To improve the data detection performance, a half-rate convolutional code and the maximum likelihood sequence estimation are applied at OTFS transceiver.
We can observe that about 1.5 dB decreases in NMSE can be achieved by the first data-aided iteration, while the performance improvement is marginal for the remaining iterations.
In fact, the first iteration can retrieve most of the unknown data symbols and provides the most significant information to reconstruct the measurement matrix in \eqref{SSRDModel}.
The following iterations can further improve the accuracy of data detection based on the refined channel estimates.
However, the additional information provided from the refined channel estimation is marginal compared with the first iteration.
Considering the high computational complexity of data detection and channel estimation, we only need the first iteration to achieve an acceptable channel estimation performance in practice.

\vspace{-3mm}
\section{Conclusions}
In this paper, we proposed an off-grid SBL-based channel estimation scheme, which estimates the original DD domain channel response instead of the effective DD domain channel.
In particular, the channel estimation problem was reformulated as 1D and 2D off-grid SSR problems adopting linear approximation, where the on-grid and off-grid components of both the delay and Doppler shifts are separated for estimation.
Based on the SBL framework, we developed 1D and 2D off-grid SBL-based channel estimation algorithms, where the on-grid components of the delay and Doppler shifts are estimated based on the recovered channel vector and the corresponding off-grid components are estimated by the proposed EM algorithm.
Extensive simulations have been conducted to demonstrate the superior channel estimation performance of our proposed schemes and revealed some interesting insights.
Particularly, (1) compared with estimating the effective DD domain channel, estimating the original DD domain channel response can avoid channel spreading due to fractional delay and Doppler shifts and thus can effectively exploit the DD domain channel sparsity; 
(2) compared with the on-grid methods, the proposed off-grid channel compressed sensing scheme can further improve the channel estimation accuracy of OTFS systems even with a low virtual sampling resolution while only introduces a slightly higher computational complexity;
(3) the proposed 1D off-grid SBL-based channel estimation scheme can achieve a superior performance while the proposed 2D off-grid scheme enjoys a much lower computational complexity via decoupling the delay and Doppler estimations;
(4) further adopting the data-aided channel estimation in our proposed scheme can improve its robustness against the interference caused by unknown data symbols and thus can further improve the channel estimation performance even in the absence of guard space.

\vspace{-2mm}
\section*{Appendix: Proof of Proposition \ref{Propo1}}\label{AppendixA}
To update ${\boldsymbol{\alpha }}$, the objective function is obtained by substituting the  items related to ${\boldsymbol{\alpha }}$ in \eqref{JointDistribution} into \eqref{MaximizationStepI}, yielding
\vspace{-2mm}
\begin{align}
&Q\left( {{\boldsymbol{\alpha }}\left| {{{\boldsymbol{\alpha }}^{\left( t \right)}},{\boldsymbol{\kappa }}_\nu ^{\left( t \right)},{\boldsymbol{\iota }}_\tau ^{\left( t \right)},\beta _0^{\left( t \right)}} \right.} \right) = {E_{\overline {\bf{h}} \left| {{\bf{y}}_{\mathrm{T}};{{\boldsymbol{\alpha }}^{\left( t \right)}},{\boldsymbol{\kappa }}_\nu ^{\left( t \right)},{\boldsymbol{\iota }}_\tau ^{\left( t \right)},\beta _0^{\left( t \right)}} \right.}}\left\{ {\ln \left[ {p\left( {\overline {\bf{h}} \left| {\boldsymbol{\alpha }} \right.} \right)p\left( {\boldsymbol{\alpha }} \right)} \right]} \right\} \label{ObjectiveFunctionAlpha}\\[-1mm]
=& {E_{\overline {\bf{h}} \left| {{\bf{y}}_{\mathrm{T}};{{\boldsymbol{\alpha }}^{\left( t \right)}},{\boldsymbol{\kappa }}_\nu ^{\left( t \right)},{\boldsymbol{\iota }}_\tau ^{\left( t \right)},\beta _0^{\left( t \right)}} \right.}}\left\{ { - \sum\limits_{k'' = 0}^{N_{\nu} - 1} {\sum\limits_{l'' = 0}^{M_{\tau} - 1} {\ln \left( {{\alpha _{k''M_{\tau} + l''}}} \right)}  - {{\overline {\bf{h}} }^{\rm{H}}}{{\boldsymbol{\Lambda }}^{ - 1}}\overline {\bf{h}}  - \rho \sum\limits_{k'' = 0}^{N_{\nu} - 1} {\sum\limits_{l'' = 0}^{M_{\tau} - 1} {{\alpha _{k''M_{\tau} + l''}}} } } } \right\}.\notag
\end{align}
\vspace{-8mm}\par\noindent
We can observe that the objective function in \eqref{ObjectiveFunctionAlpha} is neither convex nor concave w.r.t. ${\boldsymbol{\alpha }}$.
In general, there is no systematic and computational efficient solution to maximize $\eqref{ObjectiveFunctionAlpha}$.
As a compromise, we propose a closed-from updating rule of ${\boldsymbol{\alpha }}$, which is computational efficient and is guaranteed to converge to a stationary point of \eqref{ObjectiveFunctionAlpha}, as commonly done in the literature\cite{YangZaiOffgridCE}.
In particular, taking the derivative of $Q\left( {{\boldsymbol{\alpha }}\left| {{{\boldsymbol{\alpha }}^{\left( t \right)}},{\boldsymbol{\kappa }}_\nu ^{\left( t \right)},{\boldsymbol{\iota }}_\tau ^{\left( t \right)},\beta _0^{\left( t \right)}} \right.} \right)$ w.r.t. ${\alpha _{k''M_{\tau} + l''}}$, we have
\vspace{-2mm}
\begin{align}
\frac{{\partial Q\left( {{\boldsymbol{\alpha }}\left| {{{\boldsymbol{\alpha }}^{\left( t \right)}},{\boldsymbol{\kappa }}_\nu ^{\left( t \right)},{\boldsymbol{\iota }}_\tau ^{\left( t \right)},\beta _0^{\left( t \right)}} \right.} \right)}}{{\partial {\alpha _{k''M_{\tau} + l''}}}} &=  - \frac{1}{{{\alpha _{k''M_{\tau} + l''}}}} \notag\\[-1mm]
&+ \frac{1}{{\alpha _{k''M_{\tau} + l''}^2}}\{{{\left| {\{{{\boldsymbol{\mu }}^{\left(t\right)}_{\overline {\bf{h}} }}\}_{k''M_{\tau} + l''}} \right|}^2} + \{{{\boldsymbol{\Sigma }}^{\left(t\right)}_{\overline {\bf{h}} }}\}_{k''M_{\tau} + l'',k''M_{\tau} + l''}\} - \rho.
\end{align}
\vspace{-6mm}\par\noindent
Imposing $\frac{{\partial Q\left( {{\boldsymbol{\alpha }}\left| {{{\boldsymbol{\alpha }}^{\left( t \right)}},{\boldsymbol{\kappa }}_\nu ^{\left( t \right)},{\boldsymbol{\iota }}_\tau ^{\left( t \right)},\beta _0^{\left( t \right)}} \right.} \right)}}{{\partial {\alpha _{k''M_{\tau} + l''}}}} = 0$, the updating rule of ${\alpha _{k''M_{\tau} + l''}}$ can be obtained by \eqref{HyperParameterI}.

To update ${\beta _0}$, the objective function in \eqref{MaximizationStepIV} can be obtained by
\vspace{-2mm}
\begin{align}
& Q\left( {\beta _0\left| {{{\boldsymbol{\alpha }}^{\left( t \right)}},{\boldsymbol{\kappa }}_\nu ^{\left( t \right)},{\boldsymbol{\iota }}_\tau ^{\left( t \right)},\beta _0^{\left( t \right)}} \right.} \right) = {E_{\overline {\bf{h}} \left| {{\bf{y}}_{\mathrm{T}};{{\boldsymbol{\alpha }}^{\left( t \right)}},{\boldsymbol{\kappa }}_\nu ^{\left( t \right)},{\boldsymbol{\iota }}_\tau ^{\left( t \right)},\beta _0^{\left( t \right)}} \right.}}\left\{ {\ln \left[ {p\left( {{\bf{y}}_{\mathrm{T}}\left| {\overline {\bf{h}} ,{{\boldsymbol{\kappa }}_\nu^{\left( t \right)} },{{\boldsymbol{\iota }}_\tau^{\left( t \right)} },{\beta _0}} \right.} \right)p\left( {{\beta _0}} \right)} \right]} \right\}\notag\\[-1mm]
=& {E_{\overline {\bf{h}} \left| {{\bf{y}}_{\mathrm{T}};{{\boldsymbol{\alpha }}^{\left( t \right)}},{\boldsymbol{\kappa }}_\nu ^{\left( t \right)},{\boldsymbol{\iota }}_\tau ^{\left( t \right)},\beta _0^{\left( t \right)}} \right.}}\left\{ \ln \left[ {{\cal C}{\cal N}\left( {{\bf{y}}_{\mathrm{T}}\left| {\overline {\boldsymbol{\Phi }}_{\mathrm{T}} \left( {{{\boldsymbol{\kappa }}_\nu^{\left( t \right)} },{{\bf{\iota }}_\tau^{\left( t \right)} }} \right)\overline {\bf{h}} ,\beta _0^{ - 1}{{\bf{I}}_{M_{\mathrm{T}}N_{\mathrm{T}}}}} \right.} \right)} \right] + \ln \left[ {\Gamma \left( {{\beta _0}\left| {c,d} \right.} \right)} \right] \right\} \notag\\[-1mm]
=& \left( {c - 1 + M_{\mathrm{T}}N_{\mathrm{T}}} \right)\ln \left( {{\beta _0}} \right) - \left( {d + {A_{{\beta _0}}}} \right){\beta _0},
\end{align}
\vspace{-8mm}\par\noindent
where ${A_{{\beta _0}}}$ is given by \eqref{Abeta0}.
Taking derivative of $Q\left( {\beta _0\left| {{{\boldsymbol{\alpha }}^{\left( t \right)}},{\boldsymbol{\kappa }}_\nu ^{\left( t \right)},{\boldsymbol{\iota }}_\tau ^{\left( t \right)},\beta _0^{\left( t \right)}} \right.} \right)$ w.r.t. $\beta_0$ and imposing $\frac{{\partial Q\left( {\beta _0\left| {{{\boldsymbol{\alpha }}^{\left( t \right)}},{\boldsymbol{\kappa }}_\nu ^{\left( t \right)},{\boldsymbol{\iota }}_\tau ^{\left( t \right)},\beta _0^{\left( t \right)}} \right.} \right)}}{{\partial {\alpha _{k''M_{\tau} + l''}}}} = 0$, we can obtain the updating rule for $\beta_0$ as \eqref{HyperParameterIV}.

\bibliographystyle{IEEEtran}
\bibliography{OTFS}

\begin{thebibliography}{10}
\providecommand{\url}[1]{#1}
\csname url@samestyle\endcsname
\providecommand{\newblock}{\relax}
\providecommand{\bibinfo}[2]{#2}
\providecommand{\BIBentrySTDinterwordspacing}{\spaceskip=0pt\relax}
\providecommand{\BIBentryALTinterwordstretchfactor}{4}
\providecommand{\BIBentryALTinterwordspacing}{\spaceskip=\fontdimen2\font plus
\BIBentryALTinterwordstretchfactor\fontdimen3\font minus
  \fontdimen4\font\relax}
\providecommand{\BIBforeignlanguage}[2]{{%
\expandafter\ifx\csname l@#1\endcsname\relax
\typeout{** WARNING: IEEEtran.bst: No hyphenation pattern has been}%
\typeout{** loaded for the language `#1'. Using the pattern for}%
\typeout{** the default language instead.}%
\else
\language=\csname l@#1\endcsname
\fi
#2}}
\providecommand{\BIBdecl}{\relax}
\BIBdecl

\bibitem{wei2020orthogonal}
Z.~Wei, W.~Yuan, S.~Li, J.~Yuan, G.~Bharatula, R.~Hadani, and L.~Hanzo,
  ``Orthogonal time-frequency space modulation: A full-diversity next
  generation waveform,'' \emph{arXiv preprint arXiv:2010.03344}, 2020.

\bibitem{Hadani2017orthogonal}
R.~Hadani, S.~Rakib, M.~Tsatsanis, A.~Monk, A.~J. Goldsmith, A.~F. Molisch, and
  R.~Calderbank, ``Orthogonal time frequency space modulation,'' in \emph{Proc.
  IEEE Wireless Commun. and Networking Conf.}, 2017, pp. 1--6.

\bibitem{li2020performance}
S.~Li, J.~Yuan, W.~Yuan, Z.~Wei, B.~Bai, and D.~W.~K. Ng, ``Performance
  analysis of coded {OTFS} systems over high-mobility channels,'' \emph{arXiv
  preprint arXiv:2010.13008}, 2020.

\bibitem{RavitejaOTFSCE}
P.~{Raviteja}, K.~T. {Phan}, and Y.~{Hong}, ``Embedded pilot-aided channel
  estimation for {OTFS} in delay-doppler channels,'' \emph{IEEE Trans. Veh.
  Technol.}, vol.~68, no.~5, pp. 4906--4917, May 2019.

\bibitem{KollengodeMIMOOTFSDetectionCE}
M.~{Kollengode Ramachandran} and A.~{Chockalingam}, ``{MIMO-OTFS} in
  high-doppler fading channels: Signal detection and channel estimation,'' in
  \emph{Proc. IEEE Global Commun. Conf.}, Dec. 2018, pp. 206--212.

\bibitem{WeiWindowOTFS}
Z.~Wei, W.~Yuan, S.~Li, J.~Yuan, and D.~W.~K. Ng, ``Transmitter and receiver
  window designs for orthogonal time frequency space modulation,'' \emph{arXiv
  preprint arXiv:1709.02505}, 2020.

\bibitem{ShenCEMassiveMIMO}
W.~{Shen}, L.~{Dai}, J.~{An}, P.~{Fan}, and R.~W. {Heath}, ``Channel estimation
  for orthogonal time frequency space {(OTFS)} massive {MIMO},'' \emph{IEEE
  Trans. Signal Process.}, vol.~67, no.~16, pp. 4204--4217, Jul. 2019.

\bibitem{LiPDMAOTFS}
M.~{Li}, S.~{Zhang}, F.~{Gao}, P.~{Fan}, and O.~A. {Dobre}, ``A new path
  division multiple access for the massive {MIMO-OTFS} networks,'' \emph{IEEE
  J. Select. Areas Commun.}, early access, 2020.

\bibitem{ZhaoSBLOTFS}
L.~{Zhao}, W.~J. {Gao}, and W.~{Guo}, ``Sparse {Bayesian} learning of
  delay-doppler channel for {OTFS} system,'' \emph{IEEE Commun. Lett.},
  vol.~24, no.~12, pp. 2766--2769, 2020.

\bibitem{YangZaiOffgridCE}
Z.~{Yang}, L.~{Xie}, and C.~{Zhang}, ``Off-grid direction of arrival estimation
  using sparse {Bayesian} inference,'' \emph{IEEE Trans. Signal Process.},
  vol.~61, no.~1, pp. 38--43, Jan. 2013.

\bibitem{BabacanBCS}
S.~D. {Babacan}, R.~{Molina}, and A.~K. {Katsaggelos}, ``Bayesian compressive
  sensing using laplace priors,'' \emph{IEEE Trans. Image Process.}, vol.~19,
  no.~1, pp. 53--63, Jan. 2010.

\bibitem{RavitejaOTFS}
P.~{Raviteja}, K.~T. {Phan}, Y.~{Hong}, and E.~{Viterbo}, ``Interference
  cancellation and iterative detection for orthogonal time frequency space
  modulation,'' \emph{IEEE Trans. Wireless Commun.}, vol.~17, no.~10, pp.
  6501--6515, Oct. 2018.

\bibitem{ZhangMUltipleAntenna}
J.~{Zhang}, E.~{Bj\"{o}rnson}, M.~{Matthaiou}, D.~W.~K. {Ng}, H.~{Yang}, and
  D.~J. {Love}, ``Prospective multiple antenna technologies for beyond {5G},''
  \emph{IEEE J. Select. Areas Commun.}, vol.~38, no.~8, pp. 1637--1660, Jun.
  2020.

\bibitem{BajwaCSC}
W.~U. {Bajwa}, J.~{Haupt}, A.~M. {Sayeed}, and R.~{Nowak}, ``Compressed channel
  sensing: A new approach to estimating sparse multipath channels,''
  \emph{Proceed. of the IEEE}, vol.~98, no.~6, pp. 1058--1076, Jun. 2010.

\bibitem{ThomasSBL}
C.~K. {Thomas} and D.~{Slock}, ``Convergence analysis of sparse bayesian
  learning under approximate inference techniques,'' in \emph{53rd Asilomar
  Conference on Signals, Systems, and Computers}, 2019, pp. 764--768.

\bibitem{MoonSPM}
T.~K. {Moon}, ``The expectation-maximization algorithm,'' \emph{IEEE Signal
  Process. Mag.}, vol.~13, no.~6, pp. 47--60, Nov. 1996.

\bibitem{CandesCS}
E.~J. Cand\`es, J.~K. Romberg, and T.~Tao, ``Stable signal recovery from
  incomplete and inaccurate measurements,'' \emph{Communications on Pure and
  Applied Mathematics}, vol.~59, no.~8, pp. 1207--1223, Mar. 2006.

\bibitem{MaDataAidedCE}
J.~{Ma} and L.~{Ping}, ``Data-aided channel estimation in large antenna
  systems,'' \emph{IEEE Trans. Signal Process.}, vol.~62, no.~12, pp.
  3111--3124, 2014.

\bibitem{YuanOTFS}
W.~{Yuan}, Z.~{Wei}, J.~{Yuan}, and D.~W.~K. {Ng}, ``A simple variational
  {Bayes} detector for orthogonal time frequency space {(OTFS)} modulation,''
  \emph{IEEE Trans. Veh. Technol.}, vol.~69, no.~7, pp. 7976--7980, Jul. 2020.

\end{thebibliography}
\end{document}